\documentclass{jfm}
\usepackage{rotating,graphicx}
\usepackage{graphicx}
\usepackage{epstopdf, epsfig}

\usepackage{amsmath}
\usepackage{amssymb}

\usepackage{lscape}

\usepackage[utf8]{inputenc}
\usepackage[T1]{fontenc}
\usepackage{mathptmx}
\usepackage{soul}

\usepackage{microtype}

\usepackage{booktabs} 

\usepackage{graphicx}
\usepackage{caption}
\usepackage{hyperref}
\usepackage{subcaption}
\DeclareMathOperator*{\argmax}{arg\,max}

\usepackage{algpseudocode}
\usepackage{algorithm}

\usepackage{placeins}
\usepackage{listings}
\usepackage[dvipsnames]{xcolor}

\usepackage{lineno}

\usepackage{cleveref}
\newcommand\myshade{85}
\colorlet{mylinkcolor}{violet}
\colorlet{mycitecolor}{Aquamarine}
\colorlet{myurlcolor}{Aquamarine}

\hypersetup{
	linkcolor  = mylinkcolor!\myshade!black,
	citecolor  = mycitecolor!\myshade!black,
	urlcolor   = myurlcolor!\myshade!black,
	colorlinks = true,
}

\definecolor{codegreen}{rgb}{0,0.6,0}
\definecolor{codegray}{rgb}{0.5,0.5,0.5}
\definecolor{codepurple}{rgb}{0.58,0,0.82}
\definecolor{backcolour}{rgb}{0.95,0.95,0.92}

\lstdefinestyle{mystyle}{
	backgroundcolor=\color{backcolour},   
	commentstyle=\color{codegreen},
	keywordstyle=\color{magenta},
	numberstyle=\tiny\color{codegray},
	stringstyle=\color{codepurple},
	basicstyle=\ttfamily\footnotesize,
	breakatwhitespace=false,         
	breaklines=true,                 
	captionpos=b,                    
	keepspaces=true,                 
	numbers=left,                    
	numbersep=5pt,                  
	showspaces=false,                
	showstringspaces=false,
	showtabs=false,                  
	tabsize=2
}

\usepackage{amsmath}
\usepackage{comment}

\shorttitle{Comparative analysis of machine learning methods for active flow control}
\shortauthor{F. Pino, L. Schena, J. Rabault and M.A. Mendez }

\title{Comparative analysis of machine learning methods for active flow control}

\author{Fabio Pino\aff{1}
	\corresp{\email{fabio.pino@vki.ac.be}},
	Lorenzo Schena\aff{1},
	Jean Rabault\aff{2},
	\and Miguel A. Mendez\aff{1}}

\affiliation{\aff{1}von Karman Institute for Fluid Dynamics, EA Department, Sint Genesius Rode, Belgium
	\aff{2} Norwegian Meteorological Institute, Oslo, Norway
}

\begin{document}
	\maketitle
	
	\begin{abstract}
		Machine learning frameworks such as Genetic Programming (GP) and Reinforcement Learning (RL) are gaining popularity in flow control. This work presents a comparative analysis of the two, bench-marking some of their most representative algorithms against global optimization techniques such as Bayesian Optimization (BO) and Lipschitz global optimization (LIPO). {\color{black} First, we review the general framework of the model-free control problem, bringing together all methods as black-box optimization problems.} Then, we test the control algorithms on three test cases. These are (1) the stabilization of a nonlinear dynamical system featuring frequency cross-talk, (2) the wave cancellation from a Burgers' flow and (3) the drag reduction in a cylinder wake flow. We present a comprehensive comparison to illustrate their differences in exploration versus exploitation and their balance between `model capacity' in the control law definition versus `required complexity'. We believe that such a comparison paves the way toward the hybridization of the various methods, and we offer some perspective on their future development in the literature of flow control problems.
	\end{abstract}
	
	
	\section{Introduction}
	
	The multidisciplinary nature of active flow control has attracted interests from many research areas for a long time, \citep{Gunzburger2002,Wang2018,gad-el-hak_2000,bewley2001flow} and its scientific and technological relevance have ever-growing proportions \citep{Brunton2015,Noack_CHAP_BOOK,bewley2001flow}. Indeed, the ability to interact and manipulate a fluid system to improve its engineering benefits is essential in countless problems and applications, including laminar to turbulent transition \citep{Schlichting,lin2002review}, drag reduction  \citep{gad-el-hak_2000,Wang2018}, stability of combustion systems \citep{lang1987active}, flight mechanics \citep{Longuski2014}, wind energy \citep{apata2020overview,munters2018dynamic}, and aeroacoustic noise control \citep{Collis2002,Kim2014}, to name just a few.
	
	The continuous development of computational and experimental tools, together with the advent of data-driven methods from the ongoing machine learning revolution, is reshaping tools and methods in the field \citep{Noack_CHAP_BOOK,noack:hal-02398734}. Nevertheless, the quest for reconciling terminology and methods from the machine learning and the control theory community has a long history (see \cite{Bersini} and \cite{Sutton1992a}) and it is still ongoing, as described in the recent review by \cite{Recht2019} and \cite{Nian2020}. This article aims at reviewing some recent machine learning algorithms for flow control, presenting a unified framework that highlights differences and similarities amidst various techniques. We hope that such a generalization opens the path to hybrid approaches.
	
	In its most abstract formulation, the (flow) control problem is essentially a functional optimization problem constrained by the (fluid) systems’ dynamics  \citep{stengel1994optimal,kirk2004optimal}. As further discussed in Section \ref{Sec:II}, the goal is to find a control function that minimizes (or maximizes) a cost (or reward) functional which measures the controller performances (e.g. drag or noise reduction). Following Wiener's metaphors \citep{Wiener}, active control methods can be classified as white, grey or black depending on how much knowledge about the system is used to solve the optimization: the whiter the approach, the more the control relies on the analytical description of the system to be controlled.

	Machine-learning-based approaches are "black-box" or "model-free" methods. These approaches rely only on input-output data, and knowledge of the system is gathered by interacting with it. By-passing the need for a model (and underlying simplifications), these methods are promising tools for solving problems that are not amenable to analytical treatment or cannot be accurately reproduced in a numerical environment. Machine learning \citep{Moustafa,Mitchell1997,VladimirCherkassky2008,brunton2020machine} is a subset of Artificial Intelligence which combines optimization and statistics to "learn" (i.e. calibrate) models from data (i.e. experience). These models can be general enough to describe any (nonlinear) function without requiring prior knowledge and can be encoded in various forms: examples are parametric models such as Radial Basis Function (RBFs, see \cite{Fasshauer2007}) expansions or Artificial Neural Networks (ANNs, see \cite{Ian}), or tree structures of analytic expressions such as in Genetic Programming (GP, {\color{black} developed by \cite{Koza1994}}). The process by which these models are "fitted" to (or "learned" from) data is an optimization in one of its many forms \citep{Sun2019}: continuous or discrete, global or local, stochastic or deterministic. Within the flow control literature, at the time of writing, the two most prominent model-free control techniques from the machine learning literature are Genetic Programming and Reinforcement Learning \citep{sutton2018reinforcement}. Both are reviewed in this article.
	
	Genetic Programming is an evolutionary computational technique developed as a new paradigm for automatic programming and machine learning \citep{Banzhaf1997,Vanneschi2012}. GP optimizes both the structure and the parameters of a model, which is usually constructed as recursive trees of predefined functions connected through mathematical operations. The use of GP for flow control has been pioneered and popularized by Noack and coworkers \citep{noack:hal-02398734,duriez2017machine}. Successful examples on experimental problems include the drag reduction past bluff bodies \citep{Li2017}, shear flow separation control \citep{Gautier2015,Debien2016,Benard2016} and many more, as reviewed by \cite{noack:hal-02398734}. More recent extensions of this "Machine Learning Control" (MLC) approach, combining genetic algorithms with the down-hill simplex method, have been proposed by \cite{Li2019_Noack} and \cite{Maceda2021}.

	Reinforcement Learning (RL) is one of the three machine learning paradigms and encompasses learning algorithms collecting data "online", in a trial and error process. {\color{black} In Deep RL (DRL), ANNs are used to parametrize the control law or to build a surrogate of the Q function, defining the value of an action at a given state.} The use of an ANN to parametrize control laws has a long history (see \cite{Lee1997}), but their application to flow control, leveraging on RL algorithms, is at its infancy (see also \cite{Li2021} for a recent review). The landscape of RL is vast and grows at a remarkable pace, fostered by the recent success in strategy board games \citep{Silver2016,Silver2018}, video games \citep{Szita2012}, robotics \citep{Kober2014}, language processing \citep{Luketina2019} and more. In the literature of flow control, RL has been pioneered by Komoutsakos and coworkers \citep{Gazzola2014,verma2018efficient} (see also \cite{Garnier2021} and \cite{Rabault_CHAP_BOOK} for more literature). The first applications of RL in fluid mechanics were focused on the study of collective behavior of swimmers \citep{Wang2018,verma2018efficient,novati2017a,novati2019a,novati2019b}, while the first applications for flow control were presented by \cite{pivotreinforcement}, \cite{gueniat2016statistical} and by \cite{rabault2019artificial,Rabault2020,Rabault2019a}. A similar flow control problem has been solved numerically and experimentally via RL by \cite{Fan2020}. \cite{bucci2019control} showcased the use of RL to control chaotic systems such as the one-dimensional Kuramoto–Sivashinsky equation; \cite{beintema2020controlling} used it to control heat transport in a two-dimensional Rayleigh–Bénard systems while \cite{belus2019exploiting} used RL to control the interface of unsteady liquid films. Ongoing efforts in the use of DRL for flow control are focused with increasing the complexity of the analyzed test cases, either by increasing the Reynolds number in academic test cases (see \cite{ren2021applying}), or by considering realistic configurations \citep{fluids7020062}.
	
	In this article, we consider the Deep Deterministic Policy Gradient (DDPG, \cite{Lillicrap2015}) as a representative deterministic RL algorithm. This is introduced in Section \ref{Sec:IIIp3}, and the results obtained for one of the investigated test cases are compared with those obtained by \cite{Tang2020e} using a stochastic RL approach, namely the Proximal Policy Optimization (PPO) \cite{Schulman2017}. 
	
	This work puts GP and RL in a global control framework and benchmarks their performances against simpler black-box optimization methods. Within this category, we include model-free control methods in which the control action is predefined and prescribed by a few parameters (e.g a simple linear controller), and the model learning is driven by global black-box optimization. This approach, using Genetic Algorithms, has a long history \citep{fleming1993genetic}. However, we here focus on more sample efficient alternatives such as the Bayesian Optimization (BO) and the LiPschitz global Optimization technique (LIPO). Both are described in Section \ref{Sec:IIIp1}.
	
	The BO is arguably the most popular "surrogate-based", derivative-free, global optimization tool, popularized by \cite{Jones1998} and their Efficient Global Optimization (EGO) algorithm. In its most classic form \citep{Forrester2008,archetti2019bayesian}, the BO uses a Gaussian process \citep{CarlEdward} for regression of the cost function under evaluation and an acquisition function to decide where to sample next. This method has been used by \cite{mahfoze2019reducing} for reducing the skin-friction drag in a turbulent boundary layer and by \cite{blanchard2022bayesian}
	for reducing the drag in the fluidic pinball and for enhancing mixing in a turbulent jet. 
	
	The LIPO algorithm is a more recent global optimization strategy proposed by \cite{malherbe2017global}. This is a sequential procedure to optimize a function under the only assumption that it has a finite Lipschitz constant. Since this method has virtually no hyper-parameters involved, variants of the LIPO are becoming increasingly popular in hyper-parameter calibration of machine learning algorithms \citep{Ahmed2020}, but to the authors' knowledge it has never been tested on flow control applications. 
	
	All of the aforementioned algorithms are analyzed on three test cases of different dimensions and complexity.
	The first test case is the 0D model proposed by \cite{duriez2017machine} as the simplest dynamical system reproducing the frequency cross-talk encountered in many turbulent flows. The second test case is the control of nonlinear travelling waves described by the 1D Burgers' equation. This test case is representative of the challenges involved in the control of advection-diffusion problems. Moreover, recent works on Koopman analysis by \cite{Page2018} and \cite{Balabane2021} have provided a complete analytical linear decomposition of the Burgers' flow and might render this test case more accessible to "white-box" control methods. Finally, the last selected test case is arguably the most well known benchmark in flow control: the drag attenuation in the flow past a cylinder. This problem has been tackled by nearly the full spectra of control methods in the literature, including reduced order models and linear control \citep{Seidel2008,Bergmann2005,Park1994}, resolvent-based feedback control \citep{Jin2020}, reinforcement learning via stochastic \citep{rabault2019artificial} and deterministic algorithms \citep{Fan2020}, reinforcement learning assisted by stability analysis \citep{Li2021} {\color{black}and recently also GP \citep{castellanos2022machine}}.
	
	We here benchmark both methods on the same test cases against classic black-box optimization. Emphasis is given to the different precautions these algorithms require, the number of necessary interactions with the environment, the different approaches to balance exploration and exploitation, and the differences (or similarities) in the derived control laws. The remaining of the article is structured as follows. Section \ref{Sec:II} recalls the conceptual transition from optimal control theory to machine learning control. Section \ref{Sec:III} briefly recalls the machine learning algorithm analyzed in this work, while Section \ref{Sec:IV} describes the introduced test cases. Results are collected in Section \ref{Sec:V} while conclusions and outlooks are given in Section \ref{Sec:VI}.

	\section{From optimal control to machine learning}
	\label{Sec:II}
	{\color{black} An optimal control problem consists in finding a \textit{control action} $\mathbf{a}(t)\in\mathcal{A}$, within a feasible set $\mathcal{A}\subseteq{{R}^{n_a}}$, which optimizes a functional measuring our ability to keep a \emph{plant} in control theory and an \emph{environment} in reinforcement learning close to the desired states or conditions. The functional is usually a cost to minimize in control theory and a payoff to maximize in reinforcement learning. We follow the second and denote the reward function as $R(\mathbf{a})$. The optimization is constrained by the plant/environment's dynamic:}
	
	\begin{equation}
		\label{optimal_control_P}
		\begin{aligned}
			\max_{\mathbf{a}(t)\in\mathcal{A}} \quad & R(\mathbf{a})=\phi(\mathbf{s}(T)) + \int_{0}^{T} \, \mathcal{L}(\mathbf{s}(\tau),\mathbf{a}(\tau),\tau)\; d\tau,\\
			\textrm{s.t.} \quad &\begin{cases}
				\dot{\textbf{s}}(t) &= \mathbf{f}(\textbf{s}(t),\textbf{a}(t),t) \quad t\in(0,T]\\ \textbf{s}(0)&= \,\textbf{s}_0
			\end{cases}\,,
		\end{aligned}
	\end{equation} where $\mathbf{f}:\mathbb{R}^{n_s}\times \mathbb{R}^{n_a}\rightarrow \mathbb{R}^{n_s}$
	is the vector field in the phase space of the dynamical system and $\mathbf{s}\in\mathbb{R}^{n_s}$ is the system's \textit{state} vector. The action is taken by an \emph{controller} in optimal control and an \emph{agent} in reinforcement learning.
	
	The functional $R(\mathbf{a})$ comprises a \textit{running cost} (or Lagrangian) $\mathcal{L}:\mathbb{R}^{n_{s}}\times\mathbb{R}^{n_{a}}\rightarrow\mathbb{R}$, which accounts for the system's states evolution, and a \textit{terminal cost} (or Mayer term) $\phi:\mathbb{R}^{n_s}\rightarrow\mathbb{R}$, which depends on the final state condition. Optimal control problems with this cost functional form are known as Bolza problem \citep{stengel1994optimal,Lawrence,kirk2004optimal}.
	
	In closed-loop control, the agent/controller selects the action/actuation from a feedback \textit{control law} or \textit{policy} $\pi: \mathbb{R}^{n_s}\rightarrow \mathbb{R}^{n_a}$ of the kind $\mathbf{a}(t)=\pi (\mathbf{s}(t)) \in \mathbb{R}^{n_a}\,$ whereas in open-loop control the action/actuation is independent from the system states, i.e. $    \mathbf{a}(t)=\pi (t) \in \mathbb{R}^{n_a}$.
	{\color{black} One could opt for a combination of the two and consider a control law/policy $\pi: \mathbb{R}^{n_s+1}\rightarrow \mathbb{R}^{n_a}$  of the kind $\mathbf{a}(t)=\pi (\mathbf{s}(t),t) \in \mathbb{R}^{n_a}$}.
	
	{\color{black} All model-free methods seek to convert the variational problem in \eqref{optimal_control_P} into an optimization problem using function approximators such as tables or parametric models. Some authors treated the machines learning control as a regression problem \citep{duriez2017machine} and others as a dynamic programming problem \citep{bucci2019control}. We here consider the more general framework of black-box optimization, which can be tackled with a direct or indirect approach (see Figure 1). }
	
	{\color{black} In the black-box optimization setting, the function to optimize is unknown and the optimization relies on the sampling of the cost function. Likewise, the equations governing the environment/plant are unknown in model-free control techniques and the controller design solely
		relies on trial and error.
		We define the discrete version of \eqref{optimal_control_P} by considering a uniform time discretization $t_k=k\Delta t$ in the interval $t\in[0,T]$, leading to $N=T/\Delta t+1$ points indexed as $k=0,\dots N-1$. Introducing the notation $\mathbf{s}_k=\mathbf{s}(t_k)$, we collect a sequence of states $\mathbf{S}:=\{\mathbf{s}_1,\mathbf{s}_2\dots \mathbf{s}_N\}$ while taking a sequence of actions $\mathbf{A}^{\pi}:=\{\mathbf{a}_1,\mathbf{a}_2\dots \mathbf{a}_N\}$. Collecting also the reward $\mathcal{L}(\mathbf{s}_k,\mathbf{a}_k,k)$, each state-action pair allows for defining the sampled reward as }
	
	\begin{equation}
		\label{REW}
		R(\mathbf{A}^{\pi})=\phi(\mathbf{s}_N)+\sum^{N-1}_{k=0}\mathcal{L}(\mathbf{s}_k,\mathbf{a}^{\pi}_k,k)\,,
	\end{equation} where $N$ is the number of interactions with the systems and defines the length of an \emph{episode}, within which performances are evaluated.
	In the RL literature, this is known as cumulative reward and the Lagrangian takes the form $\mathcal{L}(\mathbf{s}_k,\mathbf{a}^{\pi}_k,k)=\gamma^{k} r(\mathbf{s}_k,\mathbf{a}^{\pi}_k)=\gamma^{k} r^{\pi}_k$, where $\gamma\in[0,1]$ is a discount factor to prioritize immediate over future rewards.

	\begin{figure}
		\centering
		\subfloat[]{\includegraphics[width=0.4\textwidth]{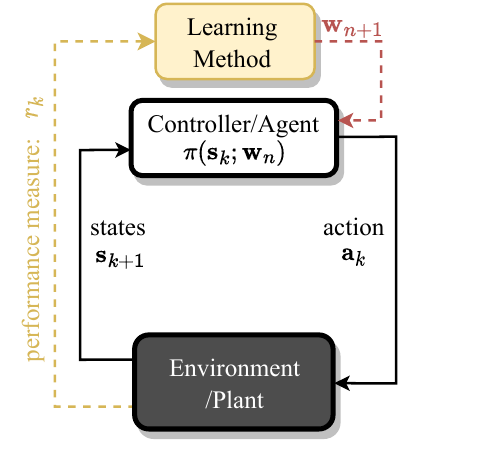}}
		\hfill
		\subfloat[]{\includegraphics[width=0.6\textwidth]{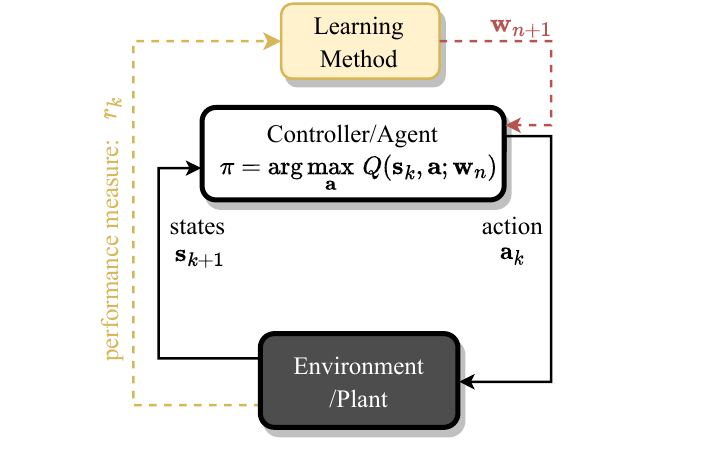}}
		\caption{General setting for a machine learning-based control problem: the learning algorithm (optimizer) improves the agent/control performances while this interacts with the environment/plant. {\color{black} Here $k$ spans the number of interactions within an episode and $n$ spans the number of episodes during the training}. {\color{black} A function approximator is used for the actuation policy in a) and for the state-value function in b)}. In both cases, the control problem is an optimization problem for the parameters $\mathbf{w}$.}
		\label{learning_scheme}
	\end{figure}

	{\color{black}
		The direct approach (Figure 1a) consists in learning an approximation of the optimal policy from the data collected. In the RL literature, these methods are referred to as `on-policy' {\color{black} if the samples are collected following the control policy and `off-policy' if these are collected following a \textit{behavioral policy} that might significantly differ from the control policy.}
		{\color{black}
			Focusing on deterministic policies, the function approximation can take the form of a parametric function $\mathbf{a}^\pi=\pi(\mathbf{s};\mathbf{w})$, where $\mathbf{w}\in\mathbb{R}^{n_w}$ is the set of (unknown) weights that must be \emph{learned}. On the other hand, in a stochastic policy the parametric function outputs the parameters of the distribution (e.g. mean and standard deviation in a Gaussian) from which the actions will be sampled. In either case, the cumulative reward is now a function of the weights controlling the policy and the learning is the iterative process that leads to larger $R(\mathbf{w}_n)$ episode after episode (cf. Figure 1a). The update of the weights can be carried out at each interaction $k$ or at each episode $n$. Moreover, one might simultaneously train multiple versions of the same parametrization (i.e. advance multiple candidates at the same time) and seek to improve the policy by learning from the experience of all candidates . In multi-agent RL, the various agents (candidates) could cooperate or compete \citep{Busoniu2010,Lowe2017}.}}
	
	{\color{black}
		In the classic GP approach to model-free control \citep{duriez2017machine}, the function approximation is built via expression trees and $\mathbf{w}$ is a collection of strings that define the operations in the tree. The GP trains a population of agents, selecting the best candidates following an \textit{evolutionary approach}.} {\color{black} Concerning the BO and LIPO implemented in this work and described in the following section, it is instructive to interpret these as single-agent and `on-policy' RL approaches, with policy embedded in a parametric function and training governed by a surrogate-based optimizer which updates the parameters at the end of each episode.}
	
	{\color{black}
		In contrast to direct methods, indirect methods (Fig 1b) do not use function approximators for the policy but seek to learn an estimation of the state-value function $Q$, also known as $Q$ function in RL. For a deterministic agent/controller and deterministic environment/plant, this is defined as }
	
	\begin{equation}
		\label{Q_pi}
		\mathbf{Q}^{\pi}(\mathbf{s}_t,{\mathbf{a}}_t)=  \phi_r(\mathbf{s}_N)+\color{black}{r(\mathbf{s}_t,{\mathbf{a}}_t)}\color{black} +\sum^N_{k=t+1}\mathcal{L}_r(\mathbf{s}_k,\mathbf{a}^{\pi}_k,k) =\color{black}{r(\mathbf{s}_t,{\mathbf{a}}_t)}\color{black}+\gamma \mathbf{V}^{\pi}(\mathbf{s}_{t+1}) \,.
	\end{equation} where 
	
	\begin{equation}
		\begin{aligned}
			\label{V_pi}
			\mathbf{V}^{\pi}(\mathbf{s}_t)= & \phi(\mathbf{s}_N)+\sum^N_{k=t}\mathcal{L}_r(\mathbf{s}_k,\mathbf{a}^{\pi}_k,k)= \phi(\mathbf{s}_N)+\sum^N_{k=t}\gamma^{k-t} r^{\pi}_k =r_k+\gamma \mathbf{V}^{\pi}(\mathbf{s}_{t+1})\,
		\end{aligned}
	\end{equation} is the \emph{value} function according to policy $\pi$, i.e. the cumulative reward one can get starting from state $\mathbf{s}_t$ and then following the policy $\pi$. The Q function gives the value of an action at a given state; if a good approximation of this function is known, the best action is simply the greedy $\mathbf{a}_k=\argmax_{a_k} Q(\mathbf{s}_t,\mathbf{a}_t)$. Then, if $Q(\mathbf{s}_k,\mathbf{a}_k;\mathbf{w}_n)$ denotes the parametric function approximating $Q(\mathbf{s}_k,\mathbf{a}_k)$, learning is the iterative process by which the approximation improves, getting closer to the definition in \eqref{Q_pi}. The black-box optimization perspective is thus the minimization of the error in the $Q$ prediction; this could be done with huge variety of tools from optimization.
	
	
	
	{\color{black}
		Methods based on the Q function {are} `off-policy' and descend from dynamic programming \citep{sutton2018reinforcement}. The most classic approach is deep Q learning \citep{Mnih2013}. `Off-policy' methods are rather uncommon in the literature of flow control and are now appearing with the diffusion of RL approaches. While the vast majority of authors use ANNs as function approximators for the $Q$ function, alternatives have been explored in other fields. For example, \cite{Kubalik2021} uses a variant of GP while
		\cite{NIPS2003_7993e112, Goumiri2020, Fan2018} use Gaussian Processes as in classic BO. We also remark that the assumption of a deterministic system is uncommon in the literature of RL, where the environment is usually treated as a Markov Decision Process (MDP). We briefly reconsider the stochastic approach in the description of the DDPG in section \ref{Sec:IIIp3}. Like many modern RL algorithms, the DDPG implemented in this work combines both `on-policy' and `off-policy' approaches. 
	}

	\section{Implemented Algorithms}\label{Sec:III}
	
	\subsection{Optimization via BO and LIPO}\label{Sec:IIIp1}
	
	We assume that the policy is a pre-defined parametric function $\mathbf{a}=\pi(\mathbf{s}_t;\mathbf{w}^{\pi})\in\mathbb{R}^{n_a}$ with a small number of parameters (say $n_w \sim \mathcal{O}(10)$). The dimensionality of the problem enables efficient optimizers such as BO and LIPO; other methods are illustrated by \cite{maceda2018taming}.
	
	\subsubsection{Bayesian Optimization (BO)}\label{sec:BO}
	
	{\color{black}
		The classic BO uses a Gaussian Process (GPr) as surrogate model of the function that must be optimized. In the `on-policy' approach implemented in this work, this is the cumulative reward function $R(\mathbf{w})$; from \eqref{Q_pi} and \eqref{V_pi}, this is $R(\mathbf{w})=V^{\pi}(\mathbf{s}_0)=Q(\mathbf{s_0},\mathbf{a}^{\pi}_0)$.}
	
	Let $\mathbf{W}^*:=\{\mathbf{w}_1,\mathbf{w}_2\dots\mathbf{w}_{n_*}\}$ be a set of $n_*$ \emph{tested} weights and $\mathbf{R}^*:=\{R_1,R_2\dots {R}_{n_*}\}$ the associated cumulative rewards. The GPr offers a probabilistic model that computes the probability of a certain reward given the observations $(\mathbf{W}^*,\mathbf{R}^*)$, i.e. $p(R(\mathbf{w})|\mathbf{W}^*,\mathbf{R}^{*})$. In a GPr, this is
	\begin{equation}
		\label{R_GP}
		p(R(\mathbf{w})|\mathbf{R}^*,\mathbf{W}^*)= \mathcal{N}(\boldsymbol{\mu},\boldsymbol{\Sigma})\,,
	\end{equation} where $\mathcal{N}$ denotes a multivariate Gaussian distribution with mean $\boldsymbol{\mu}$ and covariance matrix $\boldsymbol{\Sigma}$. In a Bayesian framework, eq \eqref{R_GP} is interpreted as a posterior distribution, conditioned to the observations ($\mathbf{W}^*,\mathbf{R}^*$). A Gaussian process is a distribution over functions whose smoothness is defined by the covariance function, computed using a kernel function. Given a set of data ${\color{black}(\mathbf{W}^*,\mathbf{R}^*)}$, this allows for building a continuous function to estimate both the reward of a possible candidate and the uncertainties associated with it.
	
	We are interested in evaluating \eqref{R_GP} on a set of $n_E$ \emph{new} samples $\mathbf{W}:=\{\mathbf{w}_1,\mathbf{w}_2\dots\mathbf{w}_{n_E}\}$ and we denote as $\mathbf{R}:=\{R_1,R_2\dots R_{n_E}\}$ the possible outcomes (treated as random variables). Assuming that the possible candidate solutions belong to the same Gaussian process (usually assumed to have zero mean \citep{CarlEdward}) as the observed data $(\mathbf{W}^*,\mathbf{R}^*)$, we have:
	\begin{equation}
		\label{N_GAUS}
		\begin{pmatrix}\mathbf{R}^* \\ \mathbf{R}\end{pmatrix} \sim \mathcal{N}
		\left(\boldsymbol{0},
		\begin{pmatrix}\mathbf{K}_{**} & \mathbf{K}_* \\ \mathbf{K}_*^T & \mathbf{K}\end{pmatrix}
		\right)\,,
	\end{equation} 
	where $\mathbf{K}_{**}=\kappa(\mathbf{W}^*,\mathbf{W}^*)\in\mathbb{R}^{n_*\times n_*}$, $\mathbf{K}_{*}=\kappa(\mathbf{W},\mathbf{W}^*)\in\mathbb{R}^{n_E\times n_*}$,  $\mathbf{K}=\kappa(\mathbf{W},\mathbf{W})\in\mathbb{R}^{n_E\times n_E}$ and $\kappa$ a kernel function.
	
	The prediction in \eqref{R_GP} can be built using standard rules for conditioning multivariate Gaussian, and the functions $\boldsymbol{\mu}$ and $\boldsymbol{\Sigma}$ in \eqref{R_GP} becomes a vector $\boldsymbol{\mu_*}$ and a matrix $\boldsymbol{\Sigma_*}$:
	\begin{align}
		\label{BO}
		\boldsymbol{\mu_*} &= \mathbf{K}_*^T \mathbf{K}_R^{-1} \mathbf{R}^*\quad\in \mathbb{R}^{n_E}\\
		\label{BO2}
		\boldsymbol{\Sigma_*} &= \mathbf{K} - \mathbf{K}_*^T \mathbf{K}_R^{-1} \mathbf{K}_*\quad\in \mathbb{R}^{n_E\times n_E}\,\,,
	\end{align} 

	where $\mathbf{K}_R=\mathbf{K}_{**}+\sigma_{R}^2\mathbf{I}$, with $\sigma_{R}^2$ the expected variance in the sampled data and $\mathbf{I}$ the identity matrix of appropriate size. The main advantage of BO is that the function approximation is sequential, and new predictions improve the approximation of the reward function (i.e. the surrogate model) episode after episode. This makes the GPr- based BO one of the most popular black-box optimization methods for expensive cost functions. 
	
	{\color{black} The BO combines the GPr model with a function suggesting where to sample next.} Many variants exist \citep{Frazier2018}, each providing their exploration/exploitation balance. The exploration seeks to sample in regions of large uncertainty, while {\color{black}exploitation} seeks to sample at the best location according to the current function approximation. The most classic function, used in this study, is the expected improvement, defined as \citep{CarlEdward} 
	
	\begin{equation}
		\label{EI}
		\operatorname{EI}({\color{black}\mathbf{w}}) =
		\begin{cases}
			(\Delta - \xi)\Phi(Z) + \sigma(\mathbf{w})\phi(Z)  &\text{if}\ \sigma(\mathbf{w}) > 0 \\
			0 & \text{if}\ \sigma({\color{black} \mathbf{w}}) = 0
		\end{cases}\,,
	\end{equation} with $\Delta=\mu(\mathbf{w}) - R(\mathbf{w}^+)$ and $\mathbf{w}^+=\argmax_{\mathbf{w}} \tilde{R}(\mathbf{w})$ the best sample so far, $\Phi(Z)$ the cumulative distribution (CDF), $\phi(Z)$ the probability density (PDF) of a standard Gaussian and 
	
	\begin{equation}
		\label{EI_z}
		Z =
		\begin{cases}
			\frac{\Delta- \xi}{\sigma(\mathbf{w})} &\text{if}\ \sigma(\mathbf{w}) > 0 \\
			0 & \text{if}\ \sigma(\mathbf{w}) = 0
		\end{cases}\,.
	\end{equation}
	
	Eq \eqref{EI} balances the desire to sample in regions where $\mu(\mathbf{w})$ is  {\color{black}larger than} $R(\mathbf{w}^+)$ (hence large and positive $\Delta$) versus sampling in regions where $\sigma(\mathbf{w})$ is large. The parameter $\xi$ sets a threshold over the minimal expected improvement that justifies the exploration.
	
	Finally, the method requires the definition of the kernel function and its hyper-parameters, as well as an estimate of $\sigma_y$. In this work, the GPr-based BO was implemented using the Python API \textit{scikit-optimize} \citep{Head2020}. The selected kernel function was a Mater kernel with $\nu=5/2$ (see Chapter 4 from \cite{CarlEdward}) which reads:
	
	\begin{equation}
		\label{Matern}
		\kappa (\mathbf{x},\mathbf{x}')=\kappa (r)=1+\frac{\sqrt{5} r}{l}+\frac{5 r^2}{3 l^2}\exp-\frac{\sqrt{5} r}{l}\,, 
	\end{equation} where $r=||\mathbf{x}-\mathbf{x}'||_2$ and $l$ the length scale of the process. {\color{black} We report a detailed description of the pseudocode we used in Appendix \ref{BO_pseudocode}.}
	

	\subsubsection{LIPschitz global Optimization (LIPO)}\label{sec:LIPO}
	
	Like BO, LIPO relies on a surrogate model to select the next sampling points \citep{malherbe2017global}. However, LIPO's surrogate function is the much simpler upper bound approximation $U(\mathbf{w})$ of the cost function $R(\mathbf{w})$ \citep{Ahmed2020}. In the \textsc{dlib} implementation by \cite{dlib09}, used in this work, this is given by:
	\begin{equation}
		\label{LIPO_SURR}
		U(\mathbf{w})= \min_{i=1\dots t}\big(R(\mathbf{w}_{i}) + \sqrt{\sigma_i + (\mathbf{w}-\mathbf{w}_i)^TK(\mathbf{w}-\mathbf{w}_i)} \big),
	\end{equation} where $\mathbf{w}_i$ are the sampled parameters, $\sigma_i$ are coefficients which account for discontinuities and stochasticity in the objective function, and $K$ is a diagonal matrix that contains the Lipschitz constants $k_i$ for the different dimensions of the input vector. We recall that a function $R(\mathbf{w}):\mathcal{W}\subseteq \mathbb{R}^{n_w}\rightarrow\mathbb{R}$ is a Lipschitz function if there exists a constant $C$ such that:
	\begin{equation}
		\lVert R(\mathbf{w}_1) - R(\mathbf{w}_2)\rVert \leq C\lVert\mathbf{w}_1 - \mathbf{w}_2\rVert, \quad \forall\; \mathbf{w}_1,\mathbf{w}_2\in\mathcal{W},
	\end{equation}
	
	where $\lVert\cdot\rVert$ is the Euclidean norm on $\mathbb{R}^{n_w}$. The Lipshitz constant $k$ of $R(\mathbf{w})$ is the smallest $C$ that satisfies the above condition \citep{davidson2009real}. In other terms, this is an estimate of the largest possible slope of the function $R(\mathbf{w})$. The values of $K$ and $\mathbf{\sigma}_i$ are found by solving the optimization problem:
	\begin{equation}
		\begin{aligned}
			\min_{K,\sigma} \quad & \lVert K\rVert_F^{2} + 10^6\sum_{i=1}^{t}\,\sigma_i^2\\
			\textrm{s.t.} \quad & U(\mathbf{w}_i)\geq R(\mathbf{w}_i),\quad \forall i\in[1\cdots t]\\
			& \sigma_i \geq 0,\quad \forall i\in[1\cdots t]\\
			& K_{i,j} \geq 0,\quad \forall i,j\in[1\cdots d]\\
			& \text{K = }\{k_1,k_2,\cdots,k_{n_w}\},\\
		\end{aligned}
	\end{equation}
	where $10^6$ is a penalty factor and $\lVert\cdot\rVert_F$ is the Frobenius norm. 
	
	To compensate for the poor convergence of LIPO in the area around local optima, the algorithm alternates between a global and a local search. If the iteration number is even, it  selects the new weights by means of the maximum upper bounding position (MaxLIPO):
	\begin{equation}
		\label{EXP_LIPO}
		\mathbf{w}_{k+1} = \argmax_{\mathbf{w}}(U(\mathbf{w})),
	\end{equation}
	otherwise, it relies on a Trust Region (TR) method \citep{powell2006newuoa} based on a quadratic approximation of $R(\mathbf{w})$ around the best weights obtained so far $\mathbf{w}^*$, i.e:
	\begin{equation}
		\begin{aligned}
			\mathbf{w}_{k+1} =\arg\max_{\mathbf{w}} \,& \overbrace {\Big(\mathbf{w}^* + g(\mathbf{w}^*)^{T}\mathbf{w} + \frac{1}{2}\mathbf{w}^T\mathbf{H} (\mathbf{w}^*)\mathbf{w}\Big)}^{m(\mathbf{w};\mathbf{w}^*)}\\
			\textrm{s.t.} \, & ||\mathbf{w}_{k+1}||<d(\mathbf{w}^*)\\
		\end{aligned}
		\label{TR_approx}
	\end{equation}
	where $g(\mathbf{w}^*)$ is the approximation of the gradient at $\mathbf{w}^*$ ($g(\mathbf{w}^*)\approx\nabla R(\mathbf{w}^*))$, $\mathbf{H}(\mathbf{w}^*)$ is the approximation of the Hessian matrix $(\mathbf{H}(\mathbf{w}^*))_{ij} \approx \frac{\partial^2 R(\mathbf{w}^*)}{\partial\mathbf{w}_i\partial\mathbf{w}_j}$ and $d(\mathbf{w}^*)$ is the radius of the trust region.
	If the TR-method converges to a local optimum with an accuracy smaller than $\epsilon$: 
	\begin{equation}
		|R(\mathbf{w}_{k})-R(\mathbf{w}^*)|<\varepsilon, \quad \forall\; \mathbf{w}_{k},
		\label{disc_LIPO}
	\end{equation}
	the optimization goes on with the global search method until it finds a better optimum. {\color{black} A detailed description of the pseudocode we used can be found in Appendix \ref{LIPO_pseudocode}.}

	\subsection{Genetic Programming}\label{Sec:IIIp2}
\begin{figure}\center
		\includegraphics[width=0.4\textwidth]{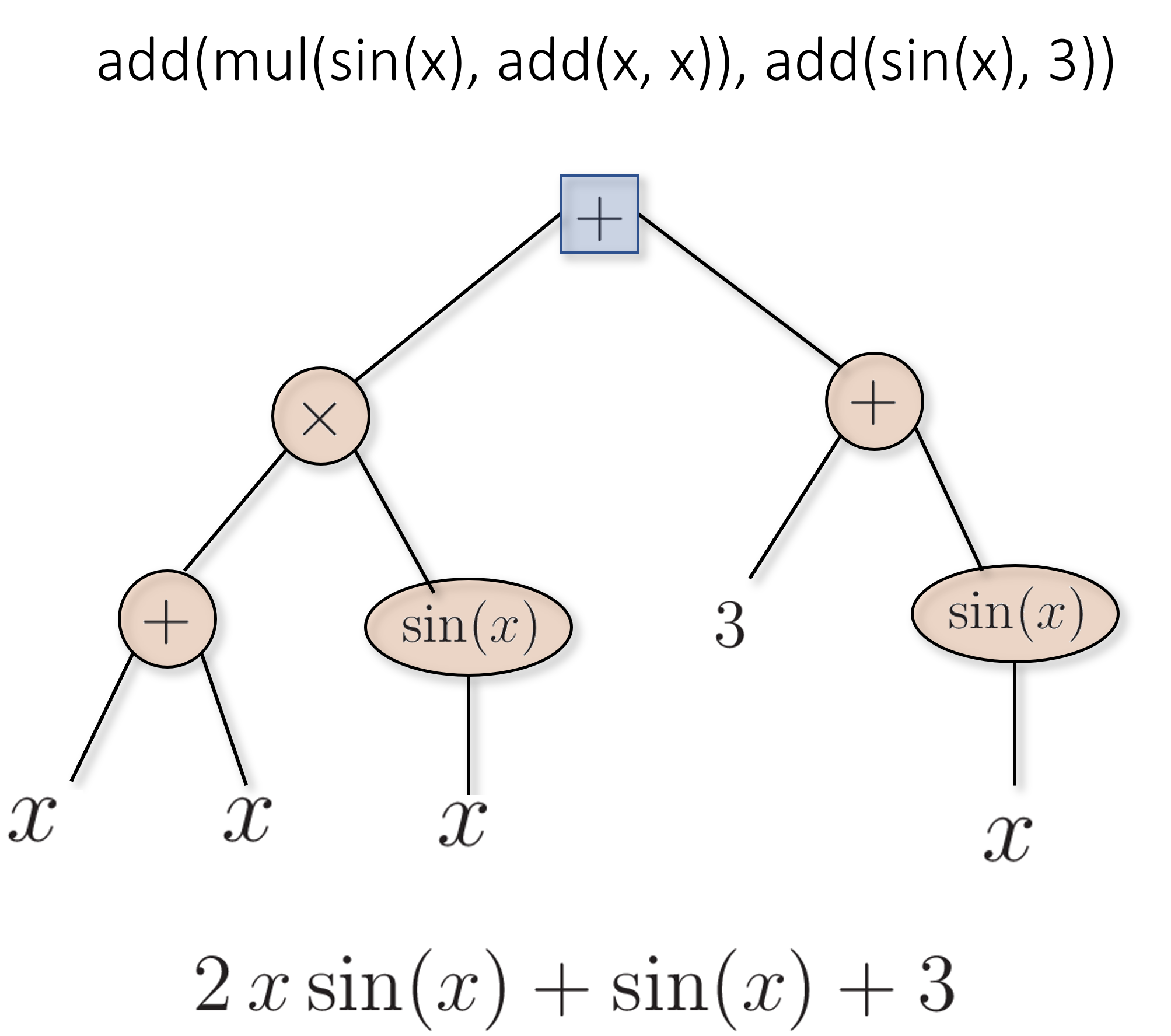}
		\centering
		\caption{Syntax tree representation of the function $2 x\sin(x)+\sin(x)+3$. This tree has a root '+' and a depth of two. The nodes are denoted with orange circles while the last entries are leafs.}
		\label{GP_Tree}
	\end{figure}

	In the Genetic Programming (GP) approach to optimal control, the policy $\mathbf{a}=\pi(\mathbf{s};\mathbf{w})$ is encoded in the form of a syntax tree. The parameters are lists of numbers and functions which can include arithmetic operations, mathematical functions, Boolean operations, conditional operations or iterative operations. An example of a syntax tree representation of a function is shown in Figure \ref{GP_Tree}. A tree (or \emph{program} in GP terminology) is composed of a root that branches out into nodes (containing functions or operations) throughout various levels. The number of levels defines the \emph{depth} of the tree, and the last nodes are called \emph{terminals} or \emph{leaves}. These contain the input variables or constants. Any combination of branches below the root is called \emph{sub-tree} and can generate a tree if the node becomes a root.
	
	Syntax trees allow encoding complex functions by growing into large structures. The trees can adapt during the training: the user provides a \emph{primitive set}, i.e. the pool of allowed functions, the maximum depth of the tree, and set the parameters of the training algorithm. Then, the GP operates on a population of possible candidate solutions (individuals) and evolves it over various steps (generations) using genetic operations in the search for the optimal tree. Classic operations include elitism, replication, cross-over and mutations, as in Genetic Algorithm Optimization \citep{haupt2004practical}. 
	The implementation of GP in this work was carried out in the \textbf{D}istributed \textbf{E}volutionary \textbf{A}lgorithms in \textbf{P}ython (DEAP) \citep{DEAP_JMLR2012} framework. This is an open-source Python library allowing for the implementation of various evolutionary strategies.

	We used a primitive set of four elementary operations ($+,-,/,\times$) and four functions ($\exp,\log,\sin,\cos$). {\color{black} In the second test case, as described in Section \ref{sec4p3}, we also include an ephemeral random constant.}
	The initial population of individuals varied between $n_I=10$ and $n_I=80$ candidates depending on the test case and the maximum depth tree was set to $17$. In all test cases, the population was initialized using the "half-half" approach, whereby half the population is initialized with the \emph{full method} and the rest with the \emph{growth method}. In the full method, trees are generated with a predefined depth and then filled randomly with nodes and leafs. In the growth method, trees are randomly filled from the roots: because nodes filled with variables or constant are terminals, this approach generates trees of variable depth.
	
	{\color{black}
		Among the optimizers available in DEAP, in this work we used the $(\mu+\lambda)$ algorithm for the first two test cases and \textit{eaSimple} \citep{Banzhaf1997,Vanneschi2012,Kober2014,easimple} for the third one. These differ in how the population is updated at each iteration. In the $(\mu+\lambda)$ both the off-springs and parents participate to the tournament while in eaSimple no distinction is made between parents and off-springs and the population is entirely replaced at each iteration.}
	
	{\color{black}Details about the algorithmic implementation of this approach can be found in Appendix \ref{GP_pseudocode}.}
	
	\subsection{Reinforcement Learning via DDPG}\label{Sec:IIIp3}
	
	The Deep Deterministic Policy gradient (DDPG) by \cite{Lillicrap2015} is an off-policy actor-critic algorithm using an ANN to learn the policy {\color{black}(direct approach, in Fig 1a)} and an ANN to learn the Q function {\color{black}(indirect approach, in Fig 1b)}. In what follows, we call $\Pi$- network the first (i.e. the actor) and $Q$-network the second (i.e. the critic).

	The DDPG combines the DPG by \cite{Silver_1} and the Deep Q learning (DQN) by \cite{Mnih2013,Mnih2015}. The algorithm has evolved into more complex versions such as the Twin Delayed DDPG \citep{Fujimoto2018}, but in this work we focus on the basic implementation. 
	
	The policy encoded in the $\Pi$ network is deterministic and acts according to the set of weights and biases $\mathbf{w}^{\pi}$, i.e. $\mathbf{a}=\pi(\mathbf{s}_t,\mathbf{w}^{\pi})$.
	The environment is assumed to be stochastic and modelled as a Markov {\color{black} Decision} Process. Therefore, \eqref{Q_pi} must be modified to introduce an expectation operator:
	\begin{equation}
		\label{Exp_Q}
		Q^{\pi}(\mathbf{s}_t,\mathbf{a}_t)=\mathbb{E}_{\mathbf{s}_{t},\mathbf{s}_{t+1}\sim E}\bigl [r(\mathbf{s}_t,\mathbf{a}_t)+\gamma Q^{\pi}(\mathbf{s}_{t+1},\mathbf{a}^{\pi}_{t+1})\bigr]\,,
	\end{equation} where the policy is intertwined in the action state relation, i.e. $Q^{\pi}(\mathbf{s}_{t+1},\mathbf{a}_{t+1})=Q^{\pi}(\mathbf{s}_{t+1},\mathbf{a}^{\pi}(\mathbf{s}_{t+1}))$ and having used the shorthand notation $\mathbf{a}^{\pi}_{t+1}=\pi(\mathbf{s}_{t+1},\mathbf{w}^{\pi})$. Because the expectation operator in \eqref{Exp_Q} solely depends on the environment ($E$ in the expectation operator), it is possible to decouple the problem of learning the policy $\pi$ from the problem of learning the function $Q^{\pi}(\mathbf{s}_t,\mathbf{a}_t)$. Concretely, let $Q(\mathbf{s}_t,\mathbf{a}_t;\mathbf{w}^{Q})$ denote the prediction of Q function by the Q network, defined with weights and biases $\mathbf{w}^{Q}$ and let $\mathcal{T}$ denote a set of $N$ transitions $(\mathbf{s}_t,\mathbf{a}_t,\mathbf{s}_{t+1},r_{t+1})$ collected through (any) policy. The performances of the Q-network can be measured as 
	\begin{equation}
		\label{q_loss}
		J^Q(\mathbf{w}^{Q})=\mathbb{E}_{\mathbf{s}_t,\mathbf{a}_t, \mathbf{r}_t\sim \mathcal{T}}\Bigl[\Bigr(Q\bigr(\mathbf{s}_t,\mathbf{a}_t;\mathbf{w}^{Q})-y_t\Bigl)^2\Bigr]\,,
	\end{equation} where the term in the squared brackets, called temporal difference, is the difference between the old Q value and the new one $y_t$, known as temporal difference target: 
	\begin{equation}
		\label{y_T}
		y_t=r(\mathbf{s}_t,\mathbf{a}_t)+\gamma Q (\mathbf{s}_{t+1},\mathbf{a}_{t+1};\mathbf{w}^{Q})\,.
	\end{equation}
	Equation \eqref{q_loss} measures how closely the prediction of the Q network satisfies the discrete Bellman equation \eqref{Q_pi}. The training of the Q network can be carried out using standard stochastic gradient descent methods using the back-propagation algorithm \citep{kelley1960gradient} to evaluate $\partial_{\mathbf{w}^{Q}} J^Q$.
	
	The training of the $Q$-network gives the off-policy flavor to the DDPG because it can carried out with an exploratory policy that largely differ from the final policy. Nevertheless, because the training of the $Q$-network is notoriously unstable, \cite{Mnih2013,Mnih2015} introduced the use of a \emph{replay buffer} to leverage accumulated experience (previous transitions) and a \emph{target network} to under-relax the update of the weights during the training. Both the computation of the cost function in \eqref{q_loss} and its gradient are performed over a random batch of transitions $\mathcal{T}$ in the replay buffer $\mathcal{R}$.
	
	The DDPG combines the Q-network prediction with a policy gradient approach to train the $\Pi$-network. This is inherited from the DPG by \cite{Silver_1}, who have shown that, given
	\begin{equation}
		\label{pi_loss}
		J^{\pi}(\mathbf{w}^{\pi})=\mathbb{E}_{\mathbf{s}_t\sim E,\mathbf{a}_t\sim \pi}\bigl[(r(\mathbf{s}_t,\mathbf{a}_t))\bigr]
	\end{equation} the expected return from the initial condition, the gradient with respect to the weights in the $\Pi$ network is:
	\begin{equation}
		\label{Pi_Grad}
		\partial_{\mathbf{w}^{\pi}}J^{\pi}=\mathbb{E}_{\mathbf{s}_t\sim E,\mathbf{a}_t\sim \pi}\bigl[\partial_{\mathbf{a}} Q(\mathbf{s}_t,\mathbf{a}_t;\mathbf{w}^Q)\,\partial_{\mathbf{w}^{\pi}} \mathbf{a}(\mathbf{s}_t;\mathbf{w}^{\pi})\bigr]\,.
	\end{equation} 
	
	Both $\partial_{\mathbf{a}} Q(\mathbf{s}_t,\mathbf{a}_t;\mathbf{w}^Q)$ and $\partial_{\mathbf{w}^{\pi}} \mathbf{a}(\mathbf{s}_t;\mathbf{w}^{\pi})$ can be evaluated via back-propagation, on the Q network and the $\Pi$ network respectively. The main extension of DDPG over DPG is the use of DQN for the estimation of the Q function.  
	
	In this work, we implement the DDPG using \textsc{Keras} API in \textsc{Python} with three minor modifications to the original algorithm. The first is a clear separation between the exploration and the exploitation phases. In particular, we introduce a number of exploratory episodes $n_{Ex}<n_{Ep}$ and the action is computed as 
	\begin{equation}
		\label{act}
		\mathbf{a}(\mathbf{s}_t)=\mathbf{a}(\mathbf{s}_t;\mathbf{w}^{\pi})+\eta(\mbox{ep})\mathcal{E}(t;\theta,\sigma^2)\,,
	\end{equation} where $\mathcal{E}(t;\theta,\sigma)$ is an exploratory random process characterized by a mean $\theta$ and variance $\sigma^2$. This could be the time-correlated \citep{Uhlenbeck1930} noise or white noise, depending on the test case at hand (see Sec. \ref{Sec:IV}). The transition from exploration to exploitation is governed by the parameter $\eta$, which is taken as $\eta(\mbox{ep})=1$ if $\mbox{ep}<n_{Ex}$ where $d^{\text{\small{ep}}-n_{Ex}}$ if $\mbox{ep}>n_{Ex}$. This decaying term for $\mbox{ep}>n_{Ep}$ progressively reduces the exploration and the coefficient $d$ controls how rapidly this is done.
	
	The second modification is in the selection of the transitions from the replay buffer $\mathcal{R}$ that are used to compute the gradient $\partial_{\mathbf{w}^{Q}} J^Q$. While the original implementation selects these randomly, we implement a simple version of the prioritized experience replay from \cite{Schaul2015}. The idea is to prioritize, while sampling from the replay buffer, those transitions which led to the largest improvement in the network performances. These can be measured in terms of Temporal Difference Error (or TD-Error):
	\begin{equation}
		\label{TD}
		\delta = r_t + \gamma Q(\mathbf{s}_{t+1}, \mathbf{a}^{\pi}_{t+1};\mathbf{w}^{Q})- Q(\mathbf{s}_t, a_t;\mathbf{w}^{Q})\,.
	\end{equation}
	This quantity measures how much a transition was \emph{unexpected}. 
	{\color{black} The rewards stored in the replay buffer ($r_{t}^{RB}$) and used in the TD computation are first scaled using a dynamic vector $r_{log}=[r^{RB}_1,r^{RB}_2,\cdots,r^{RB}_t]$ as:
		\begin{equation}
			r_{t}^{RB} = \frac{r_t - \bar{r}_{log}}{std(r_{log})+1e-10}
		\end{equation} where $\bar{r}_{log}$ is the mean value and $std(r_{log})$ is the standard deviation. The normalization makes the gradient steeper far from the mean of the sampled rewards, without changing its sign, and is found to speed-up the learning (see also \cite{van2016learning}).}
	
	As discussed by \cite{Schaul2015}, it can be shown that prioritizing unexpected transitions leads to the steepest gradients $\partial_{\mathbf{w}^{Q}} J^Q$, and thus helps overcome local minima. The sampling is performed following a triangular distribution which assigns the highest probability $p(n)$ to the transition with the largest TD error $\delta$.
	
	The third modification, extensively discussed in previous works on reinforcement learning for flow control \citep{Rabault2019a,Tang2020e,Rabault2020}, is the implementation of a sort of moving average of the actions. In other words, an action is performed for $K$ consecutive interactions with the environment, which in our work occur at every simulation's time step.
	\begin{figure*}
		\center
		\includegraphics[width=0.80\textwidth]{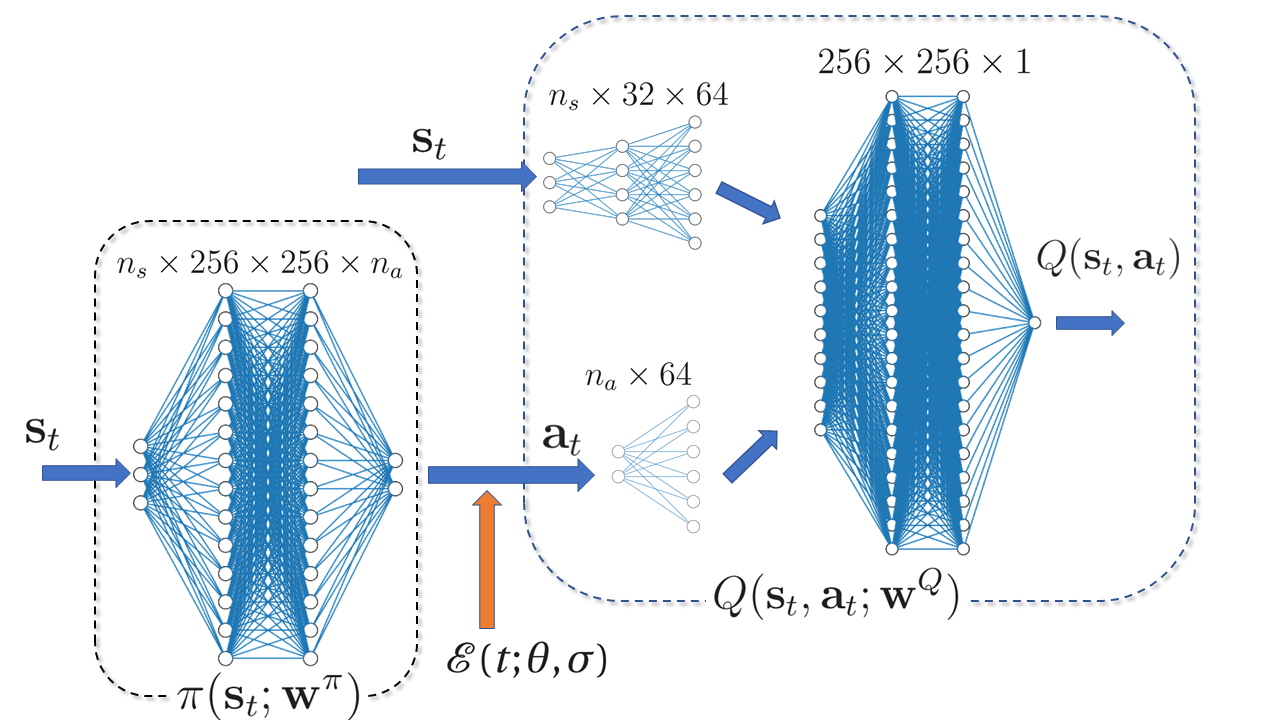}
		\centering
		\caption{ANN Architecture of the DDPG implementation analyzed in this work. The illustrated architecture is the one used for the test case in section \ref{Sec:IVp3}. During the exploration phase, the two networks are essentially decoupled by the presence of the stochastic term $\mathcal{E}$ that leads to exploration of the action space. }
		\label{DDPG_FIG}
	\end{figure*}
	
	We illustrate the neural network architecture employed in this work in Figure \ref{DDPG_FIG}. The scheme in the figure shows how the $\Pi$ network and the Q network are interconnected: intermediate layers map the current state and the action (output by the $\Pi$ network) to the core of the Q network. For plotting purposes, the number of neurons in the figure is much smaller than the one actually used and indicated in the figure. The $\Pi$ network has two hidden layers with $128$ neurons each, while the input and output depends on the test cases considered (see Sec. \ref{Sec:IV}). Similarly, the Q network has two hidden layers with $128$ neurons each and intermediate layers as shown in the figure. During the exploration phase, the presence of the stochastic term in the action selection decouples the two networks. 
	
	{\color{black}We detail the main steps of the implemented DDPG algorithm in Appendix \ref{DDPG_pseudocode}. It is important to notice that, by construction, the weights in this algorithm are updated at each interaction with the system. Hence $k=n$ and $N=1$ in the terminology of Section \ref{Sec:II}. The notion of episode remains relevant to control the transition between various phases of the learning process and to provide a comparable metrics between the various algorithms.}
	
	\section{Test Cases}\label{Sec:IV}
	
	\subsection{A 0D Frequency Cross-Talk Problem}\label{Sec:IVp1}
	
	The first selected test case is a system of nonlinear ODEs reproducing one of the main features of turbulent flows: the frequency cross-talk. This control problem was proposed and extensively analysed by \cite{duriez2017machine}. It essentially consists in stabilizing two coupled oscillators, described by a system of four ODEs, which describe the time evolution of four leading Proper Orthogonal Decomposition (POD) modes of the flow past a cylinder. The model is known as generalized mean field model \citep{LUCHTENBURG2009} and was used to describe the stabilizing effect of low frequency forcing on the wave flow past a bluff body \citep{Aleksic2010,PASTOOR2008}. The set of ODEs in the states $\mathbf{s}(t)=[s_1(t),s_2(t),s_3(t),s_4(t)]^T$, where ($s_1,s_2$) and ($s_3,s_4$) are the first and second oscillator, reads:
	
	\begin{equation}
		\dot{\textbf{s}} = \mathbf{F}(\mathbf{s})\,\textbf{s} + {\color{black}\mathbf{A}}\textbf{a},
		\label{gov_eqn_0D}
	\end{equation}
	where \textbf{a} is the forcing vector with a single scalar component interacting with the second oscillator (i.e., $\textbf{a}=[0,0,0,a]^T$) and the matrix $\mathbf{F}(\mathbf{s})$ {\color{black} and $\mathbf{A}$ are} given by:
	
	\begin{equation}
		\mathbf{F}(\mathbf{s}) =   \begin{bmatrix}
			\sigma(\mathbf{s}) &  -1 & 0 & 0\\
			1 & \sigma(\mathbf{s}) & 0 & 0 &\\
			0 & 0 & -0.1 & -10 \\
			0 & 0 & 10 & -0.1 \\
		\end{bmatrix},\qquad
		{\color{black}
			\mathbf{A} =   \begin{bmatrix}
				0 &  0 & 0 & 0\\
				0 & 0 & 0 & 0 &\\
				0 & 0 & 0 & 0 \\
				0 & 0 & 0 & 1 \\
			\end{bmatrix}.}
	\end{equation}
	The term $\sigma(\mathbf{s})$ models the coupling between the two oscillators:
	\begin{equation}
		\label{Sigma}
		\sigma(\mathbf{s}) = 0.1 - E_1 - E_2,
	\end{equation} where $E_1$ and $E_2$ are the energy of the first and the second oscillator given by: 
	\begin{equation}
		\label{Energy}
		E_1 = s_1^2+s_2^2 \quad E_2 = s_3^2 + s_4^2.
	\end{equation}
	
	This nonlinear link is the essence of the frequency cross-talk and challenges linear control methods based on linearization of the dynamical system. {\color{black} To excite the second oscillator, the actuation must introduce energy to the second oscillator, as one can reveal from the associated energy equation. This is obtained by multiplying the last two equations of the system by $s_3$ and $s_4$ respectively and summing them up to obtain:
		\begin{equation}
			\frac{1}{2}\dot{E_2} = -0.2E_2 + s_4 \,u,
			\label{0D_energy_second_oscillator_eq}\,,
		\end{equation}
		where $u\,s_4$ is the production term associated to the actuation.}

\begin{figure}
		\centering
		\begin{subfigure}{.5\textwidth}
			\centering
			\includegraphics[width=0.9\textwidth]{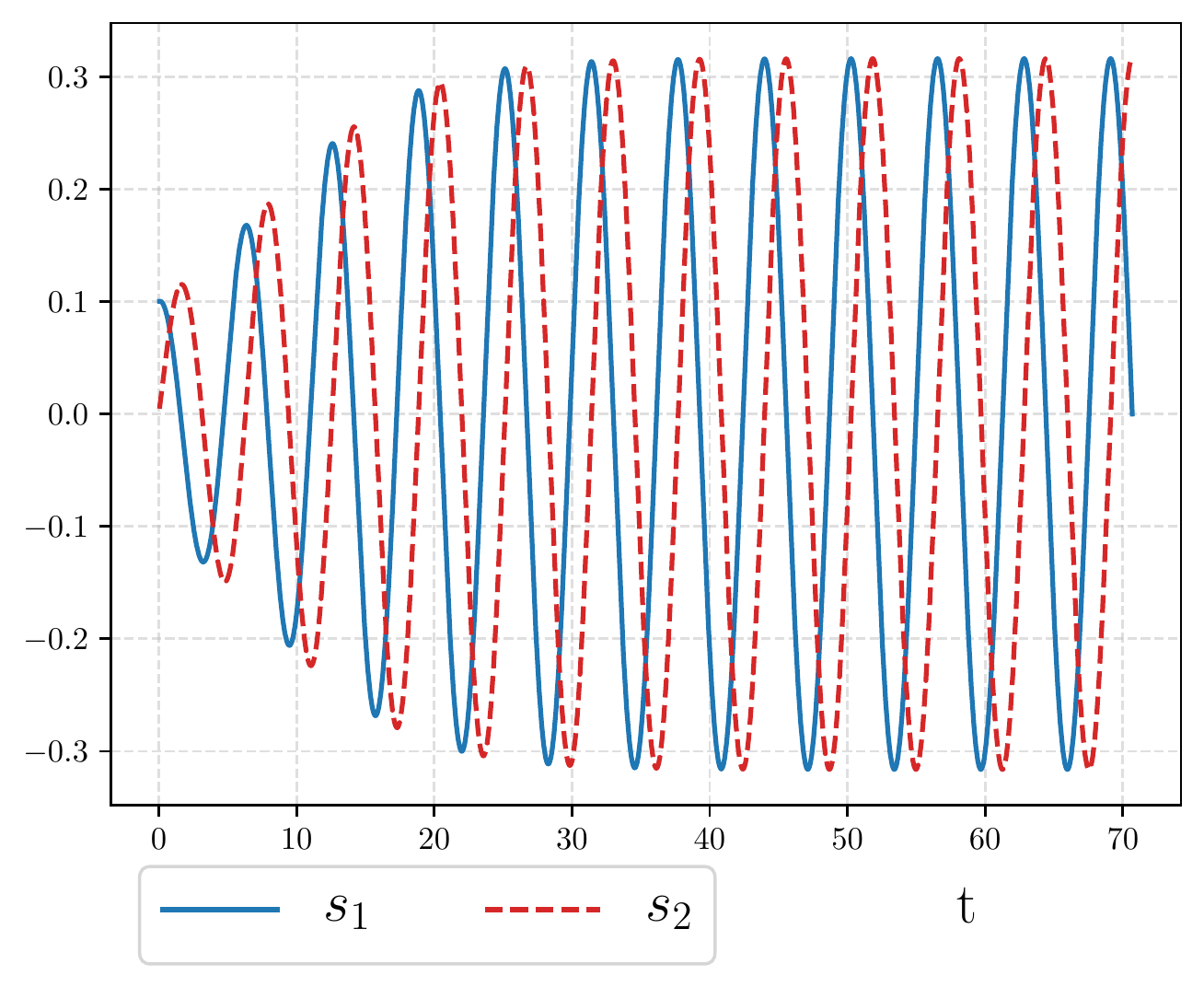}
			\caption{}
			\label{noak_des_1}
		\end{subfigure}%
		\begin{subfigure}{.5\textwidth}
			\centering
			\includegraphics[width=0.9\textwidth]{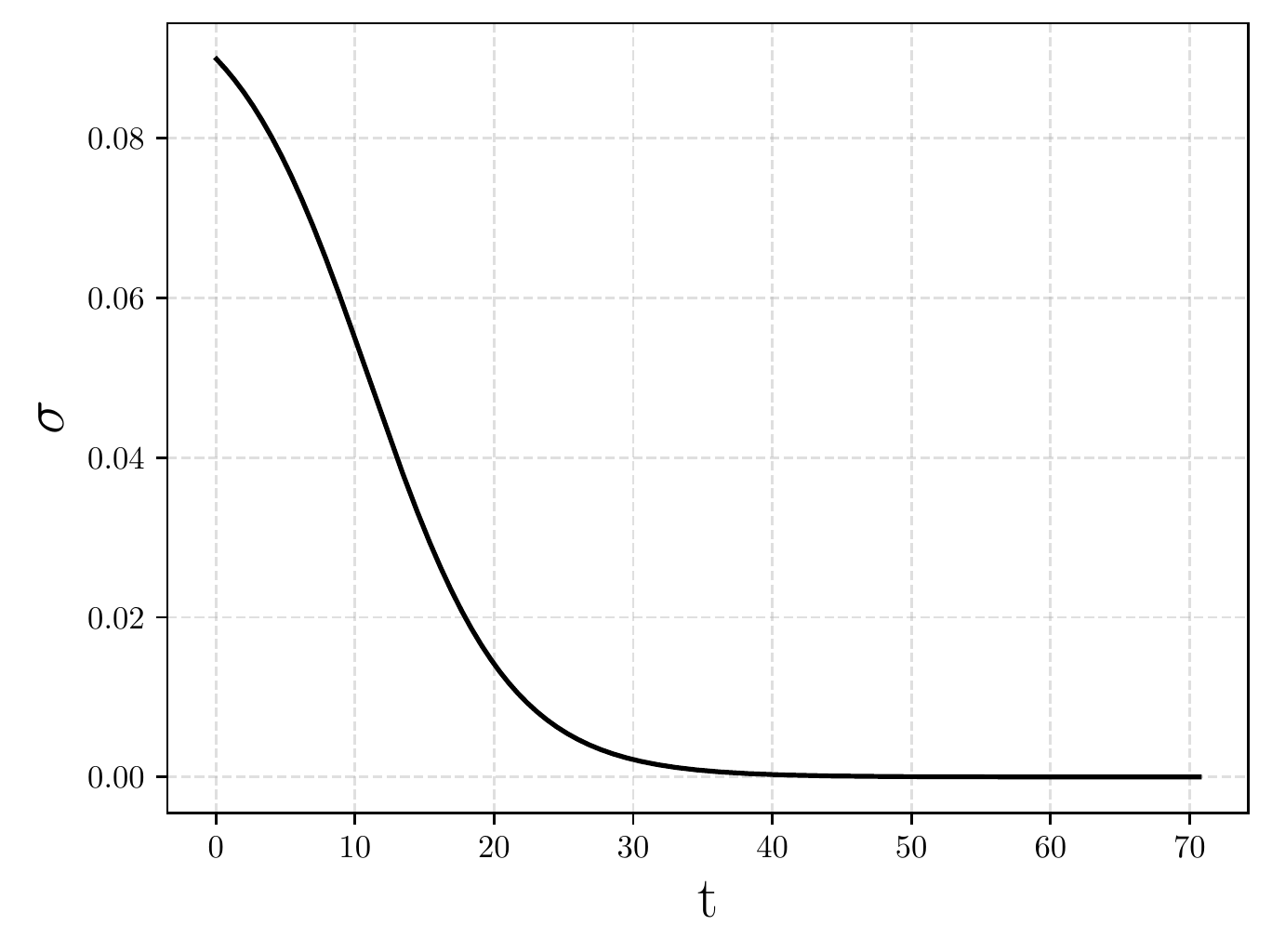}
			\caption{}
			\label{noak_des_2}
		\end{subfigure}
		\caption{Evolution of the oscillator $(s_1,s_2)$ (a) of the variable $\sigma$ \eqref{Sigma} (b) in the 0D test case in absence of actuation (${\color{black}a}=0$). As $\sigma\approx 0$, the system naturally evolves towards a `slow' limit cycle.}
	\end{figure}

	The initial conditions are set to $\mathbf{s}(0)=[0.01,0,0,0]^T$. Without actuation, the system reaches a `slow' limit cycle involving the first oscillator {\color{black}$(s_1,s_2)$}, while the second vanishes ($(s_3,s_4)\rightarrow 0$). The evolution of the oscillator $(s_1,s_2)$ with no actuation is shown in Figure \ref{noak_des_1}; Figure \ref{noak_des_2} shows the time evolution of $\sigma$, which vanishes as the system naturally reaches the limit cycle.
	{\color{black} Regardless of the state of the first oscillator, the second oscillator is essentially a linear second order system with eigenvalues $\lambda_{1,2}=-0.1\pm 10\mathrm{i}$, hence a natural frequency $\omega=10$ rad/s.}
	
	The governing equations \ref{gov_eqn_0D} were solved using scipy's package odeint with a time step of $\Delta\,t=\pi/50$. This time step is smaller than the one by \cite{duriez2017machine} ($\Delta\,t=\pi/10$), as we observed this had an impact on the training performances (aliasing in LIPO and BO optimization). 
	
	The actuators' goal is to bring to rest the first oscillator {\color{black} while exiting the second}, leveraging on the non-linear connection between the two {\color{black} and using } the least possible actuation. In this respect, the optimal control law, similarly to \cite{duriez2017machine}, is the one that minimizes the cost function:
	\begin{equation}
		\begin{gathered}
			\label{eqn_cost_function}
			J = J_a + \gamma\,J_b = \overline{s_1^2 + s_2^2} + \alpha\,\overline{a^2}\\
			\text{where} \qquad \overline{f(t)} = \frac{1}{40\pi}\int\displaylimits_{20\pi}^{60\pi} f(t') dt',
		\end{gathered}\,.
	\end{equation}
	where $\alpha$, set to $\alpha=10^{-2}$, is a coefficient set to penalize large actuations. {\color{black} Like the original problem in \cite{duriez2017machine}, the actions are clipped to the range $a_k\in[-1,1]$. }
	
	The time interval of an episode is set to $t\in[20\pi,60\pi]$, thus much shorter than the one used by \cite{duriez2017machine}. This duration was considered sufficient, as it allows the system to reach the limit cycle and to observe approximately $20$ periods of the slow oscillator. To reproduce the same cost function in a reinforcement learning framework, we rewrite \eqref{eqn_cost_function} as a cumulative reward, replacing the integral mean with the arithmetic average and setting:
	\begin{equation}
		J = \frac{1}{n_t}\sum_{k=0}^{n_t-1} s^2_{1k}+s^2_{2k}+\alpha a^2_k= -\sum_ {k=0}^{n_t-1}r_t=-R,
		\label{cost_fun_0D_dis}
	\end{equation} with $r_t$ the environment's reward at each time step. 
	For the BO and LIPO optimizers, the control law is defined as a quadratic form of the four system's states:
	\begin{equation}
		\pi(\mathbf{s};\mathbf{w}) := \mathbf{g}_w^T\mathbf{s} + \mathbf{s}^T\mathbf{H}_w\mathbf{s},
		\label{nonlinear_control_law_0D}
	\end{equation}
	with $\mathbf{g}_w\in\mathbb{R}^4$ and $\mathbf{H}_w\in\mathbb{R}^{4x4}$. The weight vectors associated to this policy is thus $\mathbf{w}\in\mathbb{R}^{20}$ and it collects all the entries in $\mathbf{g}_w$ and $\mathbf{H}_w$. For later reference, the labelling of the weights is as follows:
	\begin{equation}
		\label{weights_0D}
		\mathbf{g}_{{\color{black}w}}= \begin{bmatrix}w_1\\w_2\\w_3\\w_4 \end{bmatrix} \,\,\,\mbox{and}\,\,\, \mathbf{H}_{{\color{black}w}}= \begin{bmatrix}w_5 & w_9&w_{13}& w_{17}\\w_6 & w_{10}&w_{14}& w_{18}\\w_7 & w_{11}&w_{15}& w_{19}\\w_8 & w_{12}&w_{16}& w_{20} \end{bmatrix}.
	\end{equation}
	Both LIPO and BO seek for the optimal weights in the range [-3,3]. The BO was set up with a Matern kernel (see \eqref{Matern}) with a smoothness parameter $\nu=1.5$, a length scale of $l=0.01$, an acquisition function based on the expected improvement and an exploitation-exploration (see \eqref{EI}) trade-off parameter $\xi=0.1$. {\color{black}Regarding the learning, 100 episodes were taken for BO, LIPO and DDPG. For the GP, the upper limit is set to 1200, considering 20 generations with $\mu=30$ individuals, $\lambda=60$ off-springs and a ($\mu+\lambda$) approach.}
	
	The DDPG experiences are collected with an exploration strategy structured into three parts. The first part (until episode 30) is mostly explorative. Here the noise is clipped in the range [-0.8,0.8] with $\eta=1$ (see \eqref{act}). The second phase (between episode 30 and 55) is an off-policy exploration phase with a noise signal clipped in the range [-0.25,0.25], with $\eta=0.25$. The third phase (from episode 55 onward) is completely exploitative (with no noise). As explorative signal, we used a white noise with a standard deviation of 0.5. 
	
	\subsection{Control of the viscous Burgers's equation}\label{Sec:IVp2}
	
	We consider Burger's equation because it offers a simple 1D problem combining nonlinear advection and diffusion. The problem set is:
	\begin{linenomath*}
		\begin{equation}
			\begin{split}
				\partial_tu + u\partial_xu &= \nu\partial_{xx}u + f(x,t) + c(x,t),\\
				u(x,0) &= u_0\\
				\partial_xu(0,t)&= \partial_xu(L,t)=0
			\end{split}    
			\label{viscous_burgers_eqn}
		\end{equation}
	\end{linenomath*}
	where $(x,t)\in(0,L)\times(0,T]$ with $L=20$ and $T=15$ is the episode length, $\nu=0.9$ is the kinematic viscosity {\color{black} and $u_0$ is the initial condition, defined as the developed velocity field at $t=2.4$ starting from $u(x,0)=0$. The term $f(x,t)$ represents the disturbance and the term $c(x,t)$ is the control actuation, which are} both Gaussian functions in space, modulated by a time varying amplitude:
	\begin{linenomath*}
		\begin{align}
			\label{Burger_DEF}
			f(x,t)&={\color{black}A_f}\sin{(2\pi f_p t)}\cdot\mathcal{N}(x-x_f,\sigma),\\
			c(x,t)&=a(t){\color{black}A_c}\cdot\mathcal{N}(x- {\color{black} x_c},\sigma),
		\end{align}
	\end{linenomath*}
	taking {\color{black}$A_f=$} $100$ and $f_p=0.5$ for the disturbance's amplitude and frequencies and being {\color{black}$A_c=300$ the amplitude of the control and} {\color{black}$a(t)\in[-1,1]$} the action provided by the controller. The disturbance and the controller action are centred at $x_f=6.6$ and $x_c=13.2$ respectively and have $\sigma=0.2$. The uncontrolled system produces a set of nonlinear waves propagating in both directions at approximately constant velocities. The objective of the controller is to neutralize the waves downstream the control location, i.e. for $x>x_c$, using three observations at $x=8,9,10$. {\color{black} Because the system's characteristic is such that perturbations propagate in both directions, the impact of the controller propagates backwards towards the sensors and risks being retrofitted in the loop.}
	
	{\color{black} To analyze how the various agents deal with the retrofitting problem, we consider two scenarios: a `fully closed' loop approach and a `hybrid' approach, in which agents are allowed to produce a constant action. The constant term allows for avoiding (or at least limiting) the retrofitting problem. For the BO and LIPO controllers, we consider linear laws; hence the first approach is 
		\begin{equation}
			\label{action_burger}
			a_A(t;\mathbf{w})=w_0 \,u(8,t)+w_1\, u(9,t)+w_2\, u(10,t)\,,
		\end{equation} while the second is  
		\begin{equation}
			\label{action_burger_2}
			a_B(t;\mathbf{w})=w_0 \,u(8,t)+w_1\, u(9,t)+w_2\, u(10,t)+w_3.
		\end{equation}
		
		For the GP, we add the possibility of a constant action using an ephemeral constant, which is a function with no argument that returns a random value. Similarly, we refer to `A' and 'B' as agents that cannot produce a constant and those that do. For the DDPG, the ANN used to parametrize the policy naturally allows for a constant term; hence the associated agent is `hybrid' by default, and there is no distinction between A and B.}
	
	{\color{black} One can get more insights into the dynamics of the system and the role of the controller from the energy equation associated with \eqref{Burger_DEF}. This equation is obtained by multiplying Eq.\eqref{viscous_burgers_eqn} by u:
		\begin{equation}
			\label{energy}
			\partial_t \mathcal{E} + u \partial_x \mathcal{E} =  {\nu}\bigl[\partial_{xx}\mathcal{E}-\bigl(\partial_x u\bigr)^2\bigr] +2 u\,f(x,t) + 2 u\,c(x,u) 
		\end{equation}
		where $\mathcal{E}=u^2$ is the transported energy and $u\,f(x,t)$ and $u\,c(x,u)$ are the production/destruction terms associated to the forcing action and the control action. Because $f$ and $c$ do not act in the same location, the controller cannot act directly on the source but must rely either on the advection (mechanism I) or the diffusion (mechanism II). The first mechanism consists of sending waves towards the disturbing source so that they are annihilated before reaching the control area. Producing this backward propagation in a fully closed-loop approach is particularly challenging. This is why we added the possibility of an open-loop term. The second mechanism generates large wave numbers, that is waves characterized by large slopes so that the viscous term (and precisely the squared term in the brackets on the right-hand side of \eqref{energy}) provides more considerable attenuation. This second mechanism cannot be used by a linear controller whose actions cannot change the frequency from the sensors' observation.}

\begin{figure} \center
		\includegraphics[width=0.6\textwidth,trim={0.1cm 0.25cm 0.4cm 0},clip]{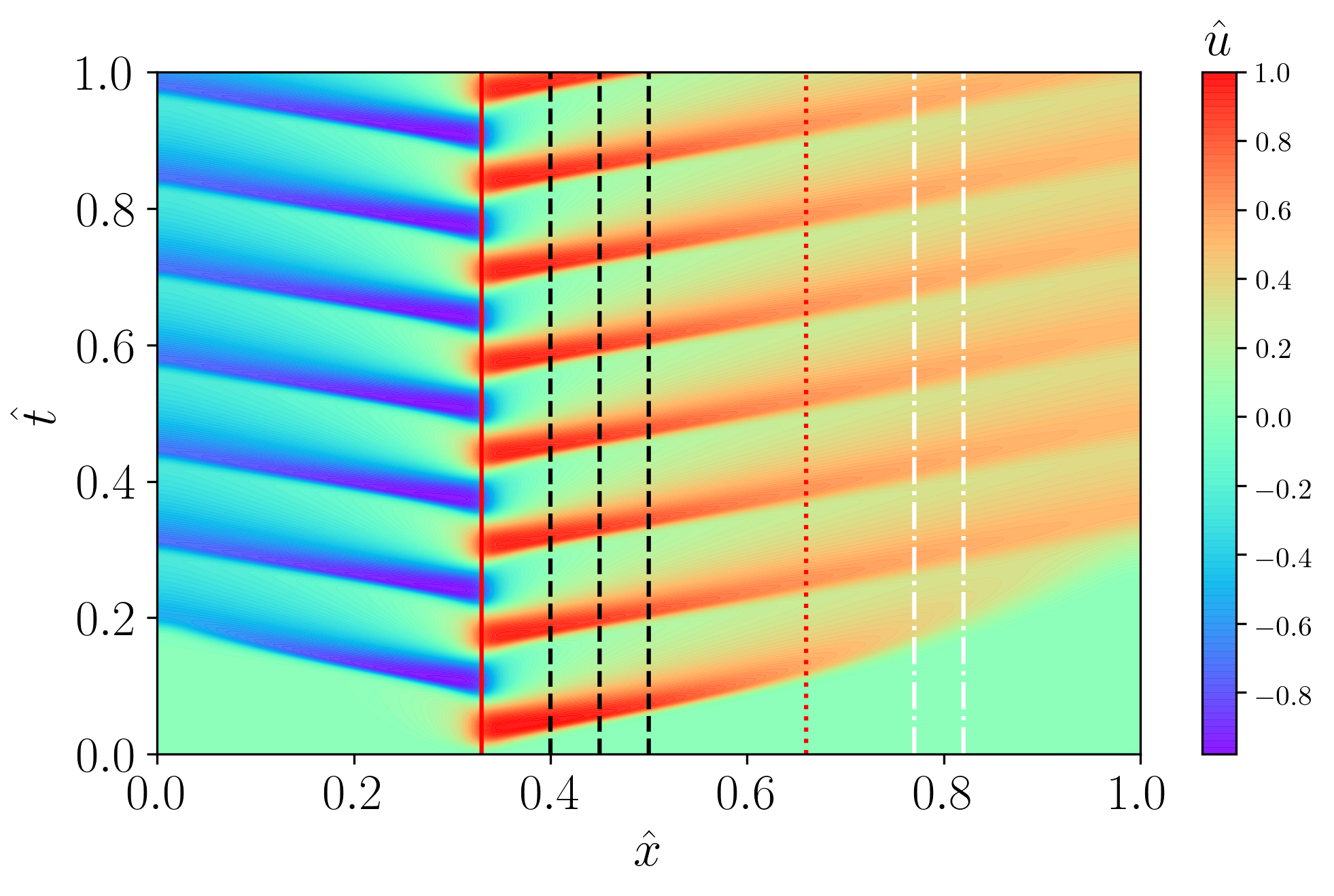}
		\caption{Contour plot of the spatio-temporal evolution of {\color{black} normalized $\hat{u}=u/max(u)$} in \eqref{viscous_burgers_eqn} for the uncontrolled problem, i.e $c(x,t)=0$ {\color{black} in the normalized space-time domain ($\hat{x}=x/L$, $\hat{t}=t/T$)}. The perturbation is centered at {\color{black}$\hat{x}=0.33$} (red continuous line) while the control law is centered at {\color{black}$\hat{x}=0.66$ (red dotted line)}. The dashed black lines visualize the location of the observation points, while the region within the white dash-dotted line is used to evaluate the controller performance.    }
		\label{viscous_burgers_env_wo_control}
	\end{figure}

	The controller's performance is measured by the reward function:
	\begin{equation}
		\label{rew_Burger}
		r(t) = -\Big(\ell_2(u_t)_{\Omega_r} + \alpha\cdot a(t)^2\Big)
	\end{equation}
	where $\ell_2(\cdot)_{\Omega_r}$ is the Euclidean norm of the displacement $u_t$ at time step $t$ over a portion of the domain $\Omega_r = \{x\in\mathbb{R}|15.4\leq x\leq 16.4\}$ called reward area, $\alpha$ is a penalty coefficient and $a_t$ is the value of the control action selected by the controller. The cumulative reward is computed with a discount factor $\gamma=1$while the penalty in the actions was set to $\alpha=100$.
	{\color{black} This penalty gives comparable importance to the two terms in \eqref{rew_Burger} for the level of wave attenuation achieved by all agents.} Figure \ref{viscous_burgers_env_wo_control} shows the evolution of the uncontrolled system in a contour plot in the space-time domain, recalling the location of perturbation, action, observation and reward area.
	
	Eq.\eqref{viscous_burgers_eqn} was solved using Crank–Nicolson’s method. The Neumann boundary conditions are enforced using ghost cells, and the system is solved at each time step via the banded matrix solver \textit{solve\_banded} from the python library \textit{scipy}. The mesh consists of $n_x=1000$ points and the time stepping is $\Delta t=0.01$, thus leading to $n_t=1500$ steps per episode.
	
	\par Both LIPO and BO optimizers operate within the bounds [-0.1, 0.1] for the weights to avoid saturation in the control action. The overall set-up of these agents is the same as the one used in the 0D test case. For the GP, the selected evolutionary strategy is $(\mu+\lambda)$, with the initial population of 10 individuals $\mu = 10$ and an offspring $\lambda = 20$ {\color{black} trained for 20 generations}. The DDPG agent set-up relies on the same reward normalization and buffer prioritization presented for the previous test case. However, the trade-off between exploration and exploitation was handled differently: the random noise term in \eqref{act} is set to zero every $N=3$ episodes to prioritize exploitation. This noise term was taken as an Ornstein-Uhlenbeck, time-correlated noise with $\theta=  0.15$ and $dt = 1e-3$ and its contribution was clipped in the range [-0.3, 0.3]. {Regarding the learning, the agent was trained for 30 episodes.}
	
	\subsection{Control of the von Kármán street behind a 2D cylinder}\label{Sec:IVp3}
	
	\begin{figure*}
		\centering 
		\includegraphics[width=0.9\textwidth]{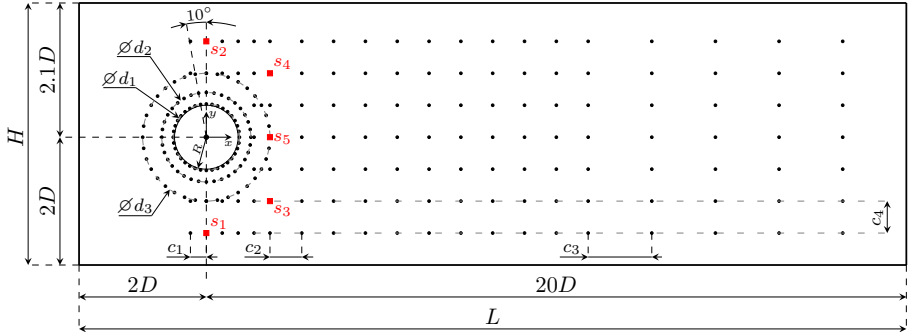}
		\caption{Geometry and observations probes for the 2D von Kármán street control test case. The 256 observations used by \cite{Tang2020e} are shown with black markers. These are organized in three concentric circles (diameters $1+0.002/D$, $1+0.02D$ and $1+0.05D$) around the cylinder and three grids (horizontal spacing $c_1 = 0.025/D$, $c_2 = 0.05/D$ and $c_3 = 0.1/D$). All the grids have the same vertical distance between adjacent points ($c_4 = 0.05/D$). The five observations used in this work (red markers) have coordinates $s_1(0,-1.5)$, $s_2(0,1.5)$, $s_3(1,-1)$ and $s_4(1,1)$ and $s_5(1,0)$. Each probe samples the pressure field.}
		\label{cylinder_des}
	\end{figure*}
	
	The third test case consists in controlling the 2D viscous and incompressible flow past a cylinder in a channel. The flow past a cylinder is a classic benchmark for bluff body wakes \citep{Zhang1995,NOACK2003}, exhibiting a supercritical Hopf bifurcation leading to the well known von Kármán vortex street. The cylinder wake configuration within a narrow channel has been extensively used for CFD benchmark purposes \citep{schafer1996benchmark} and as a test case for flow control techniques \citep{rabault2019artificial,Tang2020e,Li2021}. 
	
	We consider the same control problem as in \cite{Tang2020e}, sketched in Figure \ref{cylinder_des}. The computational domain is a rectangle of width $L$ and height $H$, with a cylinder of diameter $D=0.1$m located slightly off the symmetric plane of the channel (cf. Fig. \ref{cylinder_des}). This asymmetry triggers the development of vortex shedding.
	
	{\color{black} The channel confinement potentially leads to a different dynamics compared to the unbounded case. Depending on the blockage ratio ($b=D/H$), low frequency modes might be damped, promoting the development of high frequencies. This leads to a lower critical Reynolds and Strouhal numbers \citep{singha2010flow,kumar2006effect}, the flattening of the recirculation region and different wake lengths \citep{wiliamson1996vortex,rehimi2008experimental}. However, \cite{griffith2011vortex} and \cite{camarri2010effect} showed, through numerical simulations and Floquet stability analysis, that for $b=0.2$ ($b\approx 0.24$ in our case) the shedding properties are similar to those of the unconfined case. Moreover, it is worth stressing that the flow is expected to be fully 3D for the set of parameters here considered \cite{mathupriya2018numerical,kanaris2011three}. Therefore, the 2D test case considered in this work is a rather academic benchmark, yet characterized by a rich and complex dynamics \citep{sahin2004numerical} reproducible at a moderate computational cost.}
	

The reference system is located at the centre of the cylinder. At the inlet ($x=-2D$), as  in \cite{schafer1996benchmark}, a parabolic velocity profile is imposed:
\begin{equation}
	u_{inlet} = \frac{-4U_m}{H^2}\Big(y^2 - 0.1Dy - 4.2D^2\Big),
\end{equation}
where $U_m=1,5$m/s. This leads to a Reynolds number of $Re=\overline{U} D/\nu=400$ using the mean inlet velocity $\overline{U}=2/3 U_m$ as a reference and taking a kinematic viscosity of $\nu=2.5e-4$m$^2$/s. It is worth noticing that this is much higher than $Re=100$ considered by \cite{Jin2020}, who defines the Reynolds number based on the maximum velocity.

The computational domain is discretized with an unstructured mesh refined around the cylinder, and the incompressible Navier-Stokes equations are solved using the incremental pressure correction scheme (IPCS) method in the FEniCS platform \citep{alnaes2015fenics}. The mesh consists of 25865 elements and simulation time step is set to $\Delta t=1e-4[s]$ to respect the CFL condition. The reader is referred to \cite{Tang2020e} for more details on the numerical set-up and the mesh convergence analysis. 

In the control problem, every episode is initialized from a snapshot that has reached a developed shedding condition. This was computed by running the simulation without control for $T = 0.91$s $= 3T^*$, where $T^*= 0.303$s is the vortex shedding period. We computed $T^*$ by analyzing the period between consecutive pressure peaks observed by probe $s_5$ in an uncontrolled simulation. The result is the same as the one found by \cite{Tang2020e}, who performed a Discrete Fourier Transform (DFT) of the drag coefficient.

The instantaneous drag and lift on the cylinder are calculated via the surface integrals:
\begin{equation}
	F_D = \int\,(\sigma\cdot n)\cdot e_x\, dS, \qquad F_L = \int\,(\sigma\cdot n)\cdot e_y\, dS,
\end{equation}
where $S$ is the cylinder surface, $\sigma$ is the Cauchy stress tensor, $n$ is the unit vector normal to the cylinder surface, $e_x$ and $e_y$ are the unit vectors of the x and y axes respectively. The drag and lift coefficient are calculated as $C_D = {2F_D}/({\rho\Bar{U}^2D})$ and $C_L = {2F_L}/({\rho\Bar{U}^2D})$ respectively.

\begin{figure}
	\centering
	\includegraphics[width=0.26\textwidth]{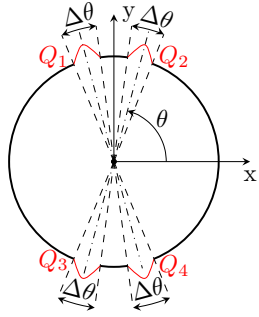}
	\caption{Location of the four control jets for the 2D von Kármán street control test case. These are located at $\theta=75^{o}, 105^{o}, 255^{o}, 285^{o}$ and have width $\Delta \theta = 15^{o}$. The velocity profile is defined as in \eqref{Eq_Jets}, with flow rate defined by the controller and shifted to have zero-net mass flow.}
	\label{cylinder_zoom}
\end{figure}

The control action consists in injecting/removing fluid from four synthetic jets positioned on the cylinder boundary as shown in Figure \ref{cylinder_zoom}. The jets are symmetric with respect to the horizontal and vertical axes. These are located at $\theta=75^{o}, 105^{o}, 255^{o}, 285^{o}$ and have the same width $\Delta \theta = 15^{o}$. The velocity profile in each of the jets is taken as:

\begin{equation}
	\label{Eq_Jets}
	u_{jet}(\theta) = \frac{\pi}{\Delta \theta D}Q_i^*\cos{\Big(\frac{\pi}{\Delta \theta}(\theta - \theta_i)\Big)}
\end{equation}
where $\theta_i$ is the radial position of the i-th jet and $Q^*_i$ is the imposed flow rate. Eq \eqref{Eq_Jets} respects the non-slip boundary conditions at the walls. To ensure a zero-net mass injection at every time step, the flow rates are mean shifted as $Q_i^* = Q_i - \Bar{Q}$ with $\Bar{Q}=\frac{1}{4}\sum_{i}^{4}\,Q_i$ the mean value of the four flow rates.

The flow rates in the four nozzle constitute the action vector, i.e. $\mathbf{a}=[Q_1,Q_2,Q_3,Q_4]^T$ in the formalism of Section \ref{Sec:II}. To avoid abrupt changes in the boundary conditions, the control action is kept constant for a period of $T_c=100\Delta t= 1e-2[s]$. This is thus equivalent to having a moving average filtering of the controller actions with impulse response of length $N=10$. The frequency modulation of such a filter is 

\begin{equation}
	\label{Filt}
	H(\omega)=\frac{1}{10}\Bigl |\frac{\sin(5\omega)}{\sin(\omega/2)}\Bigr|
\end{equation} with $\omega=2\pi f/f_s$. The first zero of the filter is located at $\omega=2\pi/5$, thus $f=f_s/5=2000 Hz$, while the attenuation at the shedding frequency is negligible. Therefore, this filtering allows the controller to act freely within the range of frequencies of interest to the control problem, while preventing abrupt changes that might compromise the stability of the numerical solver. Each episode has a duration of $T=0.91$s, corresponding to $2.73$ shedding periods in uncontrolled conditions. This allows having 91 interactions per episode (i.e. 33 interactions per vortex shedding period).

The actions are linked to the pressure measurements (observations of the flow) in various locations. In the original environment by \cite{Tang2020e}, 256 probes were used, similarly to \cite{rabault2019artificial}. The locations of these probes are shown in Figure \ref{cylinder_des} using black markers. In this work, we reduce the set of probes to $n_s=5$. A similar configuration was analyzed by \cite{rabault2019artificial} although using different locations. In particular, we kept the probes $s_1$ and $s_2$ at the same $x$ coordinate, but we moved them further away from the cylinder wall to reduce the impact of the injection on the sensing area. Moreover, we slightly move the sensors $s_3, s_4, s_5$ downstream in regions where the vortex shedding is stronger. {\color{black} The chosen configuration has no guarantee of optimality and was heuristically defined by analyzing the flow field in the uncontrolled configuration. Optimal sensor placement for this configuration is discussed by \cite{paris2021robust}.}

The locations used in this work are recalled in Figure \ref{cylinder_des}. The state vector, in the formalism of Section \ref{Sec:II}, is thus the set of pressure at the probe locations, i.e. $\mathbf{s}=[p_1,p_2,p_3,p_4,p_5]^T$. For the optimal control strategy identified via the BO and LIPO algorithms in Section \ref{sec:BO} and \ref{sec:LIPO}, a linear control law is assumed, hence $\mathbf{a}=\mathbf{W}\mathbf{s}$, with the 20 weight coefficients labelled as follows

\begin{equation}
	\begin{bmatrix}Q_1  \\Q_2\\Q_3\\Q_4 \end{bmatrix} =
	\begin{bmatrix}w_1 & w_2& w_3& w_4 & w_5 \\w_6 & w_7& w_8& w_9 & w_{10} \\w_{11} & w_{12}& w_{13}& w_{14} & w_{15} \\w_{16} & w_{17}& w_{18}& w_{19} & w_{20}  \end{bmatrix} 
	\begin{bmatrix}p_1  \\p_2\\p_3\\p_4\\p_5 \end{bmatrix}    \,\,.
\end{equation}

It is worth noticing the zero-net mass condition enforced by removing the average flow rate from each action could be easily imposed by constraining all columns of $\mathbf{W}$ to add up to zero. For example, setting the symmetry $w_1=-w_{11}$, $w_6=-w_{16}$, etc. (leading to $Q_1=-Q_3$ and $Q_2=-Q_4$) allows for halving the dimensionality of the problem and thus considerably simplifying the optimization. Nevertheless, one has infinite ways of embedding the zero-net mass condition and we do not impose any, letting the control problem act in $\mathbb{R}^{20}$.

Finally, the instantaneous reward $r_t$ is defined as
\begin{equation}
	r_t = \langle F_{D}^{base}\rangle_{T_c} -\langle F_D\rangle_{T_c} - \alpha|\langle F_L\rangle_{T_c}|\,,
	\label{eq:rew_cylinder}
\end{equation} where $\langle\bullet\rangle_{T_c}$ is the moving average over $T_c=10\Delta t$, $\alpha$ is the usual penalization parameter set to {\color{black} $0.2$} and $F_{D}^{base}$ is the averaged drag due to the steady and symmetric flow. This penalization term prevents the control strategies from relying on the high lift flow configurations \cite{rabault2019artificial} and simply blocking the incoming flow. The cumulative reward was given with $\gamma=1$.  According to \cite{Bergmann2005}, the active flow control cannot reduce the drag due to the steady flow, but only the one due to the vortex shedding. Hence, in the best case scenario, the cumulative reward is the sum of the averaged steady state drag contributions:
\begin{equation}
	R^* = \sum_{t=1}^T\,r_t = \sum_{t=1}^T\,\langle F_{D}^{base}\rangle_{T_c} = 14.5.
	\label{reward_fun_cylinder}
\end{equation}

The search space for the optimal weights in LIPO and BO was bounded to [-1, 1]. Moreover, the action resulting from the linear combination of such weights with the states collected in the $i-$th interaction was multiplied by a factor $2e-3$, to avoid numerical instabilities. The BO settings are the same as in the previous test-cases, except for the smoothness parameter that was reduced to $\nu = 1.5$. On the GP side, the evolutionary strategy applied was the \textit{eaSimple}'s \citep{easimple} implementation in Deap - with hard-coded elitism to preserve the best individuals.  {\color{black}
	To allow the GP to provide multi outputs, four populations of individuals were trained simultaneously (one for each control jet). Each population evolves independently (with no genetic operations allowed between them) although the driving reward function (Eq.\eqref{reward_fun_cylinder}) values their collective performance. This is an example of multi-agent reinforcement learning. Alternative configurations, to be investigated in future works, are the definition of a multiple-output trees or cross-population genetic operations.}


Finally, the DDPG agent was trained using the same exploration policy of the Burgers' test-case, alternating 20 exploratory episodes with $\eta=1$ and 45 exploitative episodes with $\eta=0$ (c.f eq \eqref{eq:rew_cylinder}). During the exploratory phase, an episode with $\eta=0$ is taken every $N=4$ episodes and the policy weights are saved. We used the Ornstein-Uhlenbeck time correlated noise with $\theta=  0.1$ and $dt = 1e-2$ in eq. \eqref{act}, clipped in the range [-0.5, 0.5]. 

\section{Results and Discussions}\label{Sec:V}

We present here the outcomes of the different control algorithms in terms of learning curves and control actions for the three investigate test cases. Given the heuristic nature of these control strategies, we ran several training sessions for each, using different seeding values for the random number generator. We define as \emph{learning curve} the upper bound of the cumulative reward $R(\mathbf{w})$ in \eqref{REW} obtained \emph{at each episode} within the various training sessions. Moreover, we define as \emph{learning variance} the variance of the global reward between the various training sessions \emph{at each episode}. We considered ten training sessions for all environments and for all control strategies. In the episode counting shown in the learning curves and the learning variance, it is worth recalling that the BO initially performs 10 explorative iterations. {\color{black}For the DDPG, since the policy is continuously updated at each time step, the global reward is not representative of the performances of a specific policy but is used here to provide an indication of the learning behaviour.}

For the GP, each iteration involves $n_p$ episodes, with $n_p$ the number of individuals in the population (in a jet actuation). The optimal weights found by the optimizers and the best trees found by the GP are reported in the appendix.

{\color{black} Finally, for all test cases, we perform a robustness analysis for the derived policies. This analysis consists in testing all agents in a set of 100 episodes with random initial conditions and comparing the distribution of performances with the ones obtained during the training (where the initial condition was always the same). It is worth noticing that different initial conditions could be considered during the training, as done by \cite{castellanos2022machine}, to derive the most robust control law for each method. However, in this work we were interested in the \emph{best} possible control law (at the cost of risking over-fitting) for each agent and their ability to \emph{generalize} in settings that differ from the training conditions. }

\subsection{The 0D Frequency Cross-talk problem}

We here report on the results for the four algorithms for the 0D problem in Section \ref{Sec:IVp1}. All implemented methods found strategies capable of solving the control problem, bringing to rest the first oscillator ($s_1,s_2$) while exiting the second ($s_3,s_4$). Table \ref{table_results_0D} collects the final best cumulative reward for each control method together with the confidence interval, defined as $1.96$ time the standard deviation within the various training sessions.

\FloatBarrier
\begin{table}
	\centering
	\begin{tabular}{c@{\hspace{0.3cm}}c@{\hspace{0.2cm}}c@{\hspace{0.2cm}}c@{\hspace{0.2cm}}c@{\hspace{0.2cm}}}
		\toprule
		$\cdot 10^{-3}$ & \textbf{LIPO} & \textbf{BO} & \textbf{GP} & \textbf{DDPG}\\ 
		\midrule
		\begin{tabular}[c]{@{}c@{}}Best \\ Reward\end{tabular}& 
		-\textbf{8.96}$\;\pm$0.75 & -\textbf{9.41}$\;\pm$1.33 & -\textbf{2.77}$\;\pm$1.49 & -\textbf{2.98}$\;\pm$1.37\\ 
		\bottomrule
	\end{tabular}
	\caption{Mean optimal cost function (bold) and confidence interval (over 10 training sessions with different random number generator seeds) obtained ad the end of the training for the 0D frequency cross-talk control problem.}
	\label{table_results_0D}
\end{table}

\begin{figure}
	\centering
	\begin{subfigure}{.5\textwidth}
		\centering
		\includegraphics[width=\textwidth]{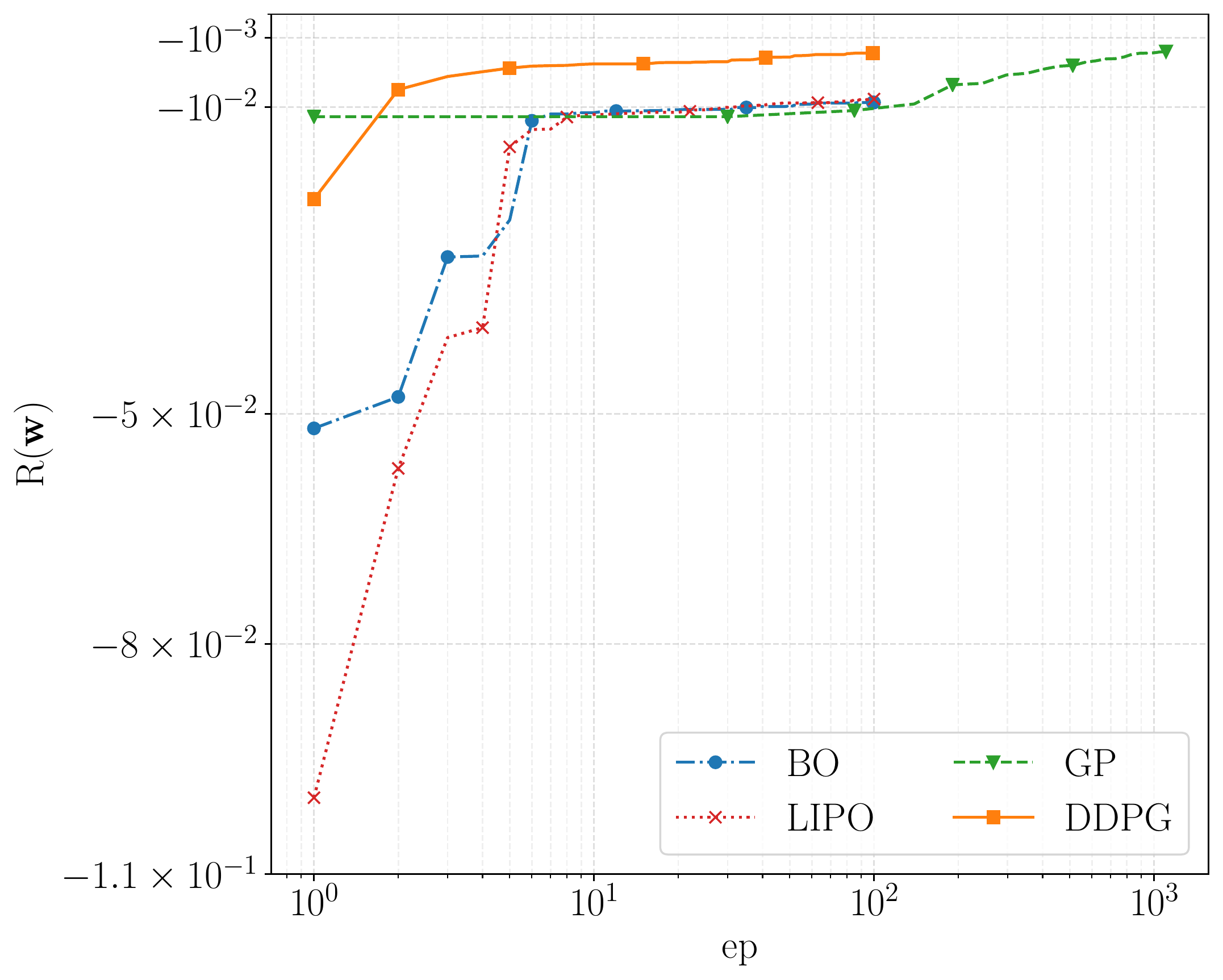}
		\caption{Learning curve }
		\label{learning_curve_0D}
	\end{subfigure}%
	\begin{subfigure}{.5\textwidth}
		\centering
		\includegraphics[width=0.92\textwidth]{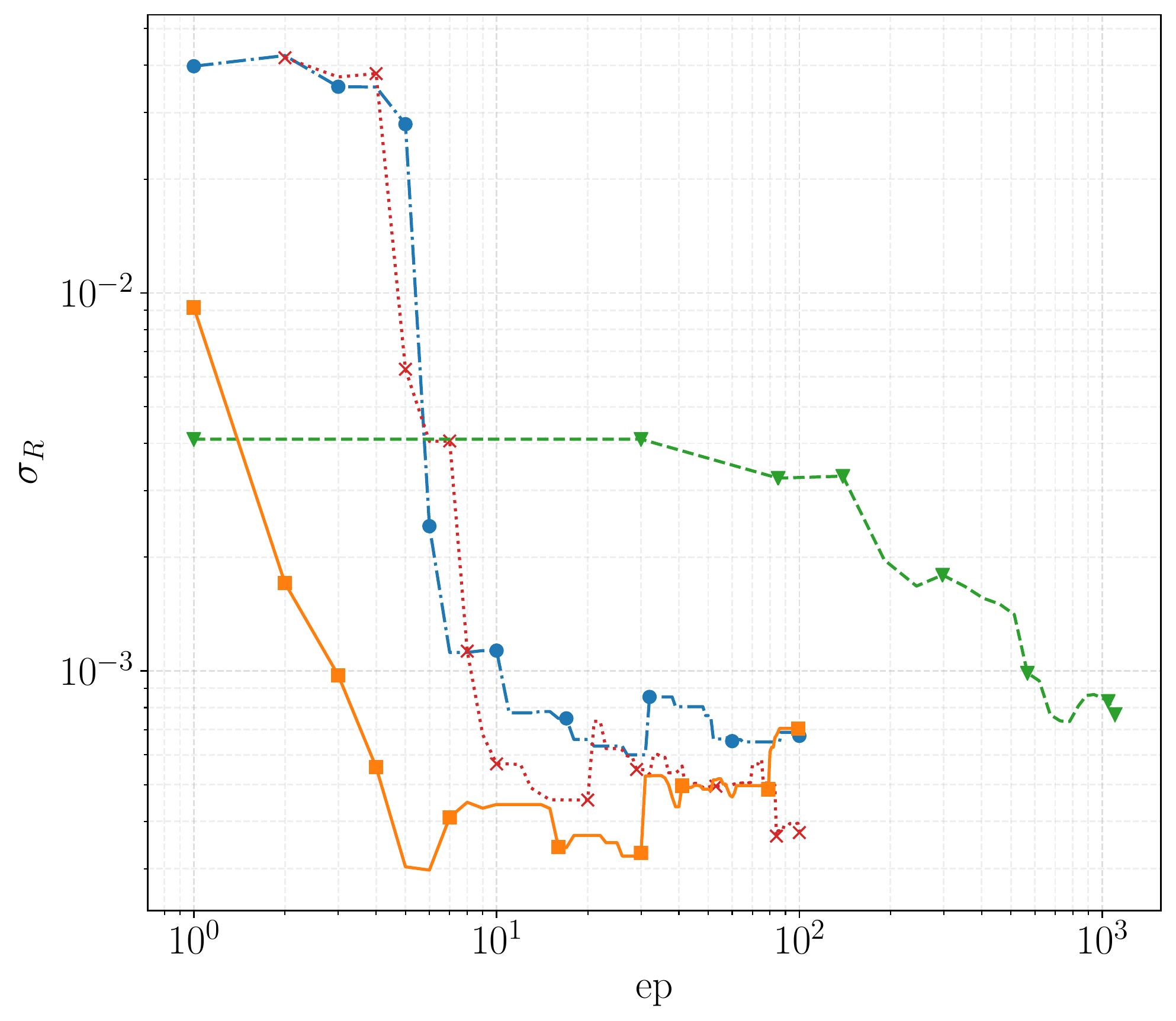}
		\caption{Learning curve variance}
		\label{standard_deviation_curves_0D}
	\end{subfigure}
	\caption{Comparison of the learning curves (a) and their variances (b) for different machine learning methods for the 0D test case (Sec. \ref{Sec:IVp1}).}
\end{figure}

The control law found by the GP yields the highest reward and the highest variance. Figures \ref{learning_curve_0D} and \ref{standard_deviation_curves_0D} show the learning curve and learning variance for the various methods.

\begin{figure}
	\centering
	\includegraphics[width=0.5\textwidth]{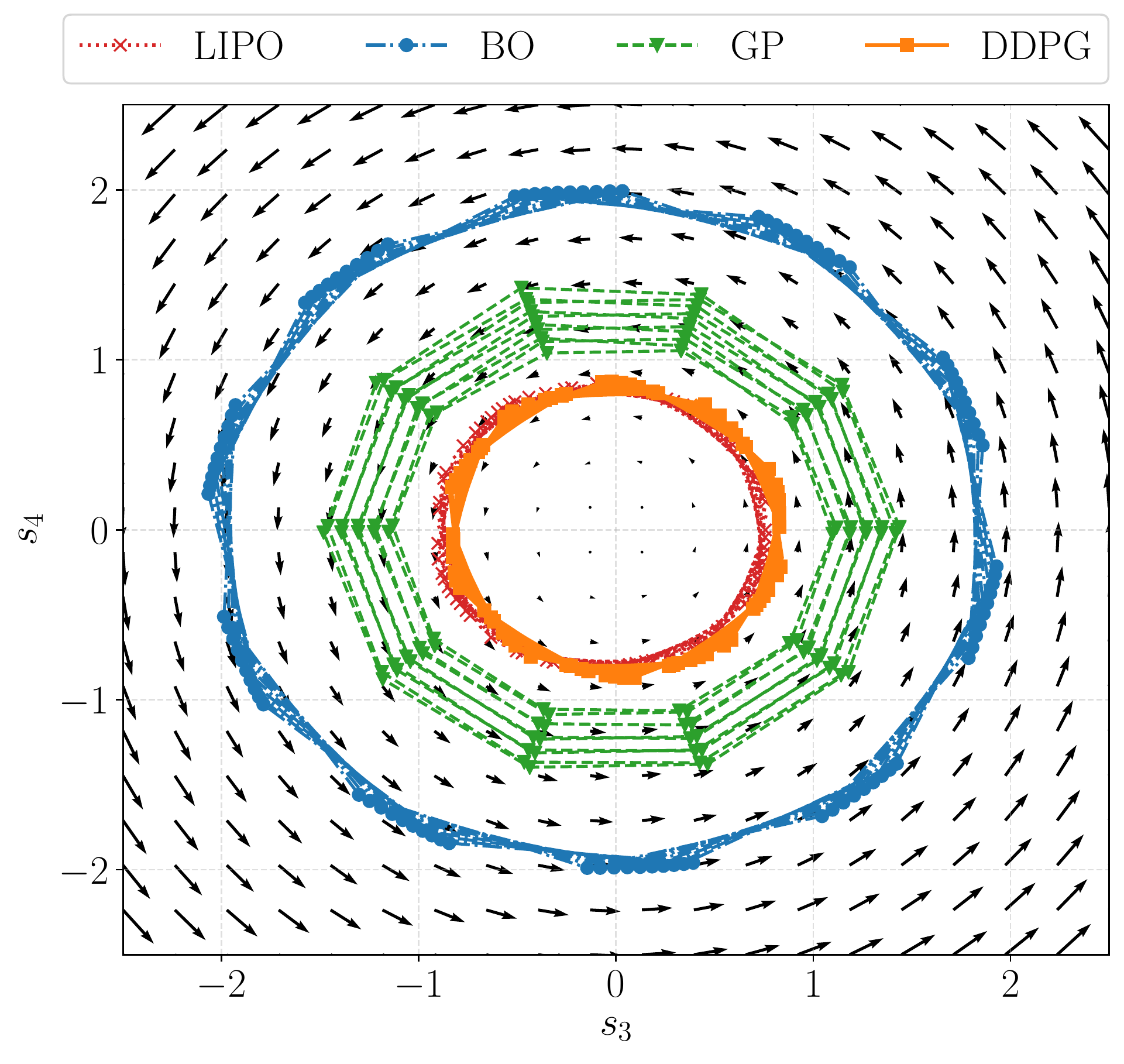}
	\caption{Orbit of the second oscillator ($s_3,s_4$) in the 0D control problem governed by Eq.\eqref{gov_eqn_0D}) ({\color{black} right} column of {\color{black}Table}~\ref{0D_results_1}) in the last part of the episode (from 194s to 200s). The colored curves corresponds to the four control methods.}
	\label{comp_s3_s4_0D_state_space}
\end{figure}

\begin{figure*}
	\centering
	\includegraphics[width=0.95\textwidth]{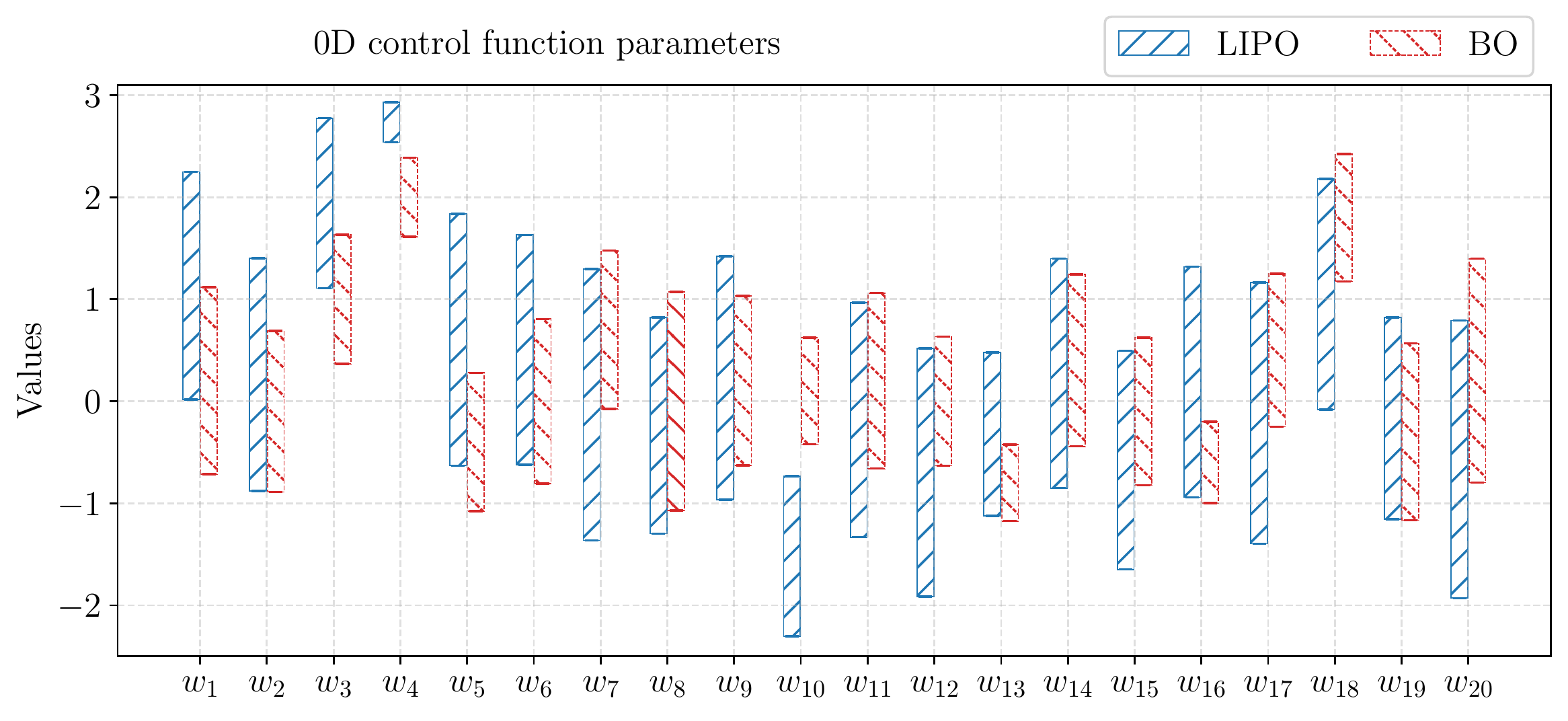}
	\caption{Weights of the control action for the 0D control problem in \eqref{0D_weights_comp_LIPO_BO}. The coloured bars represent a standard deviation around the mean value found by LIPO and BO.}
	\label{0D_weights_comp_LIPO_BO}
\end{figure*}

The learning curve for the GP is initially flat because the best reward from the best individuals of each generation is taken after all individuals have been tested. Considering that the starting population consists of 30 individuals, this shows that approximately three generations are needed before significant improvements are evident. In its simple implementation considered here, the distinctive feature of the GP is the lack of a programmatic explorative phase: exploration proceeds only through the genetic operations, and their repartition does not change over the episodes. This leads to a relatively constant (and significant) reward variance over the episodes. Possible variants to the implemented algorithms could be the reduction of the explorative operations (e.g. mutation) after various iterations (see, for example, \cite{Mendez2021_F}). Nevertheless, the extensive exploration of the function space, aided by the large room for manoeuvre provided by the tree formalism, is arguably the main reason for the success of the method, which indeed finds the control law with the best cumulative reward (at the expense of a much larger number of episodes).

In the case of the DDPG, the steep improvement in the learning curve in the first 30 episodes might be surprising, recalling that in this phase the algorithm is still in its heavy exploratory phase (see Sec. \ref{Sec:IIIp3}). This trend is explained by the interplay of two factors: (1) we are showing the upper bound of the cumulative reward and (2) the random search is effective in the early training phase since improvements over a (bad) initial choice are easily achieved by the stochastic search, but smarter updates are needed as the performances improve. This result highlights the importance of the stochastic contribution in \eqref{act}, and its adaptation during the training to balance exploration and exploitation.

\begin{table*}
	\centering
	\begin{tabular*}{\textwidth}{cc}
		\multicolumn{2}{c}{\hspace{5mm}\textbf{LIPO} }\\ \toprule
		\hspace{5mm}
		\includegraphics[width = 0.43\textwidth]{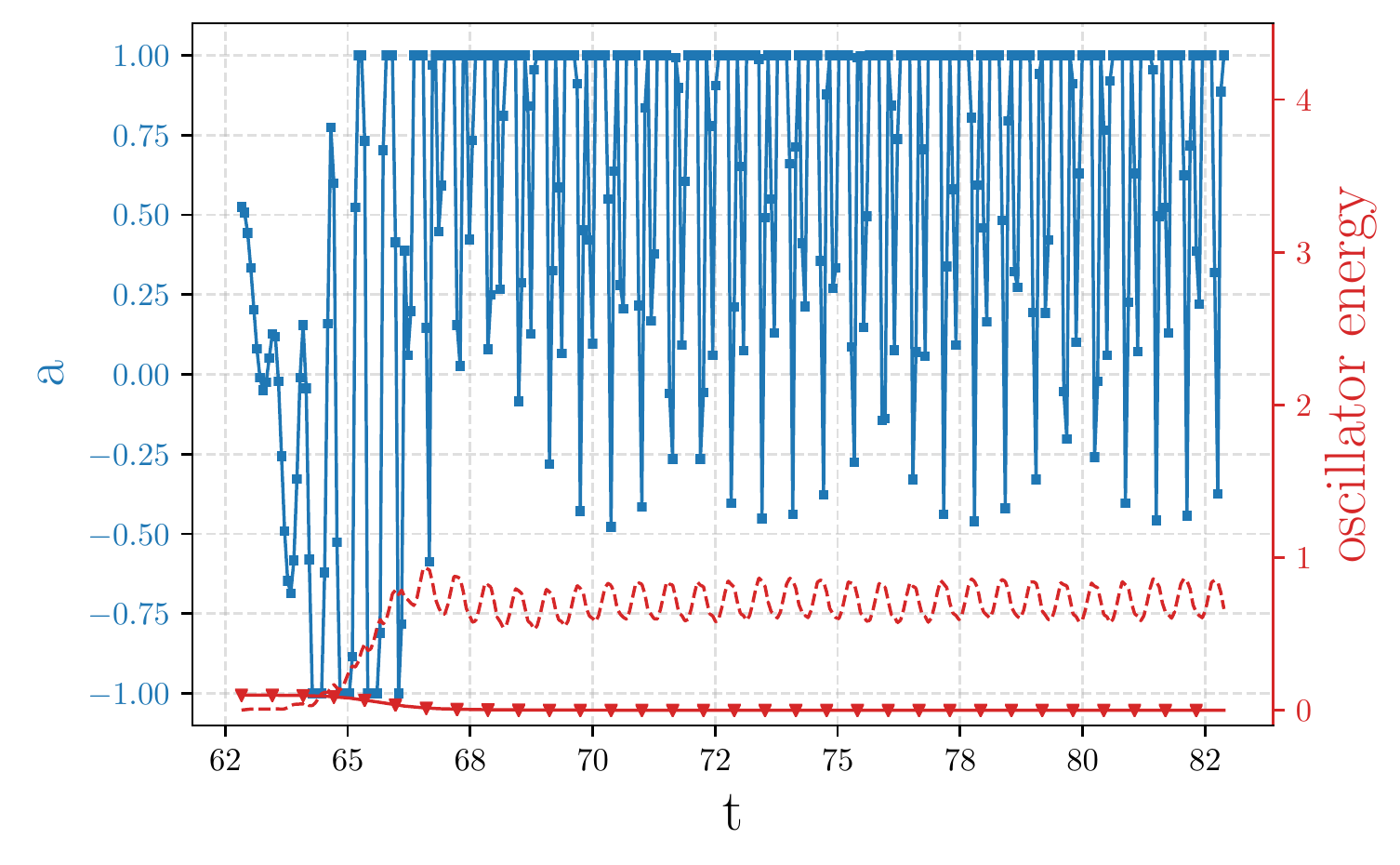}
		&\hspace{5mm} \includegraphics[width = 0.43 \textwidth]{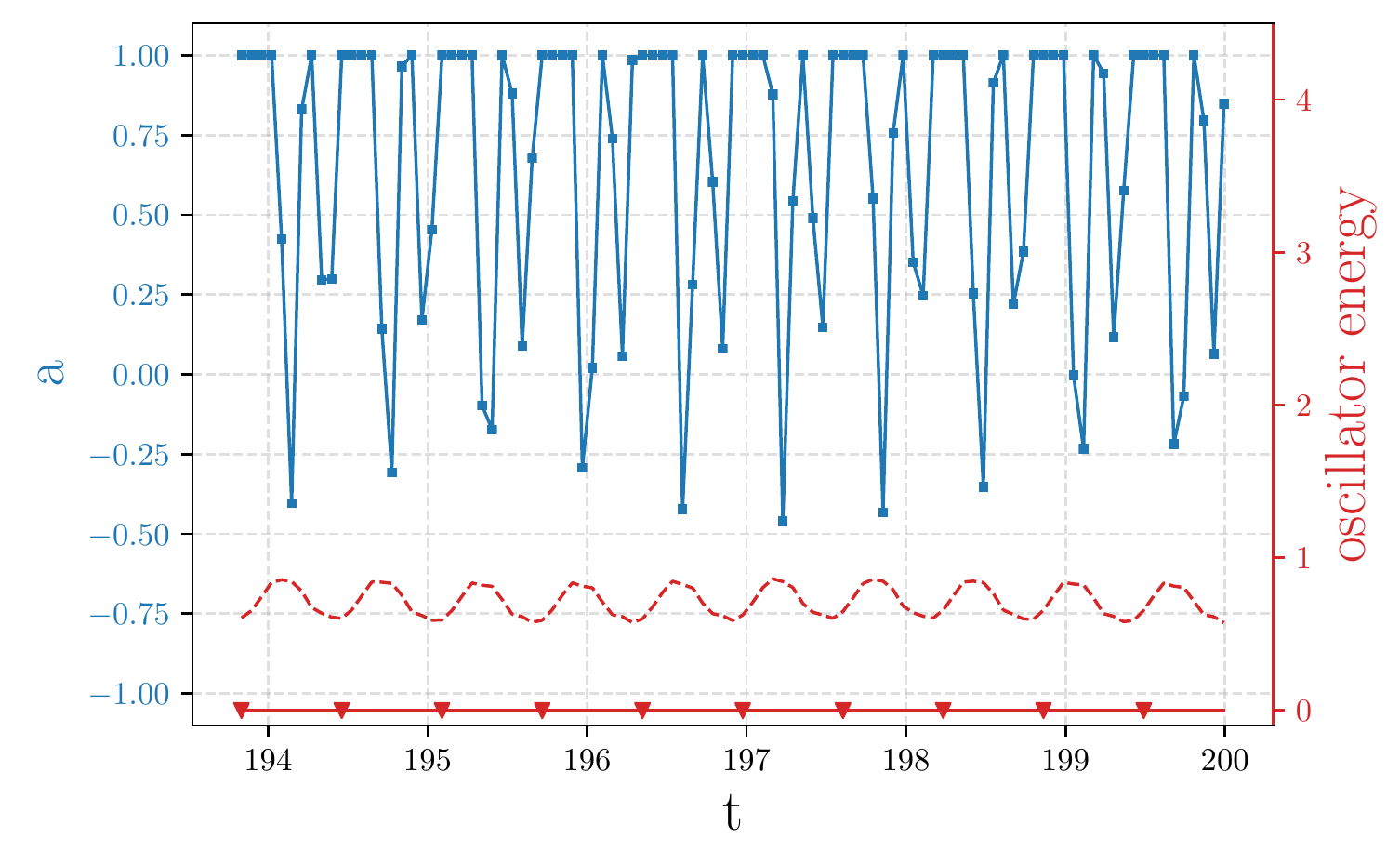}
		\\ 
		\multicolumn{2}{c}{\hspace{5mm}\textbf{BO}}              \\ \toprule
		\multicolumn{1}{c}{}\includegraphics[width = 0.43 \textwidth]{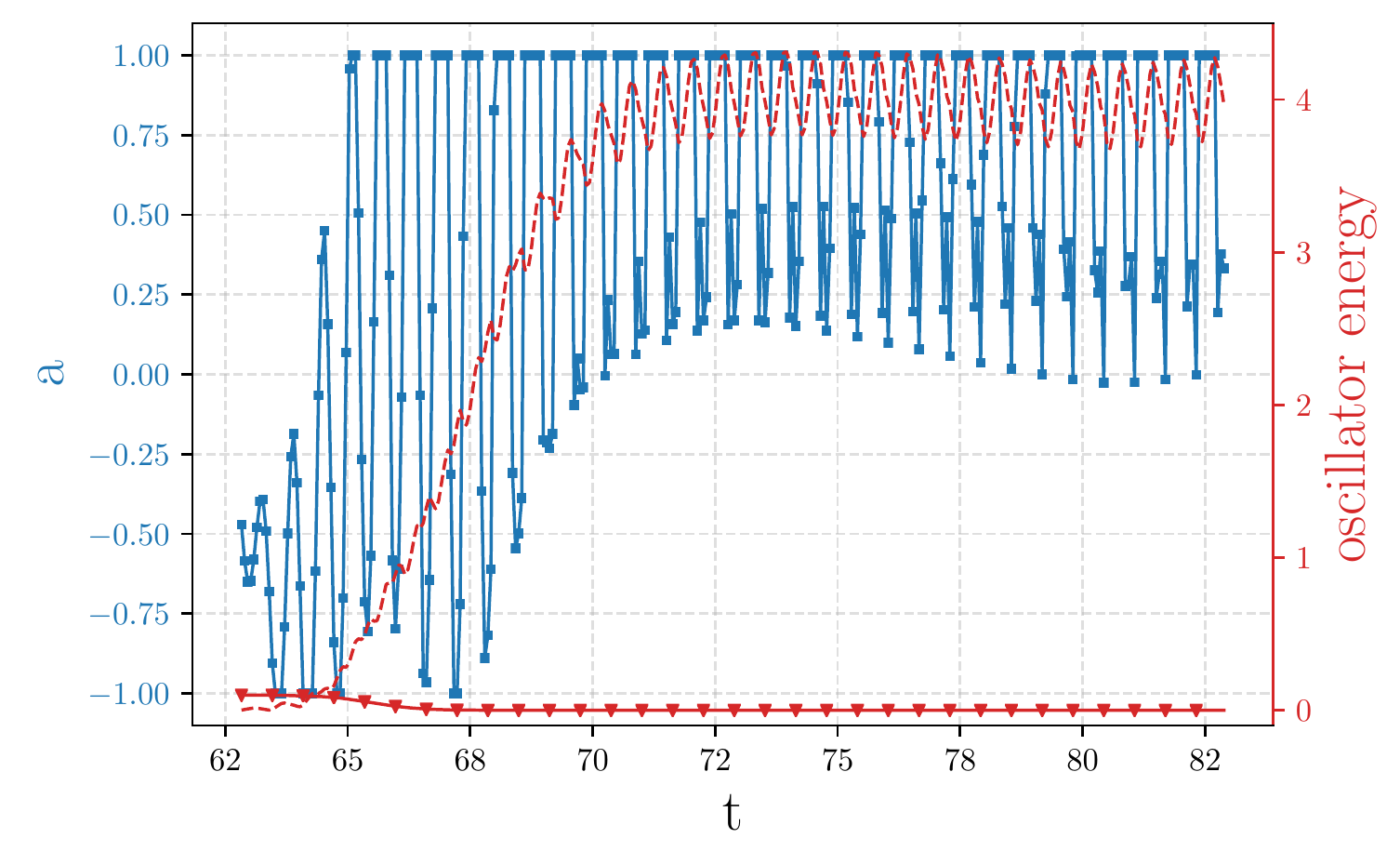} & \multicolumn{1}{c}{}\includegraphics[width = 0.43 \textwidth]{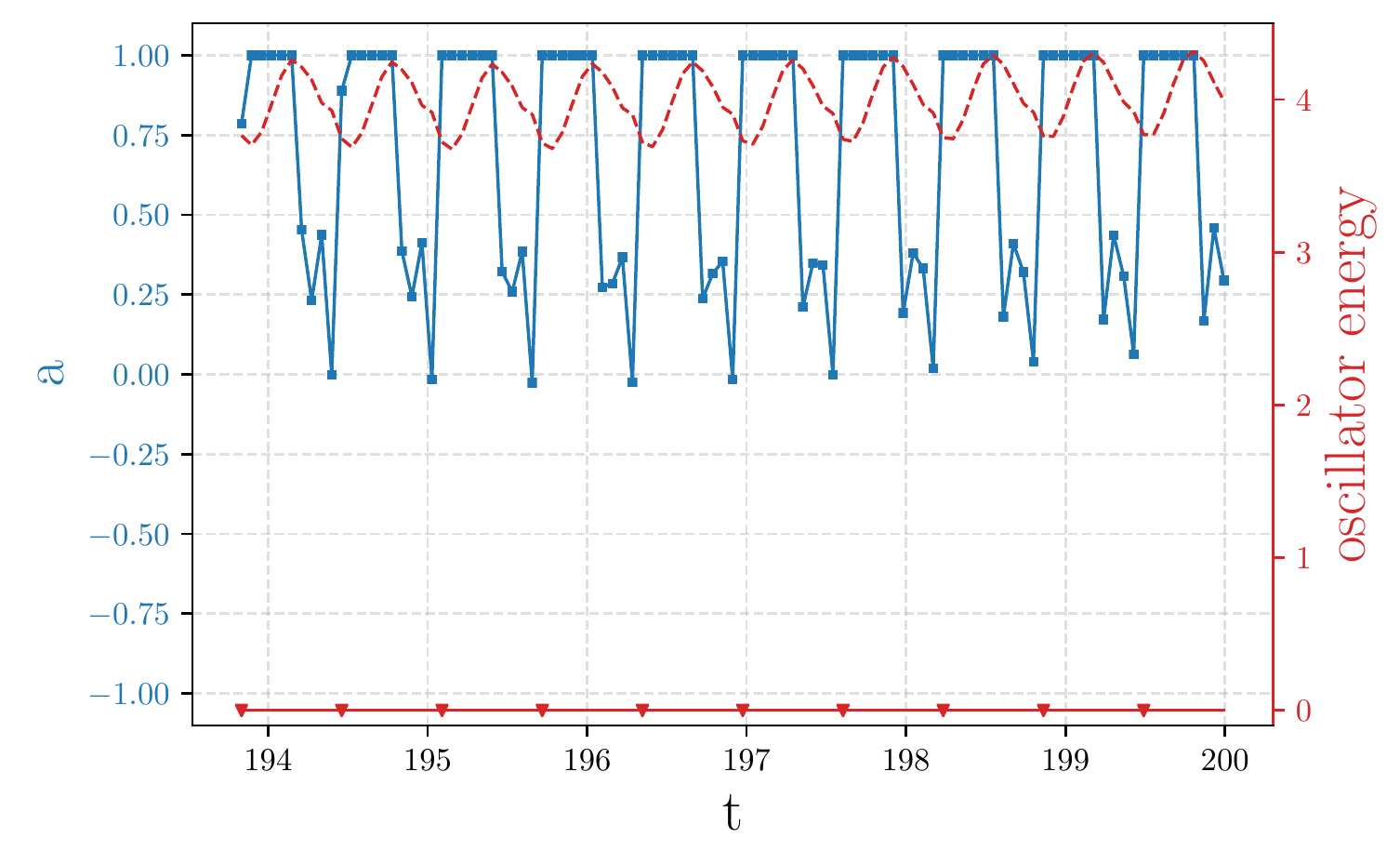} \\
		\multicolumn{2}{c}{\hspace{5mm}\textbf{GP}}              \\ \toprule
		\multicolumn{1}{c}{} \includegraphics[width = 0.43 \textwidth]{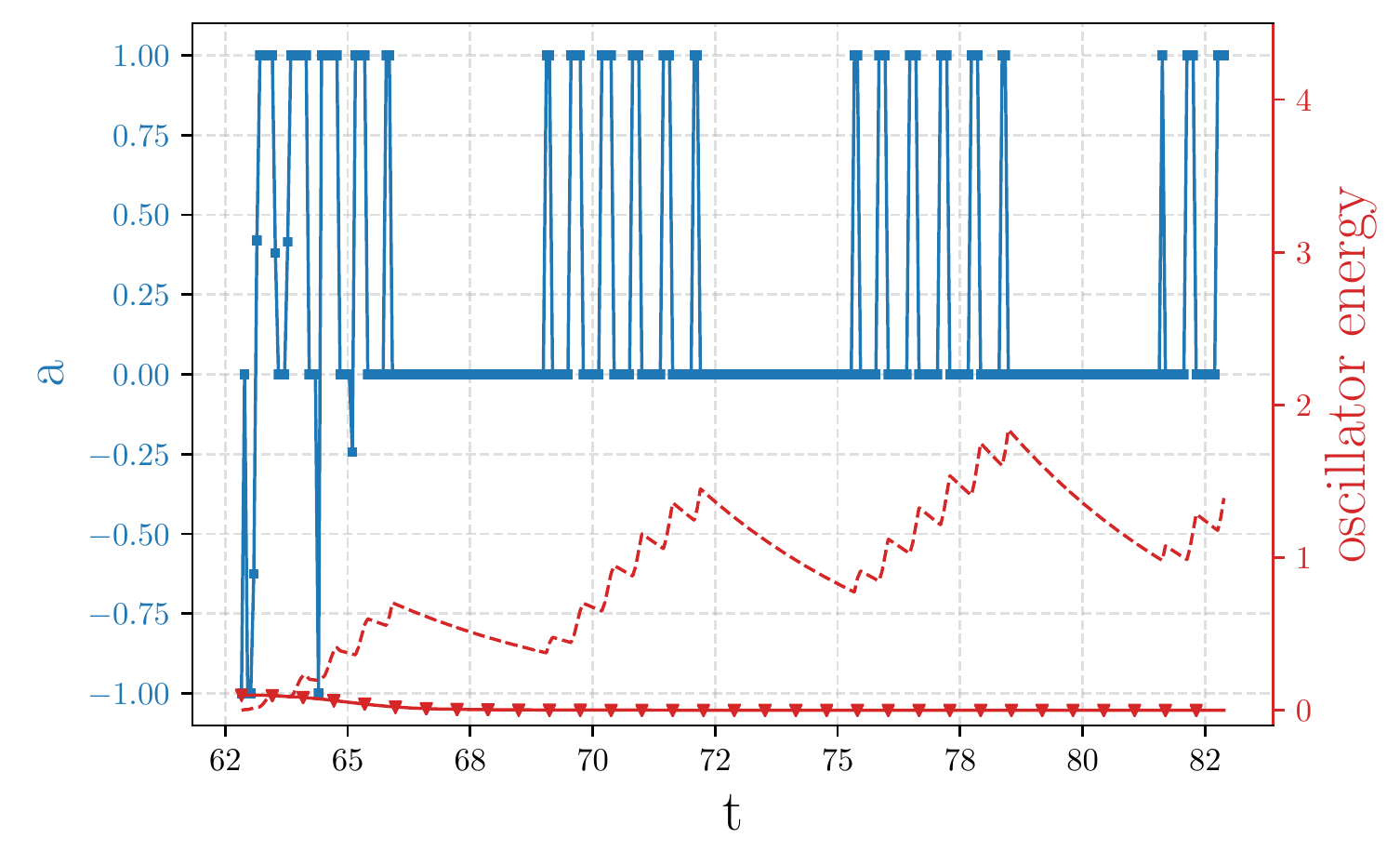} &    \multicolumn{1}{c}{}\includegraphics[width = 0.43 \textwidth]{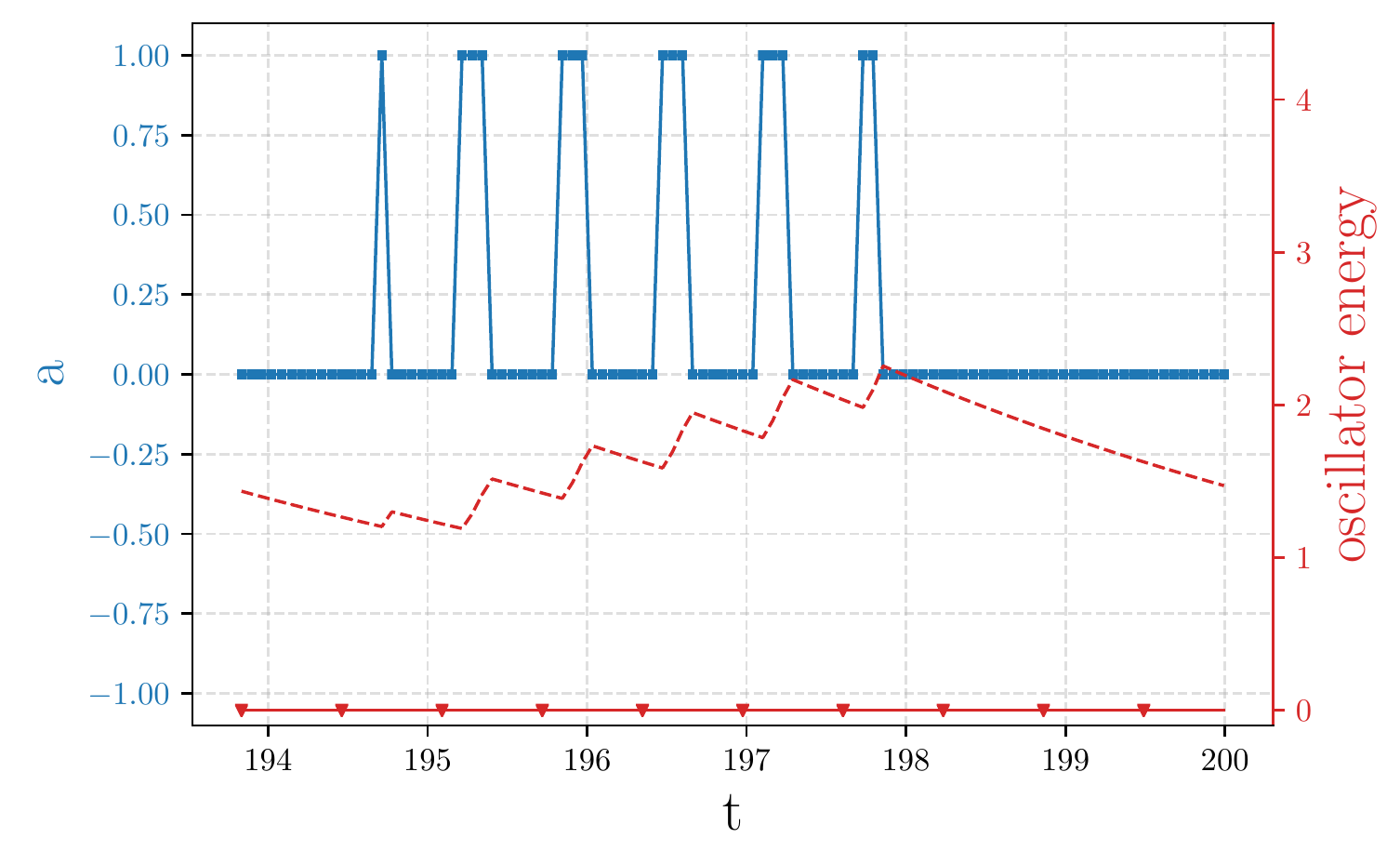}  \\
		\multicolumn{2}{c}{\hspace{5mm}\textbf{DDPG}}            \\ \toprule
		\multicolumn{1}{c}{} \includegraphics[width = 0.43 \textwidth]{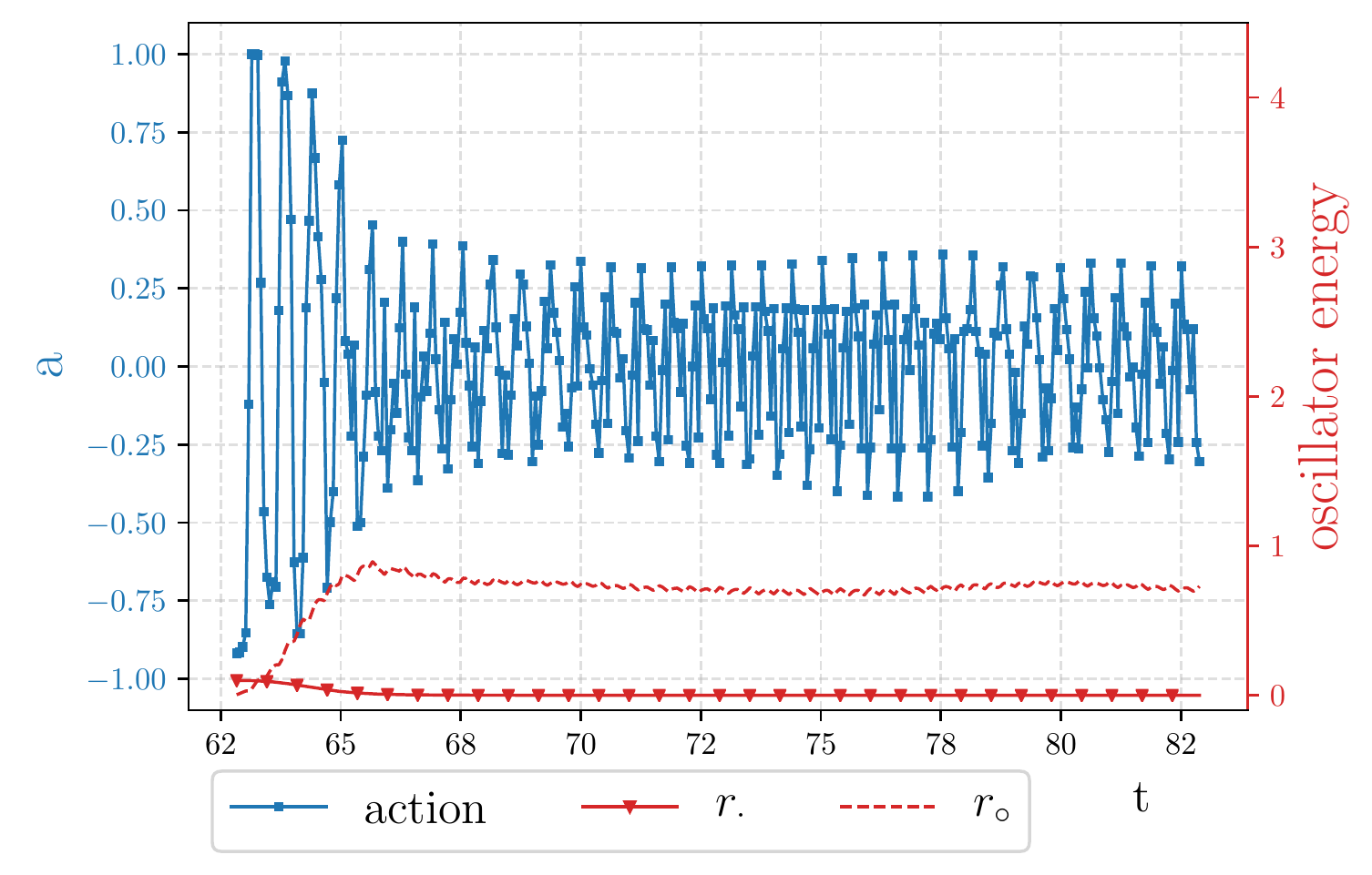} &     \multicolumn{1}{c}{}\includegraphics[width = 0.43 \textwidth]{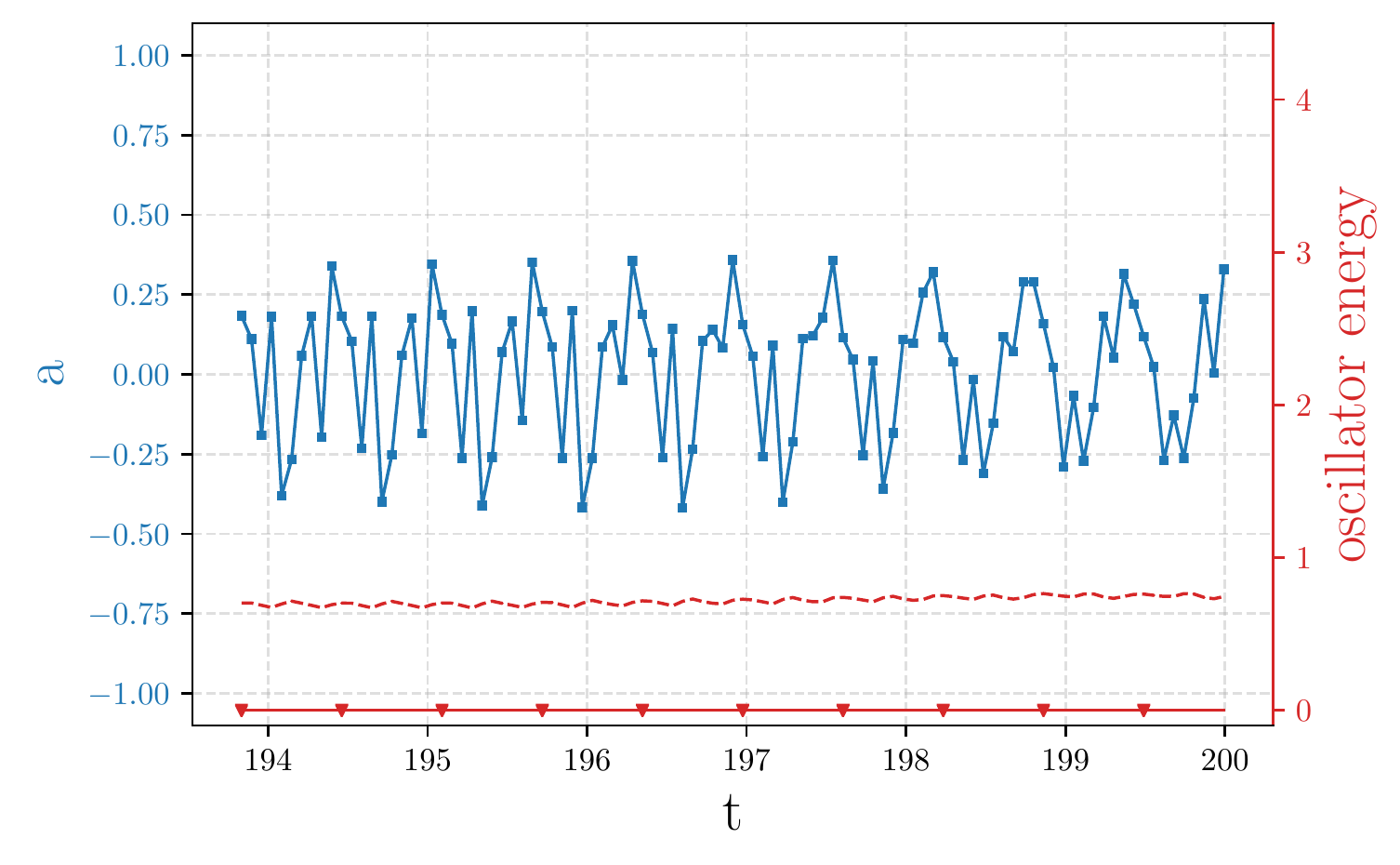} 
	\end{tabular*}
	\caption{Evolution of the best control function a (continuous blue line with squares), the energy of the first oscillator (continuous red line with triangles) and the energy of the second one (dashed red line), for the different control methods. The figures on the left report the early stage of the simulation, until the onset of a limit cycle condition, and those on the right the final time steps. }
	\label{0D_results_1}
\end{table*}

The learning behaviour of BO and LIPO is similar. Both have high variance in the early stages, as the surrogate model of the reward function is inaccurate. But both manage to obtain non-negligible improvements over the initial choice while acting randomly. The reader should notice that the variance of the LIPO at the first episode is 0 for all trainings because the initial points are always taken in the middle of the parameter space, as reported in Algorithm \ref{Alg_LIPO} ({\color{black}in Appendix \ref{algo_psedo_code}}). {\color{black} Hence the data at $\mbox{ep}=0$ is not shown for the LIPO}. For both methods, the learning curve steepens once the surrogate models become more accurate, but reach a plateau that has surprisingly low variance after the tenth episode. This behaviour could be explained by the difficulty of both the LIPO and GPr models in representing the reward function.

\begin{figure}
	\centering
	\includegraphics[width=\textwidth]{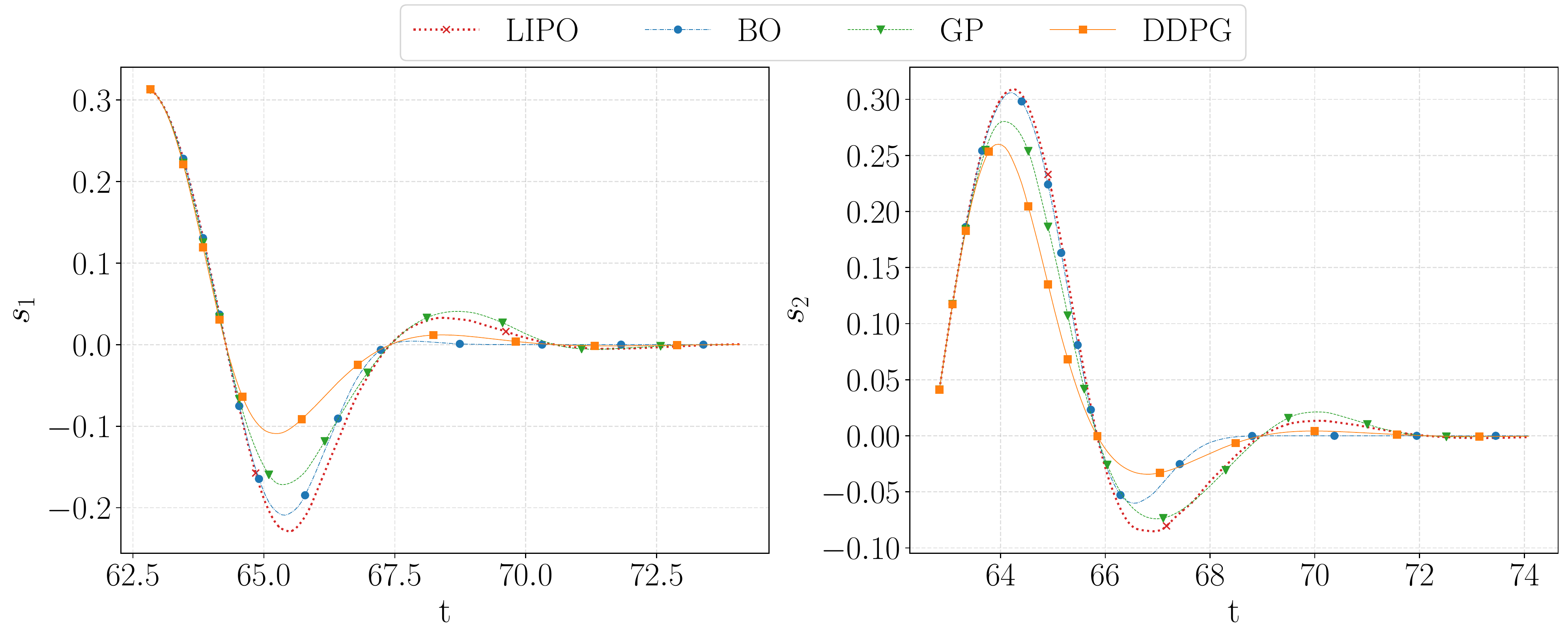}
	\caption{Evolution of the states $s_1$ and $s_2$, associated with the unstable oscillator, obtained using the optimal control action provided by the different machine learning methods.}
	\label{best_state_0D_1}
\end{figure}

Comparing the different control strategies identified by the four methods, the main difference resides in the settling times and energy consumption. Fig.\ref{best_state_0D_1} shows the evolution of $s_1$ and $s_2$ from the initial conditions to the controlled configuration for each method.

As shown in Eq.\eqref{eqn_cost_function}, the cost function accounts mainly for the stabilization of the first oscillator and the penalization of too strong actions. In this respect, the better overall performance of the GP is also visible in the transitory phase of the first oscillator, shown in Fig.\ref{best_state_0D_1}, and in the evolution of the control action. These are shown in Table \ref{0D_results_1} for all the investigated algorithms. For each algorithm, the figure on the left shows the action policy and the energy $E_1$ (continuous red line with triangles) and $E_2$ (dashed red line) (see Eq.\eqref{Energy}) of the two oscillators in the time span $t=62-82$, i.e. during the early stages of the control. The figure on the right shows a zoom in the time span $t=194-200$, once the system has reached a steady (controlled) state. 
The control actions by LIPO and BO are qualitatively similar and results in small oscillation in the energy of the oscillator. Both sustain the second oscillator with periodic actions that saturates. The periodicity is in this case enforced by the simple quadratic law that these algorithms are called to optimize. The differences in the two strategies can be well visualized by the different choice of weights (cf. equation \eqref{weights_0D}), which are shown in Figure \ref{0D_weights_comp_LIPO_BO}. While the {\color{black}LIPO} systematically gives considerable importance to the weight $w_{10}$, which governs the quadratic response to the state $s_2$, the {\color{black}BO} favors a more uniform choice of weights, resulting in a limited saturation of the action and less variance. The action saturation clearly highlight the limits of the proposed quadratic control law. Both LIPO and BO give a large importance to the weight $w_4$ because this is useful in the initial transitory to quickly energize the second oscillator. However, this term becomes a burden once the first oscillator is stabilized and forces the controller to over-react.

{\color{black} To have a better insight about this behaviour, we analyse the linear stability of the second oscillator. We linearize $\mathbf{s}_1$ around its mean value $\mathbf{s}_1^0 = \overline{\mathbf{s}_1}$ averaged over $t\in[70,60\pi]$.  We then obtain the linearized equation in terms of small perturbation, i.e. $\dot{\mathbf{s}}_2^{'}= \mathbf{K} \mathbf{s}'_2$, with  $\mathbf{s}_2=[s_3',s_4']$}.

\begin{figure}
	\centering
	\subfloat[]{\includegraphics[width=0.33\textwidth]{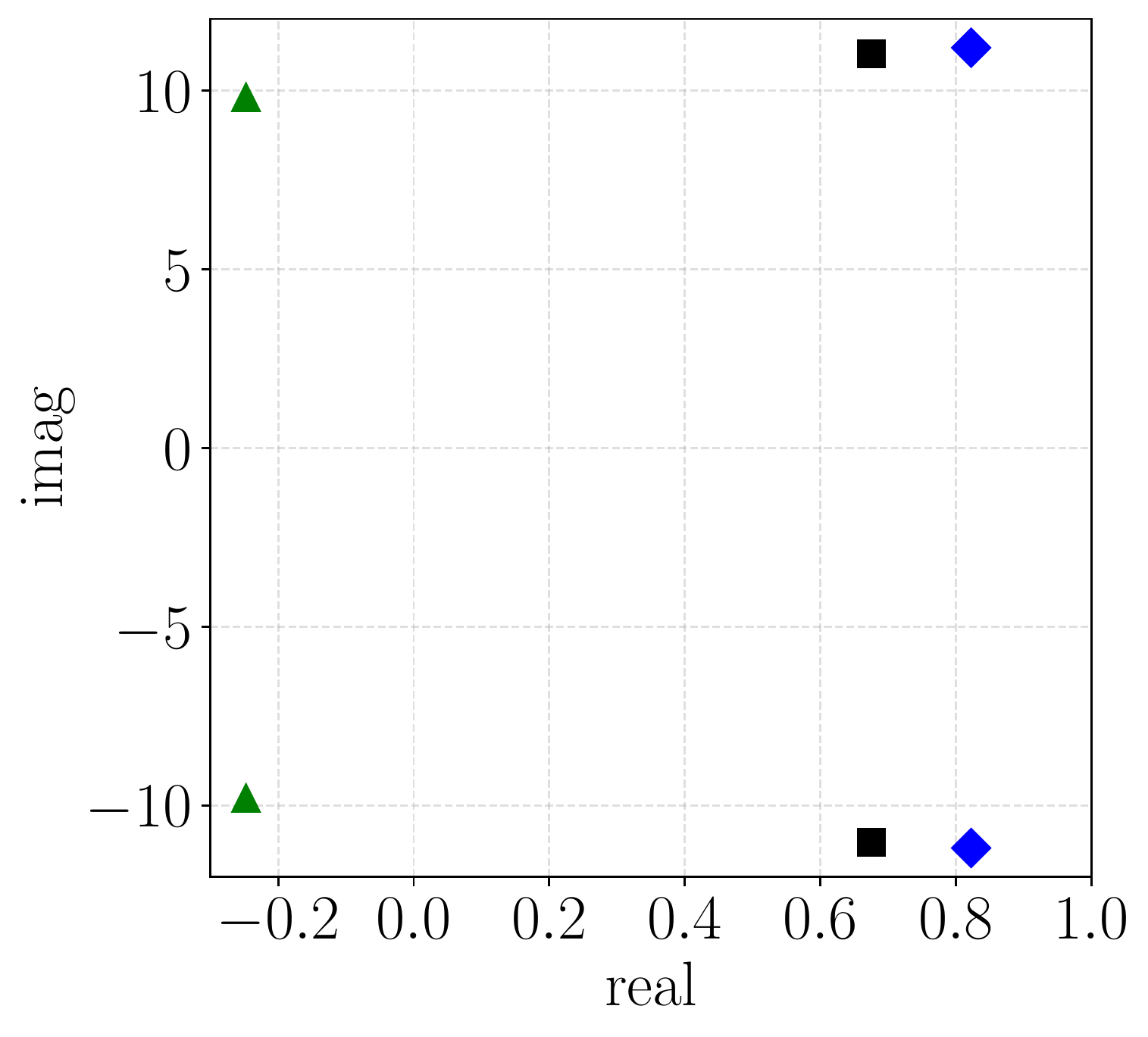}\label{fig:f1}}
	\hfill
	\subfloat[]{\includegraphics[width=0.33\textwidth]{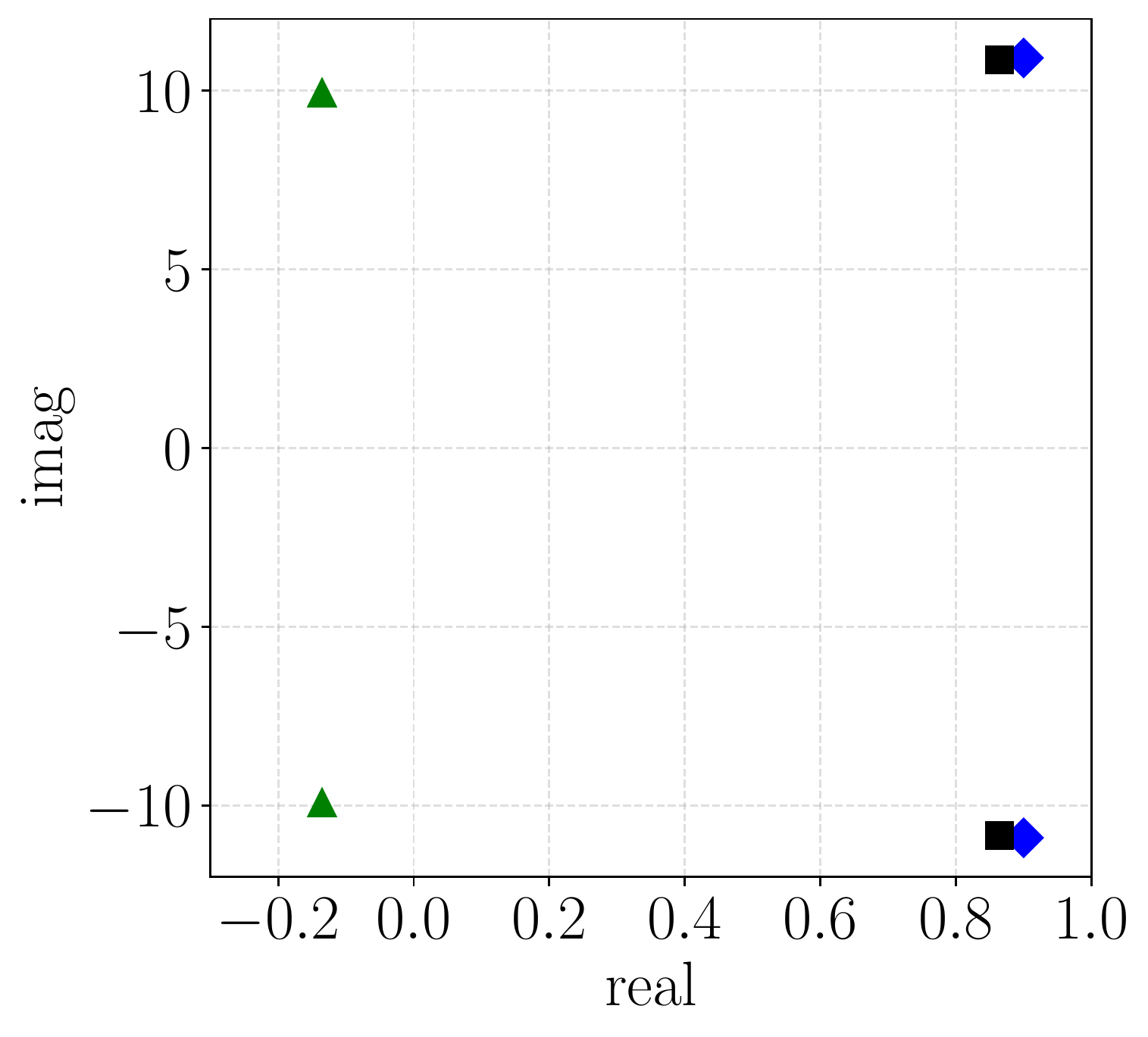}}
	\subfloat[]{\includegraphics[width=0.33\textwidth]{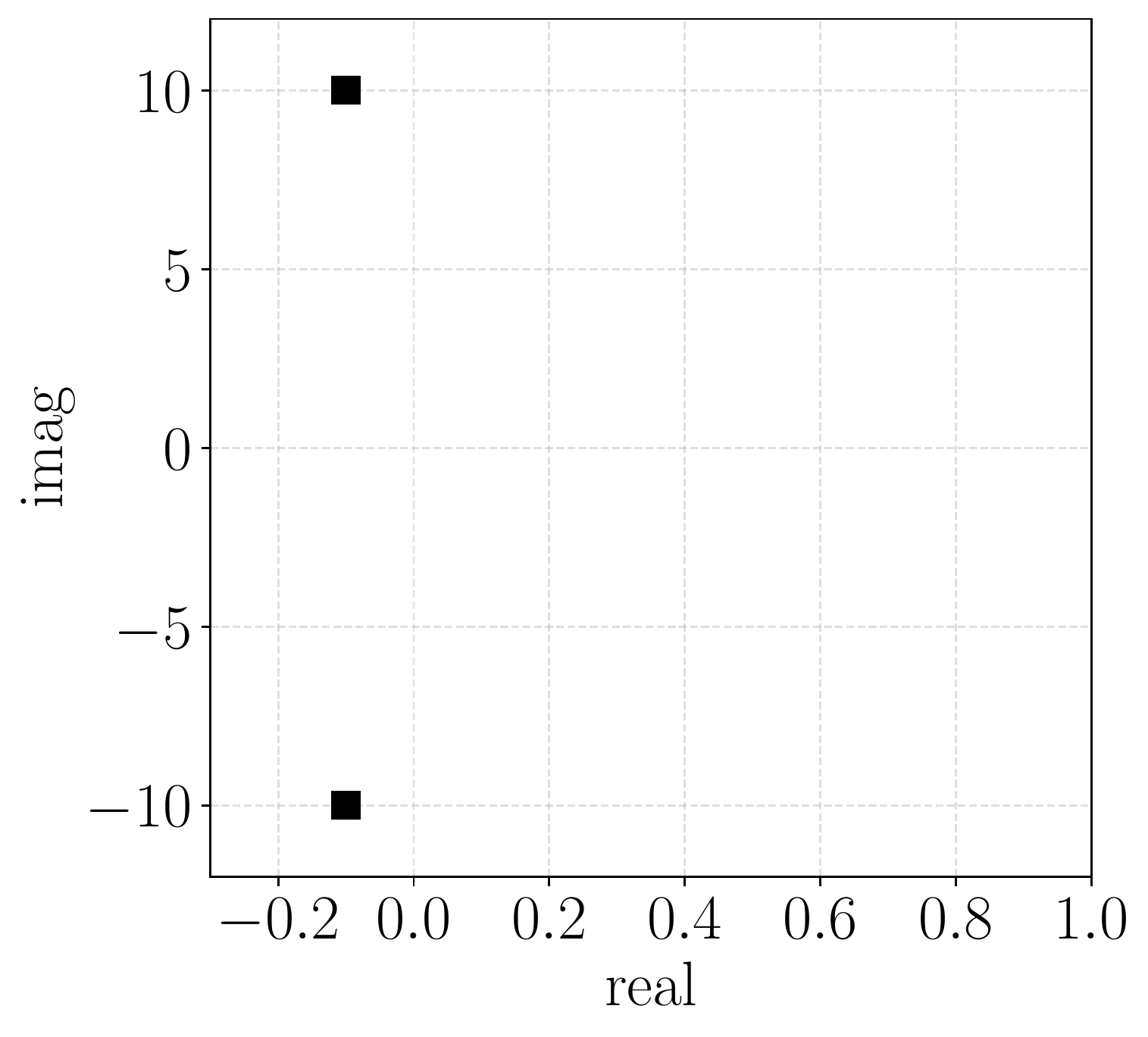}}
	\caption{{\color{black}Eigenvalues of the linearized second oscillator around its mean values in the developed case, controlled with linear combination (blue diamonds), with the nonlinear combination (green triangles) and with both linear and nonlinear terms (black squares) Eq.\eqref{nonlinear_control_law_0D} for LIPO and BO). The coefficient of the control function are those of the best solution found by LIPO (a), BO (b) and DDPG (c).}}
	\label{eigenvalues_perturbation}
\end{figure}

{\color{black} Fig.\ref{eigenvalues_perturbation} shows  the effect of the liner (blue diamonds), nonlinear (green triangles) and combined terms (black squares) over the eigenvalue of $\mathbf{K}$ of the best solution found by LIPO, BO and DDPG. It stands out that an interplay between the linear (destabilizing) and nonlinear (stabilizing) terms results in the oscillatory behaviour of $s_3$ and $s_4$ around their mean value $\mathbf{s}_0$ (averaged over $t\in[70,60\pi]$) for the optimizers, whereas DDPG is capable of keeping the system stable using only its linearized part.}

{\color{black} 
	Another interesting aspect is that }simplifying the control law {\color{black}(Eq.\eqref{weights_0D})} to the essential terms
\begin{equation}
	a = s_1w_1 + s_4w_2 + s_1 s_4 w_3,
\end{equation} allows the LIPO to identify a control law with comparable performances in less than five iterations. 

{\color{black} It is worth noticing that the cost function in \eqref{cost_fun_0D_dis} places no emphasis on the states of the oscillator $s_3,s_4$. Although the performances of LIPO and BO are similar according to this metric, the orbits in Figure \eqref{comp_s3_s4_0D_state_space} show that the BO keeps the second oscillator at unnecessarily larger amplitudes. This also shows that the problem is not sensitive to the amount of energy in the second oscillator once this has passed a certain value.} {\color{black}Another interesting aspect is the role of non-linearities in the actions of the DDPG agent. Thanks to its nonlinear policy, the DDPG immediately excites the second oscillator with strong actions around 10 rad/s, i.e. close to the oscillator's resonance frequency, even if, in the beginning, the first oscillator is moving at approximately 1 rad/s. On the other hand, the LIPO agent requires more time to achieve the same stabilization and mostly relies on its linear terms (linked to $s_1$ and $s_2$) because the quadratic ones are of no use in achieving the necessary change of frequency from sensor observation to actions.}

The GP and the DDPG use their larger model capacity to propose laws that are far more complex and more effective. The GP selects an impulsive control (also reported by \cite{duriez2017machine}) while the DDPG proposes a periodic forcing. The impulsive strategy of the GP performs better than the DDPG (according to the metrics in \ref{eqn_cost_function}) because it {\color{black} exchange more energy with the second oscillator with a smaller control effort. This is evident considering the total energy passes to the system through the actuation term in \eqref{0D_energy_second_oscillator_eq}  ($\sum_{i=0}^N\,|us_4|)$). The DDPG agent has exchanged 187 energy units, whereas the GP agent exchanged 329. In terms of control cost, defined as $\sum_{i=1}^N\,|u|$, the GP has a larger efficiency with 348 units against more than 420 for the DDPG.}  {\color{black}Moreover,} this can also be shown by plotting the orbits of the second oscillator under the action of the four controller, as done in Figure \ref{comp_s3_s4_0D_state_space}. Indeed, an impulsive control is hardly described by a continuous function and this is evident from the complexity of the policy found by the GP, which reads:

$$     a =\big(\log{(s_2+s_4)} + e^{e^{(s_4)}}\Big) +
\frac{\sin\big(\log(s_2)\big)}{
	\sin\big(\sin\big(\tanh\big( \log{\big(-e^{(s_2^2 - s_3^2)}-s_3\big)}\cdot\big(\tanh(\sin{(s_1)-s_2}) - s_2s_4\big)\big)\big)\big)}
\label{eqn_best_control_0D_GP}
$$

The best GP control strategy consists of two main terms. The first depends on $s_2$ and $s_4$ and the second takes all the states at the denominator and only $s_2$ at the numerator. This allows to moderate the control efforts once the first oscillator is stabilized. 

\begin{figure}
	\centering
	\includegraphics[width=0.6\textwidth]{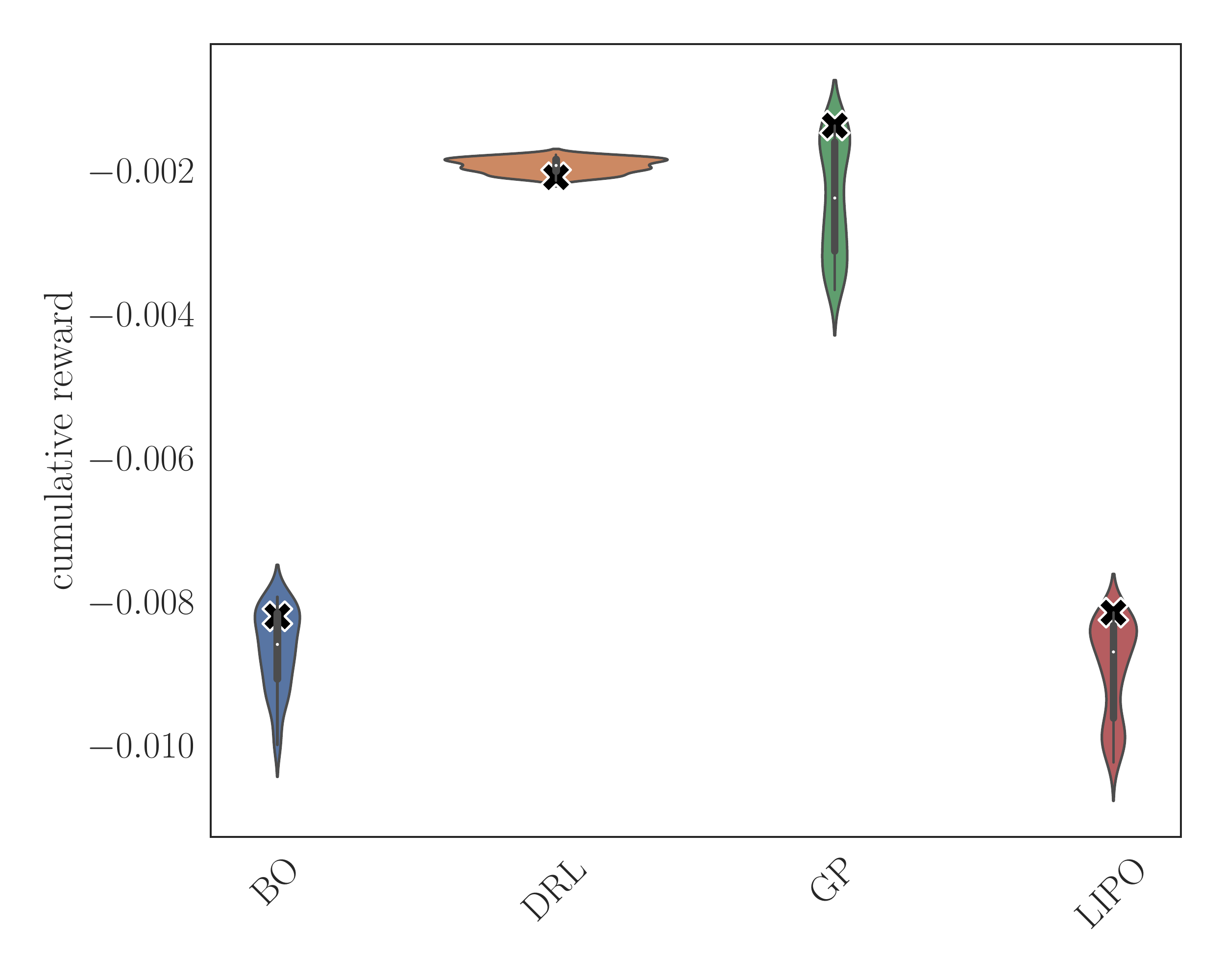}
	\caption{{\color{black}Robustness analysis of the optimal control methods with randomized initial conditions for the 0D testcase. The violin plots represent the distribution of cumulative rewards obtained, whereas the black crosses show the best result of each controller at the end of the training phase.}}
	\label{fig:0d-violin}
\end{figure}

{\color{black} Finally, the results from the robustness study are collected in Fig.\ref{fig:0d-violin}. This figure shows the distribution of the global rewards obtained for each agent while randomly changing the initial conditions 100 times. These instances were obtained by taking as an initial condition for the evaluation a random state in the range $t\in[60,66]$. The cross markers indicate the results obtained by the best agent for each method, trained while keeping the same initial condition. These violin plots can be used to provide a qualitative overview of the agents \emph{robustness} and \emph{generalization}. We consider an agent `robust' if its performances are independent of the initial conditions; thus, if the distribution in Figure \ref{fig:0d-violin} is narrow. We consider an agent `general' if its performance on the training conditions is compatible with the unseen conditions; thus, if the cross in Figure \ref{fig:0d-violin} falls \emph{within} the distribution of cumulative rewards. In this sense, the DDPG agent excels in both robustness and generalization, while the GP agent, which achieves the best performances on \emph{some} initial conditions, is less robust. On the other hand, the linear agents generalize well, and have worse control performance but robustness comparable to the GP agent.}

\FloatBarrier

\subsection{Viscous Burgers' equation test case}\label{sec4p3}

\begin{table}
	\centering
	\begin{tabular}{c@{\hspace{0.3cm}}c@{\hspace{0.2cm}}c@{\hspace{0.2cm}}c@{\hspace{0.2cm}}c@{\hspace{0.2cm}}}
		\toprule
		$\cdot 10^{3}$ & \textbf{LIPO} & \textbf{BO} & \textbf{GP} & \textbf{DDPG}\\ 
		\midrule
		\begin{tabular}[c]{@{}c@{}}Best \\ Reward\end{tabular}& 
		-\textbf{7.26}$\;\pm$0.93 & -\textbf{7.10}$\;\pm$0.32 & -\textbf{12.06}$\;\pm$12.25 & -\textbf{6.88}$\;\pm$0.58\\ 
		\bottomrule
	\end{tabular}
	\caption{Same as table \ref{table_results_0D} but for the control of nonlinear waves in the viscous Burger's equation.}
	\label{table_results_burgers}
\end{table}

\begin{figure}
	\centering
	\begin{subfigure}{.5\textwidth}
		\centering
		\includegraphics[width=\textwidth]{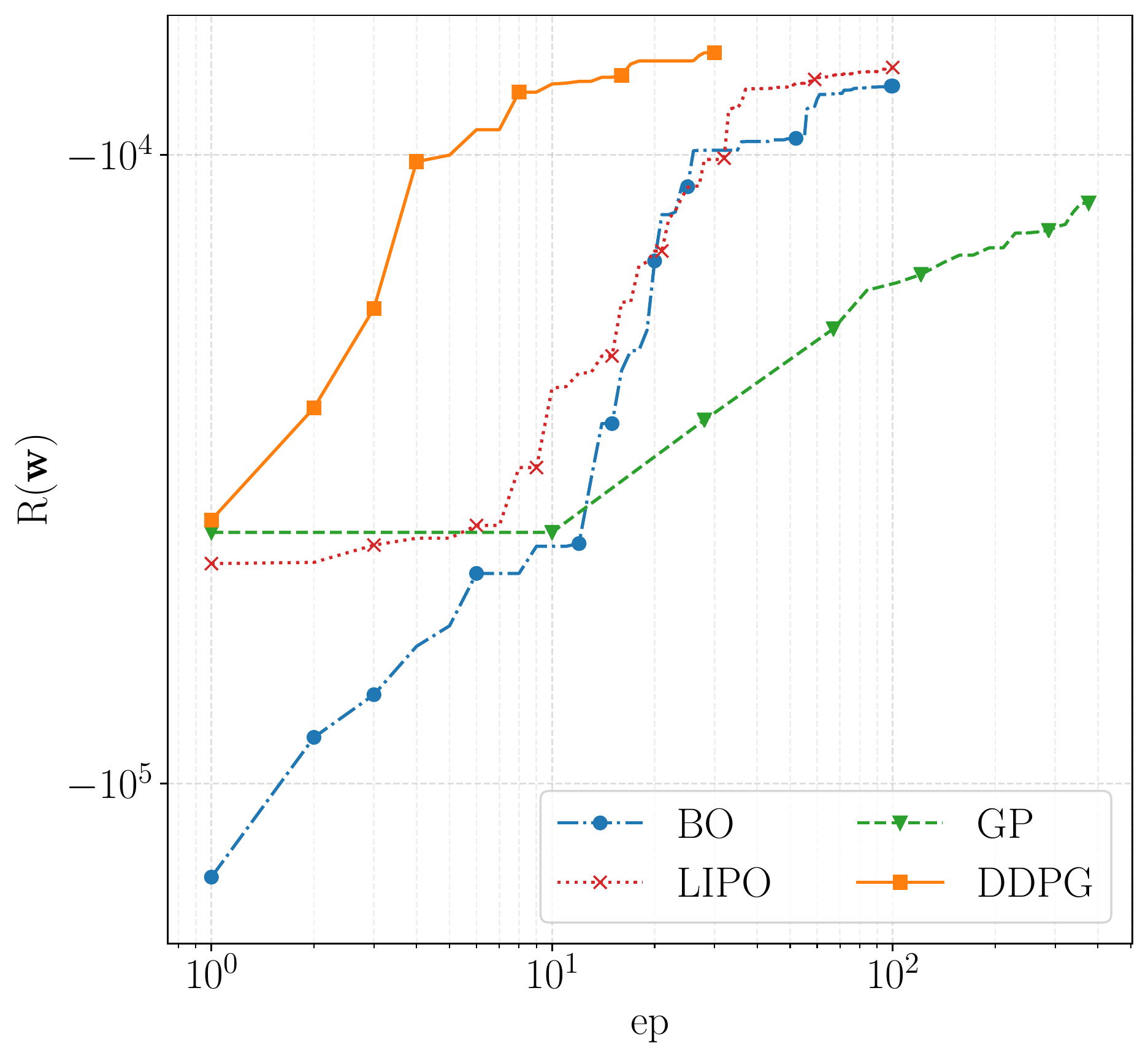}
		\caption{Learning curve}
		\label{learning_curve_burgers}
	\end{subfigure}%
	\begin{subfigure}{.5\textwidth}
		\centering
		\includegraphics[width=\textwidth]{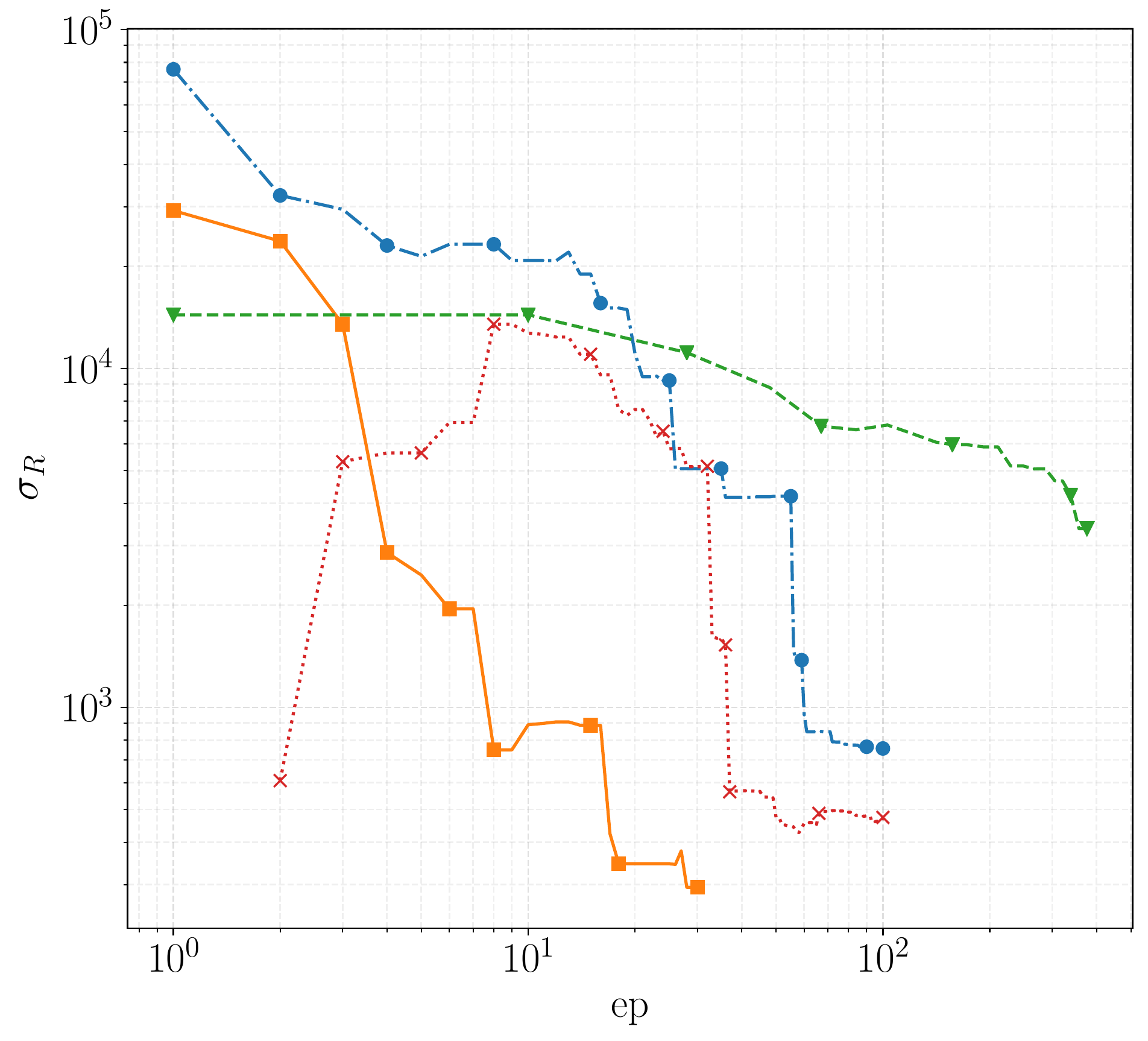}
		\caption{Learning curve variance}
		\label{learning_uncertainty_burgers_variance}
	\end{subfigure}
	\caption{Comparison of the learning curves (a) and their variances (b) for different machine learning methods for the 1D Burgers Equation test case (Sec. \ref{Sec:IVp2}).}
\end{figure}

\begin{figure*}
	\centering 
	\includegraphics[width=0.8\textwidth]{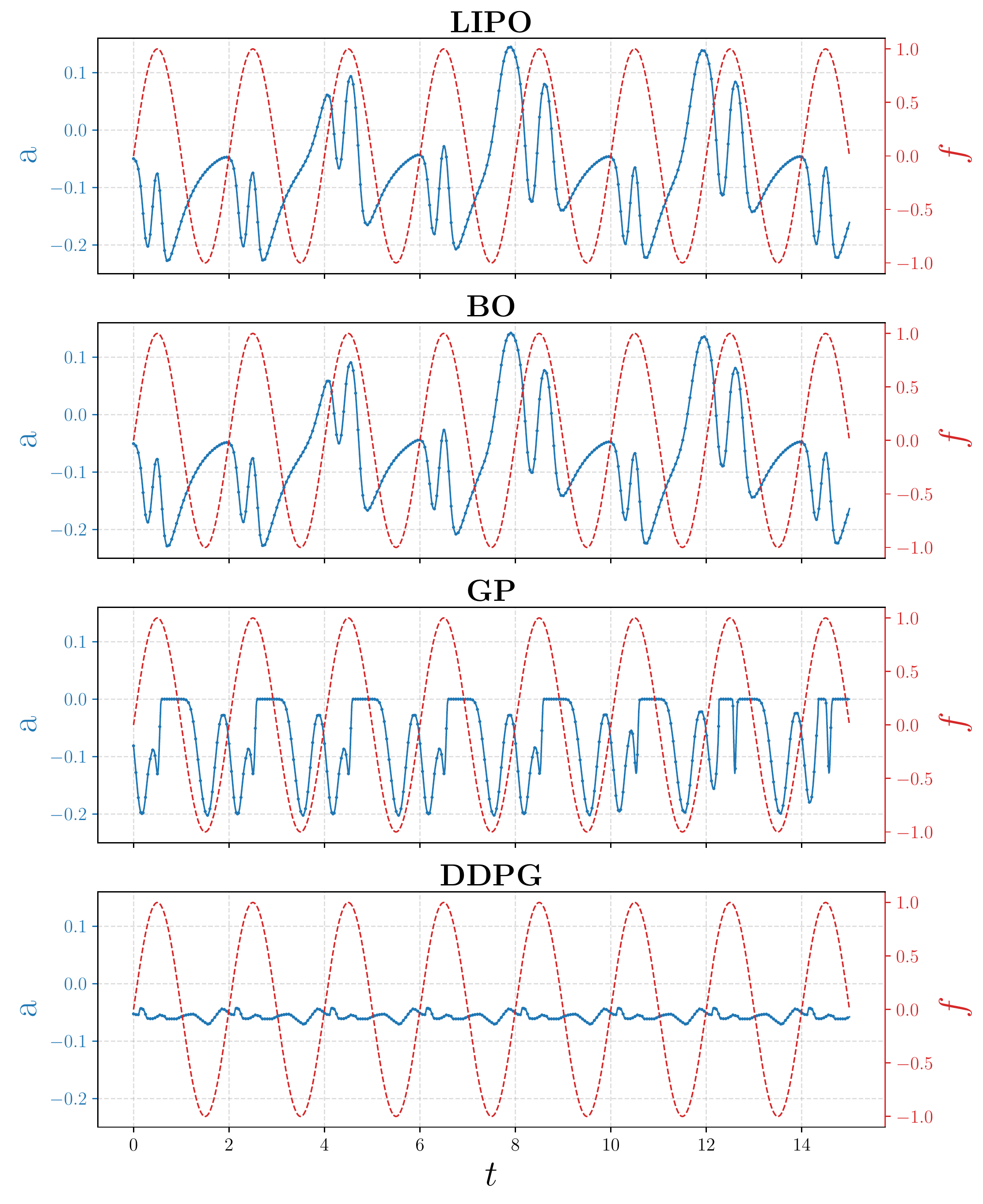}
	\centering
	\caption{Comparison of the control actions derived by the four machine learning methods. The action for each control methods are shown in blue (left axis) while  the curves in dashed red show the evolution of the introduced perturbation {\color{black} divided by $A_f$} (cf. \eqref{Burger_DEF}).}
	\label{action_comp_buergers}
\end{figure*}

\begin{figure}
	\centering
	\includegraphics[width=0.7\textwidth]{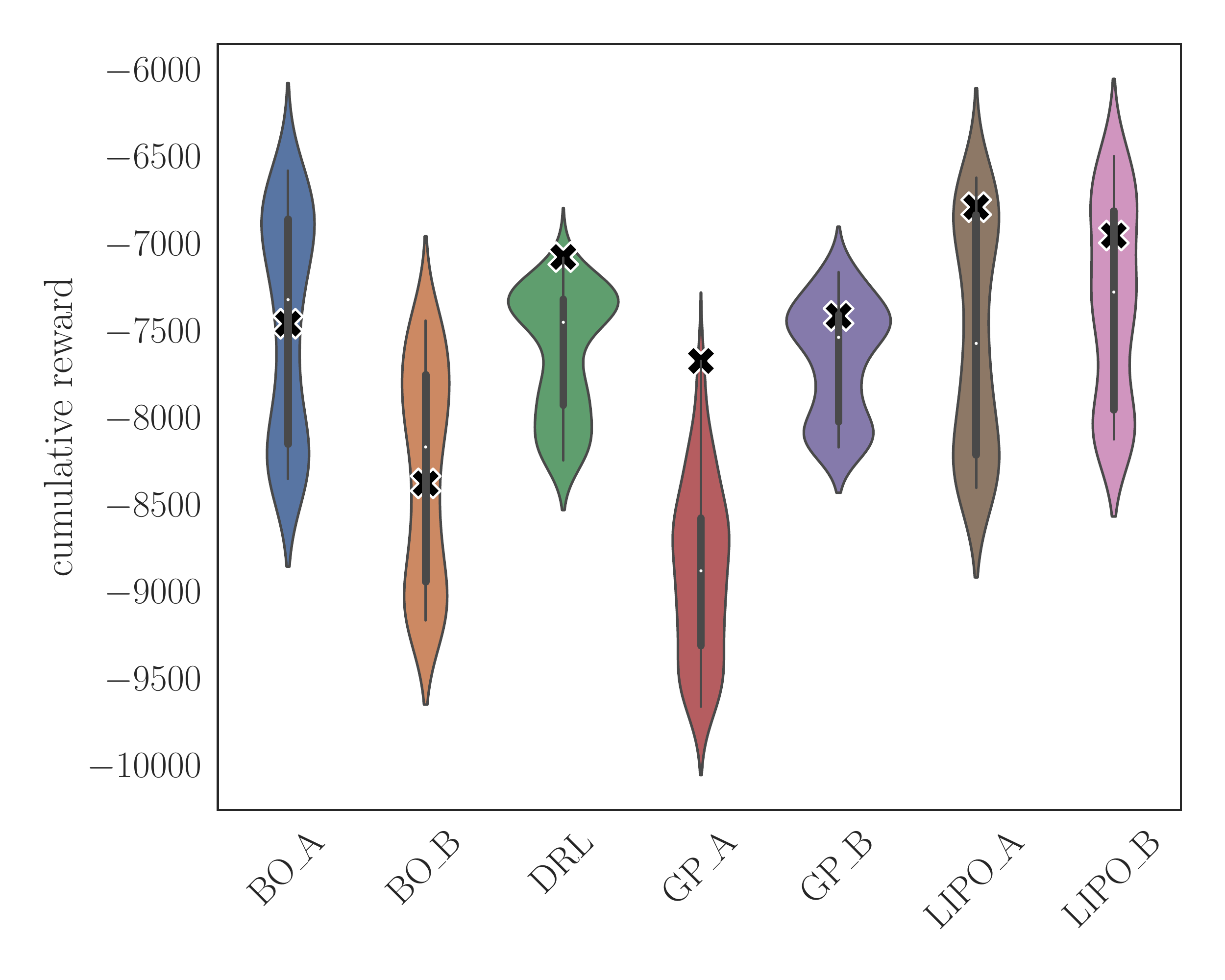}
	\caption{{\color{black}Robustness analysis of the optimal control methods with randomized initial conditions for the Burgers eq. testcase. The violin plots represent the distribution of cumulative rewards obtained, whereas the black crosses show the best result of each controller at the end of the training phase.}}
	\label{fig:my_label}
\end{figure}

We here present the results of the viscous Burgers' test case (cf Sec.\ref{Sec:IVp2}) {\color{black} focusing first on the cases for which neither the linear controllers BO and LIPO nor the GP can produce a constant action ( (laws A in section \ref{Sec:IVp2})}. As for the previous test case, Table \ref{table_results_burgers} collects the final best cumulative reward for each control method together with the confidence interval, while figures \ref{learning_curve_burgers} and \ref{learning_uncertainty_burgers_variance} show the learning curve and the learning variance over ten training sessions.  The DDPG achieved the best performance, with low variance, whereas the GP performed worse in both maximum reward and variance. LIPO and BO give comparable results. For the LIPO, the learning variance grows initially, as the algorithm randomly selects the second and third episodes' weights.

\begin{figure*}
	\centering 
	\includegraphics[width=\textwidth]{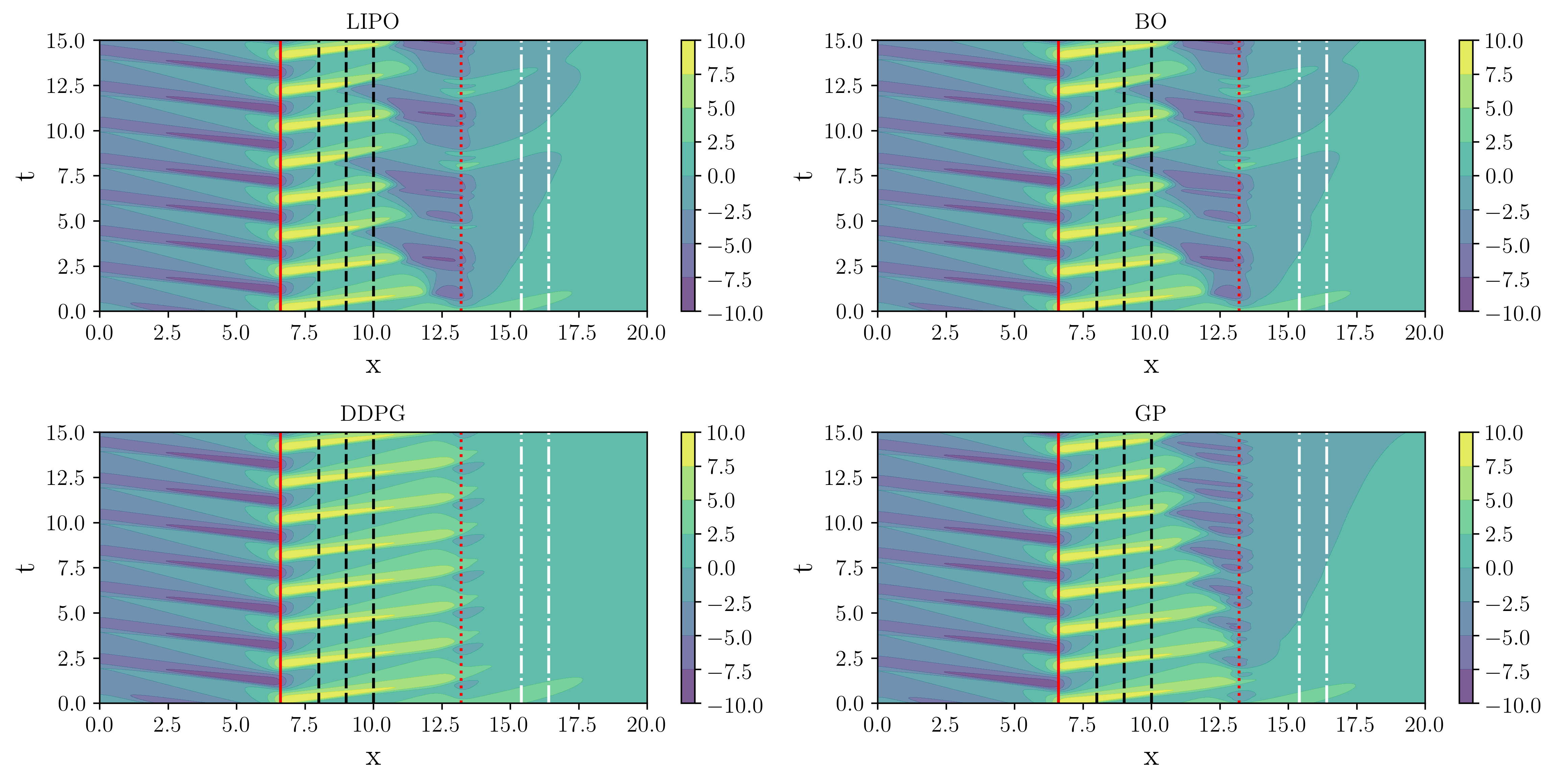}
	\centering
	\caption{Contour plot of the spatio-temporal evolution of u in {\color{black}governed by Eq. \eqref{viscous_burgers_eqn}} using the best control action of the different methods. The perturbation is centred at x = 6.6 (red continuous line) while the control law is centred at x = 13.2 {\color{black}(red dotted line)}. The dashed black lines visualize the location of the observation points, while the region within the white dash-dotted line is used to evaluate the controller performance.  An animation of the system controlled by the best method is provided in the supplemental material. }
	\label{staio_temporal_evol_optimal_burger}
\end{figure*}

For this test case, the GPr-based surrogate model of the reward function used by the BO proves to be particularly successful in approximating the expected cumulative reward. This yields steep improvements of the controller from the first iterations (recalling that the BO runs ten exploratory iterations to build its first surrogate model, which are not included in the learning curve). On the other hand, the GP does not profit from the relatively simple functional at hand and exhibits the usual stair-like learning curve since $20$ iterations were run with an initial population of $10$ individuals.

The control laws found by BO and LIPO have similar weights (with differences of the order $\mathcal{O}(10^{-2})$), although the BO has much lower variance among the training sessions. Figure \ref{action_comp_buergers} shows the best control law derived by the four controller, together with the forcing term. These figures should be analyzed together with table \ref{staio_temporal_evol_optimal_burger} which shows the spatio-temporal evolution of the variable $u(x,t)$ under the action of the best control law derived by the four algorithms.

The linear control laws of BO and LIPO are characterized by two main periods: one that seeks to cancel the incoming wave and the second that seeks to compensate for the control action's upward propagation. This upward propagation is revealed in the spatiotemporal plots in Fig. \ref{staio_temporal_evol_optimal_burger} for the BO and LIPO while it is moderate in the problem controlled via GP and absent in the case of the DDPG control. {\color{black} The advective retrofitting (mechanism I in Sec.\ref{Sec:IVp3})} challenges the LIPO and the BO agents because actions are fed back into the observations after a certain time and {\color{black} these agents, acting linearly, are unable to leverage the system diffusion by triggering higher frequencies (mechanism II in Sec. \ref{Sec:IVp3})}. By contrast, the GP, hinging on its larger model capacity, does introduce strong gradients to leverage diffusion.

\begin{figure}
	\centering
	\includegraphics[width=\textwidth]{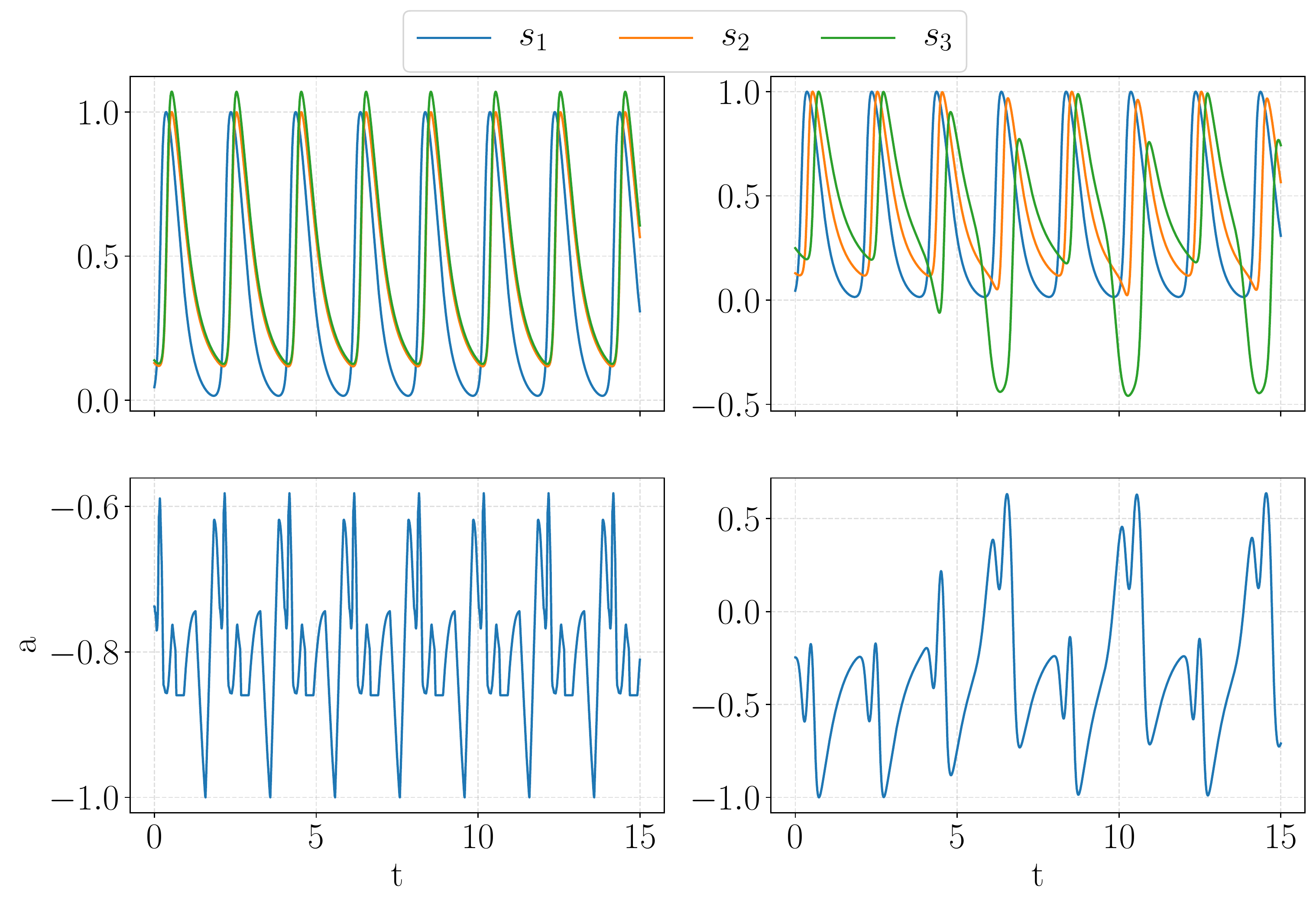}
	\caption{{\color{black}Comparison of the action and observation evolution along an episode for DDPG (left) and LIPO (right) in the second test case (Sec. \ref{Sec:IVp2}).}}
	\label{actions_LIPO_DDPG_comparison}
\end{figure}

{\color{black} An open-loop strategy such as a constant term in the policy appears useful in this problem, and the average action produced by the DDPG, as shown in Figure \ref{action_comp_buergers}}, demonstrates that this agent is indeed taking advantage of it. {\color{black} This is why we also analyzed the problem in mixed conditions, giving all agents the possibility to provide a constant term. The BO, LIPO and GP results in this variant are analyzed together with the robustness study, in which 100 randomly selected initial conditions are considered. The results are collected in Figure \ref{fig:my_label}, with the subscript A referring to agents that do not have the constant term and B to agents that do have it.
	
	Overall, the possibility of acting with a constant contribution is well appreciated by all agents, although none reach the performances of the DDPG. This shows that the success in the DDPG is not solely due to this term but also ability to generate high frequencies. This is better highlighted in Figure \ref{actions_LIPO_DDPG_comparison}, which shows a zoom on the action and the observations for the DDPG and the BO. While both agents opt for an action whose mean is different from zero, the frequency content of the action is clearly different and, once again, the available non-linearities play an important role.}

		

\subsection{von Kármán street control test case}

\begin{figure*}
	\centering 
	\includegraphics[width=\textwidth]{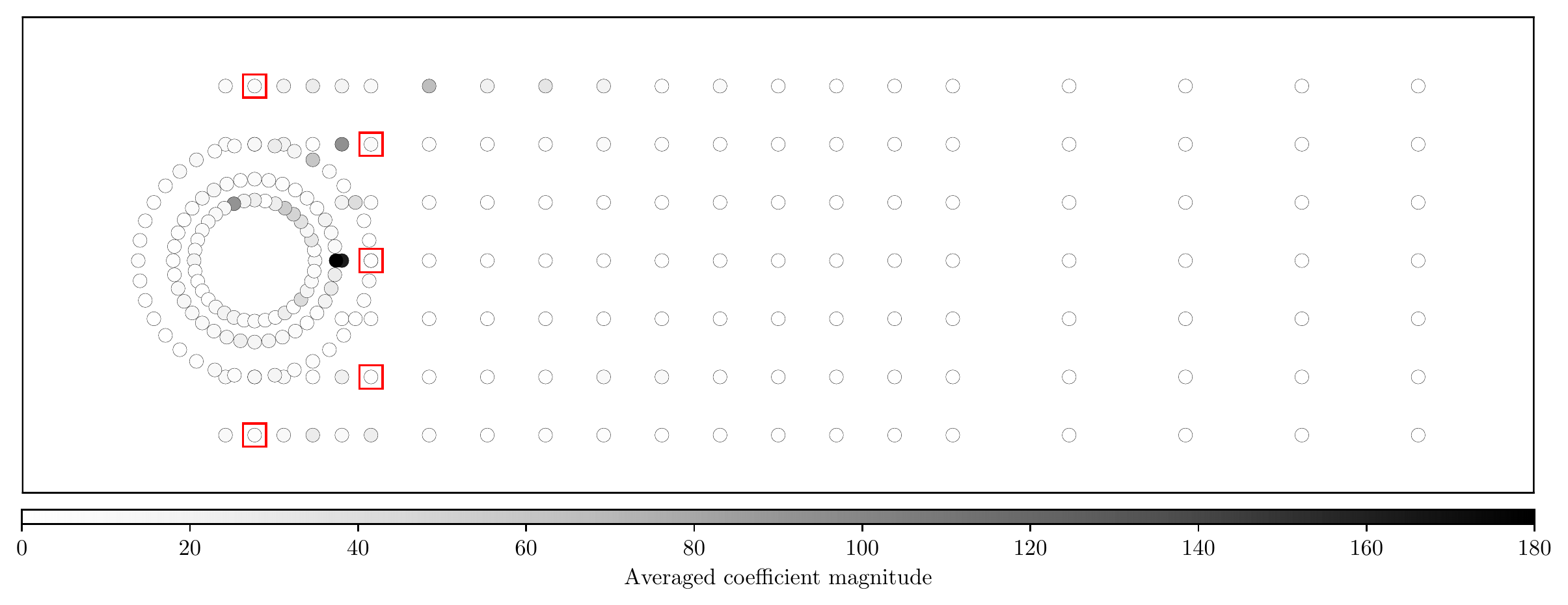}
	\caption{Scatter plot of the sensor locations, coloured by the norm of the weights $\mathbf{w}_{1j},\mathbf{w}_{2j},\mathbf{w}_{3j},\mathbf{w}_{4j}$ that link the observation at state $j$ with the action vector $\mathbf{a}=[a_1,a_2,a_3,a_4]$ in the linear regression of the policy by \cite{Tang2020e}}
	\label{cylinder_avg_coeff_pinv}
\end{figure*}

\begin{table}
	\centering
	\begin{tabular}{c@{\hspace{0.3cm}}c@{\hspace{0.2cm}}c@{\hspace{0.2cm}}c@{\hspace{0.2cm}}c@{\hspace{0.2cm}}}
		\toprule
		& \textbf{LIPO} & \textbf{BO} & \textbf{GP} & \textbf{DDPG}\\ 
		\midrule
		\begin{tabular}[c]{@{}c@{}}Best \\ Reward\end{tabular}& 
		\textbf{6.53}$\;\pm$0.34 & \textbf{6.41}$\;\pm$0.89 & \textbf{7.14}$\;\pm$0.86 & \textbf{5.66}$\;\pm$2.64\\
		\bottomrule
	\end{tabular}
	\caption{Same as table \ref{table_results_0D} but for the von Kármán street control problem.}
	\label{table_results_cylinder}
\end{table}

We begin the analysis of this test case with an investigation on the performances of the RL agent trained by \cite{Tang2020e} using the Proximal Policy Optimization (PPO) on the same control problem. As recalled in section \ref{Sec:IVp3}, these authors used $236$ probes, located as shown in Figure \ref{cylinder_des}, and a policy $\mathbf{a}=f(\mathbf{s};\mathbf{w})$ represented by an ANN with three layers with $256$ neurons each. Such a complex parametric function gives a large model capacity, and it is thus natural to analyse whether the trained agent leverage this potential model complexity.
\begin{figure}
	\centering
	\begin{subfigure}{.5\textwidth}
		\centering
		\includegraphics[width=0.95\textwidth]{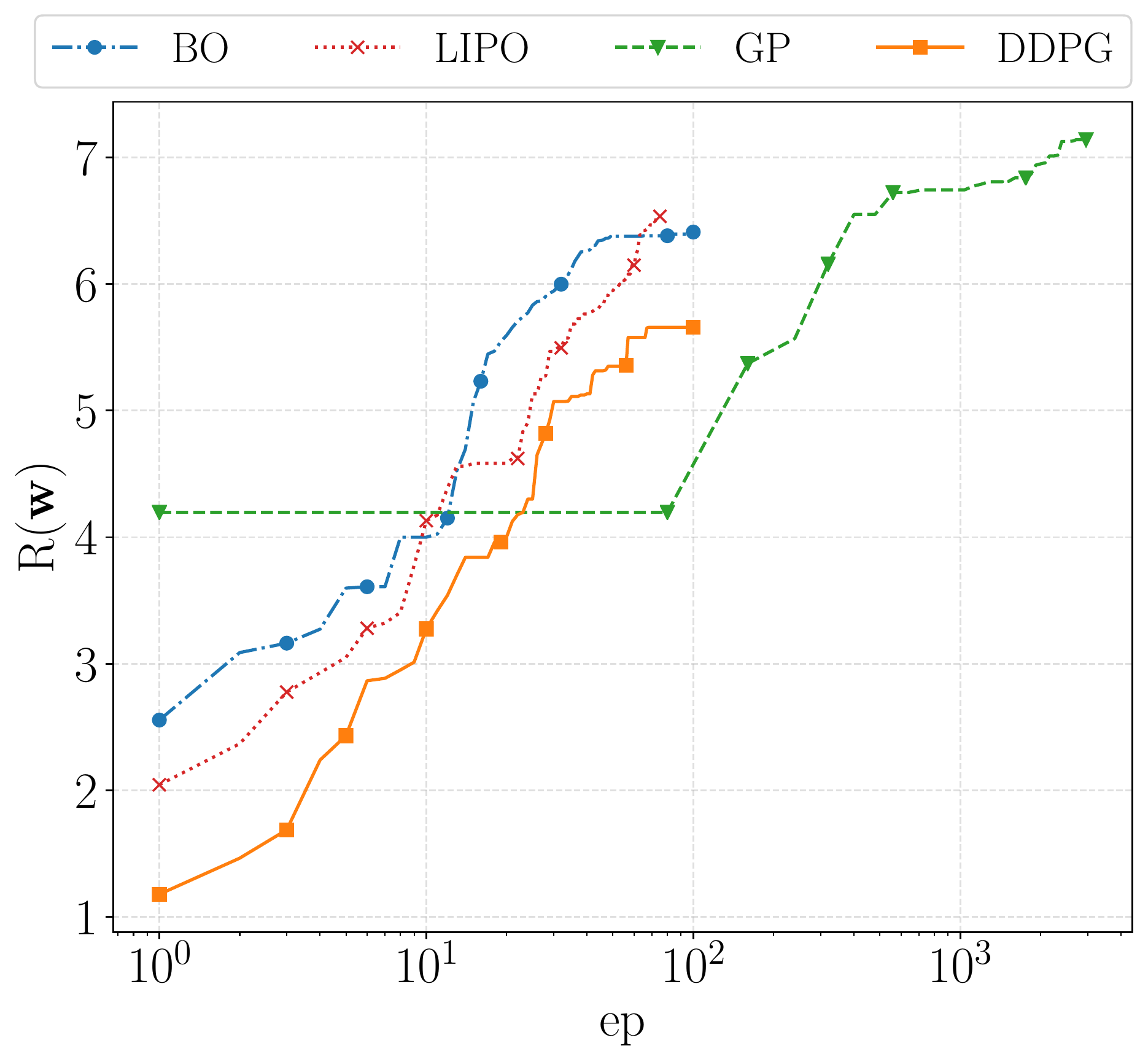}
		\caption{Learning curve}
		\label{learning_curve_cylinder}
	\end{subfigure}%
	\begin{subfigure}{.5\textwidth}
		\centering
		\includegraphics[width=\textwidth]{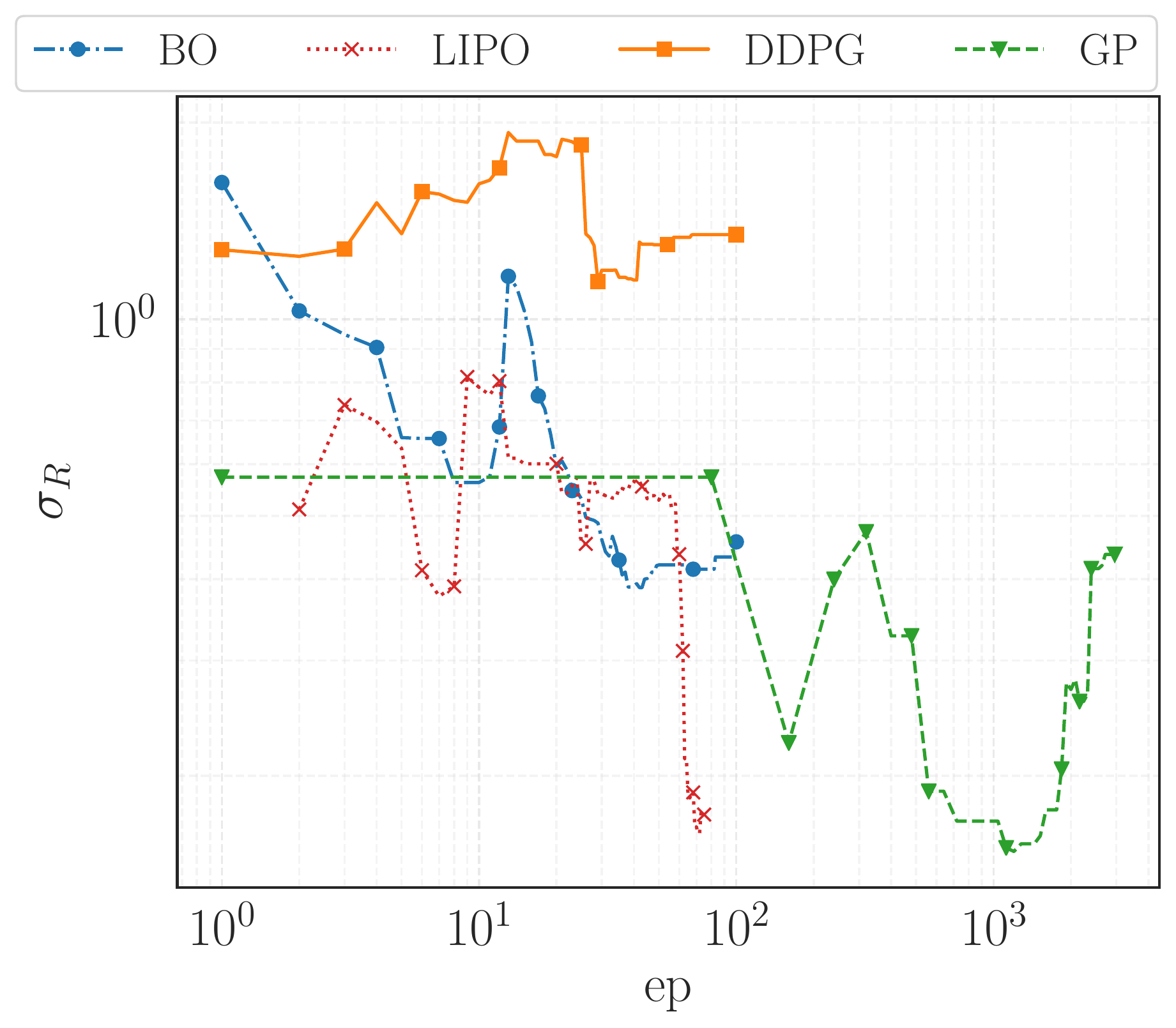}
		\caption{Learning curve variance}
		\label{uncertainty_learning_curve_cylinder}
	\end{subfigure}
	\caption{Comparison of the learning curves (a) and their variances (b) for different machine learning methods for the von Kármán street control problem (Sec. \ref{Sec:IVp3}).}
\end{figure}
To this end, we perform a linear regression of the policy identified by the ANN. Given $\mathbf{a}\in \mathbb{R}^{4}$ the action vector and $\mathbf{s}\in \mathbb{R}^{236}$ the state vector collecting information from all probes, we seek the best linear law of the form $\mathbf{a}=\mathbf{W}\mathbf{s}$, with $\mathbf{W}\in\mathbf{R}^{4\times 236}$ the matrix of weights of the linear policy. Let $\mathbf{w}_j$ denote the $j$-th raw of $\mathbf{W}$, hence the set of weights that linearly map the state $\mathbf{s}$ to the action $\mathbf{a}_j$, i.e. the flow rate in the one of the fourth injections. One thus has $\mathbf{a}_j=\mathbf{w}^T_j\mathbf{s}$. 

To perform the regression, we produce a dataset of $n_*=400$ samples of the control law, by interrogating the ANN agent trained by \cite{Tang2020e}. Denoting as $\mathbf{s}_i^*$ the evolution of the state $i$ and as $\mathbf{a}^*_j$ the vector of actions proposed by the agent at the $400$ samples, the linear fit of the control action is the solution of a linear least square problem, which using Ridge regression yields:

\begin{equation}
	\label{a_J}
	\mathbf{a}^*_j=\mathbf{S} \mathbf{w}_j \rightarrow \mathbf{w}_j=(\mathbf{S}^T\mathbf{S}+\alpha \mathbf{I})^{-1}\mathbf{S}^T \mathbf{a}^*_j
\end{equation} where $\mathbf{S}=[\mathbf{s}^*_1,\mathbf{s}^*_2,\dots \mathbf{s}^*_{236}]\in \mathbb{R}^{400\times 236}$ is the matrix collecting the $400$ samples for the $236$ observations along its columns, $\mathbf{I}$ is the identity matrix of appropriate size and $\alpha$ is a regularization term. In this regression, the parameter $\alpha$ is taken running a K=5 fold validation and looking for the minima of the out-of sample error. 

The result of this exercise is illuminating for two reasons. The first is that the residuals in the solution of \eqref{a_J} have a norm of $||\mathbf{a}^*_j-\mathbf{S} \mathbf{w}_j||=1e-5$. This means that despite the large model capacity available to the ANN, the RL by \cite{Tang2020e} is de-facto producing a linear policy.

\begin{figure*}
	\centering 
	\includegraphics[width=0.74\textwidth]{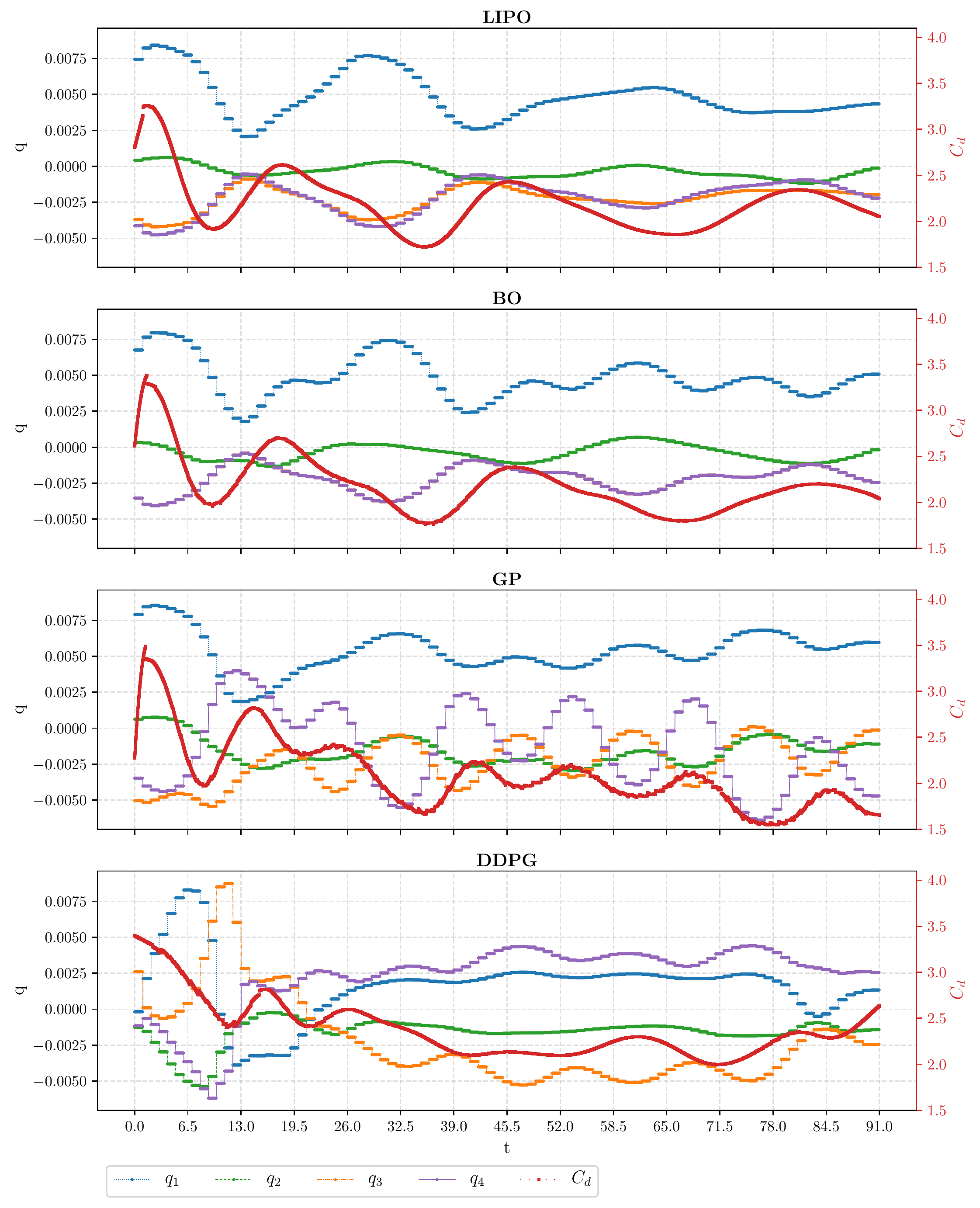}
	\caption{Evolution of the jets' flow rates(left) and the drag around the cylinder(right) for the best control action found by the different machine learning methods.}
	\label{action_comparison_cylinder}
\end{figure*}

The second reason is that analyzing the weights $w_{i,j}\in \mathbf{W}$, in the linearized policy $\mathbf{a}_j=\mathbf{W}\mathbf{s}$, allows for quickly identifying which of the sensors is more important in the action selection process. The result, in the form of a coloured scatter-plot, is shown in Figure \ref{cylinder_avg_coeff_pinv}. The markers are placed at the sensor location and coloured by the sum $\sum_i w^2_{i,j}$ for each of the $j$-th sensors.
This result shows that only a tiny fraction of the sensors play a role in the action selection. In particular, the two most important ones are placed on the rear part of the cylinder and have much larger weights than all the others.


In the light of this result with the benchmark RL agent, it becomes particularly interesting to perform the same analysis of the control action proposed by DDPG and GP, since BO and the LIPO use a linear law by construction. Figure \ref{learning_curve_cylinder} and \ref{uncertainty_learning_curve_cylinder} show the learning curves and learning variance as a function of the episodes, while table \ref{table_results_cylinder} collects the results for the four methods in terms of the best reward and confidence interval as done for the previous test cases.

The BO and the LIPO reached an average reward of $6.43$ (with the best performances of the BO hitting $7.07$) in 80 episodes while the PPO agent trained by \cite{Tang2020e} required $800$ to reach a reward of $6.21$. While \cite{Tang2020e}'s agent aimed at achieving a \emph{robust policy} across a wide range of Reynolds numbers, {\color{black}it appears that, for this specific problem, the use of an ANN-based policy with more than 65000 parameters and 236 probes drastically penalize the sample-efficiency of the learning if compared to a linear policy with 5 sensors and 20 parameters.}

\begin{figure*} \center
	\centering
	\includegraphics[width=\textwidth]{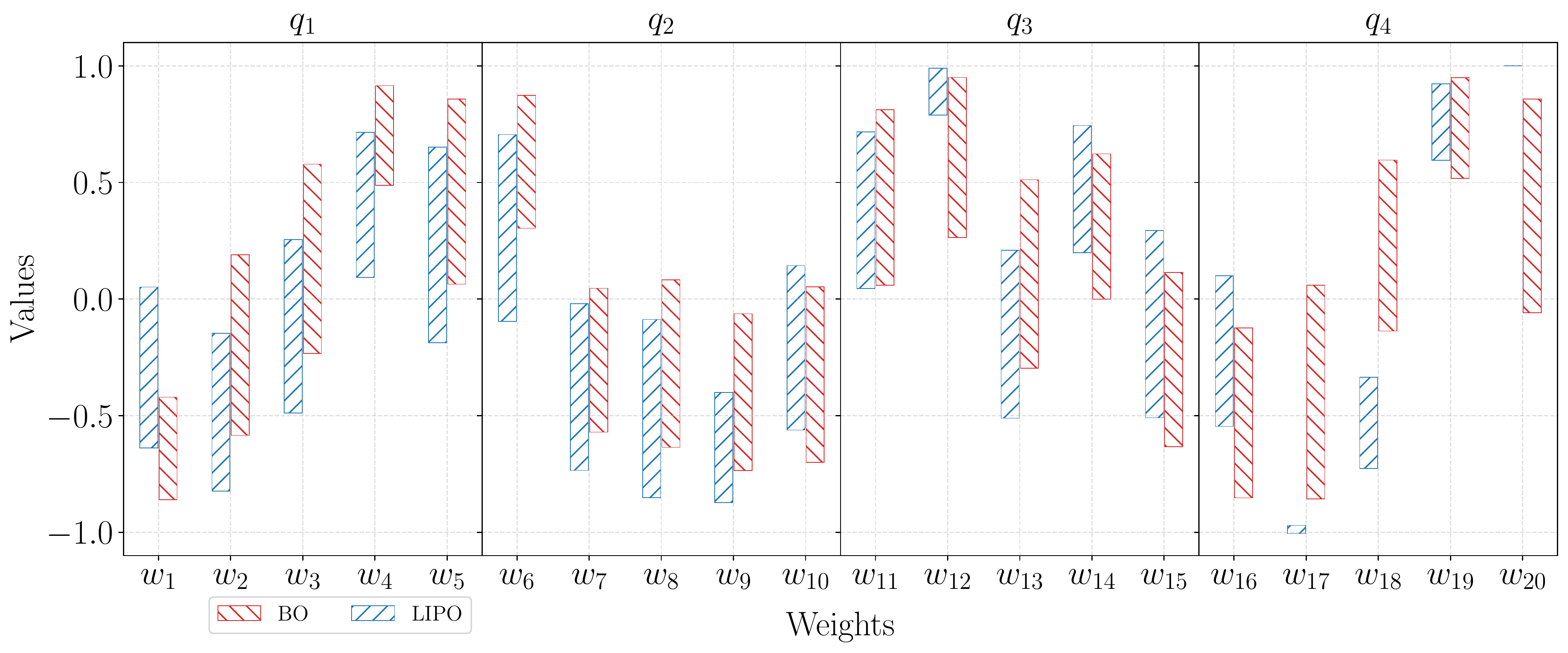}
	\caption{Weights of control action for the von Kármán street control problem, given by a linear combination of the system's states for the four flow rates. The coloured bars represent a standard deviation around the mean value found by LIPO and BO with ten random number generator seeds.}
	\label{cylinder_weights_comp_LIPO_BO}
\end{figure*}

Genetic Programming had the best mean control performance, {with \color{black} 33\%  reduction of the average drag coefficient compared to the uncontrolled case} and remarkably small variance. LIPO had the lowest standard deviation due to its mainly deterministic research strategy, which selects only two random coefficients at the second and third optimization steps.

\begin{table*}
	\centering
	\begin{tabular}{cccc}
		\multicolumn{4}{c}{\hspace{5mm}\textbf{DDPG} }\\\midrule
		\includegraphics[width = 0.22
		\textwidth]{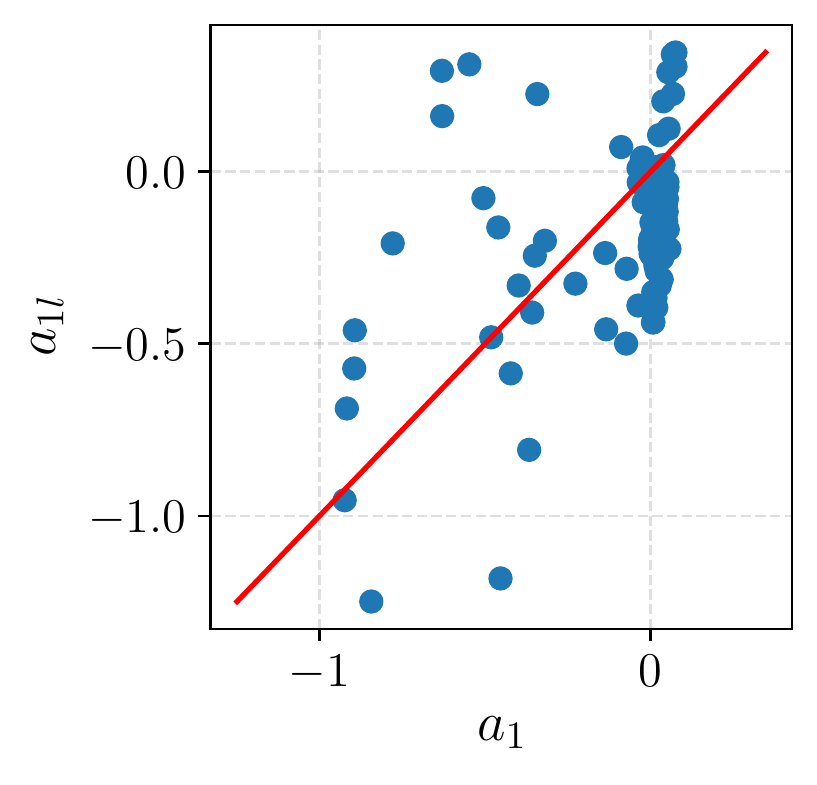} & \includegraphics[width = 0.22   \textwidth]{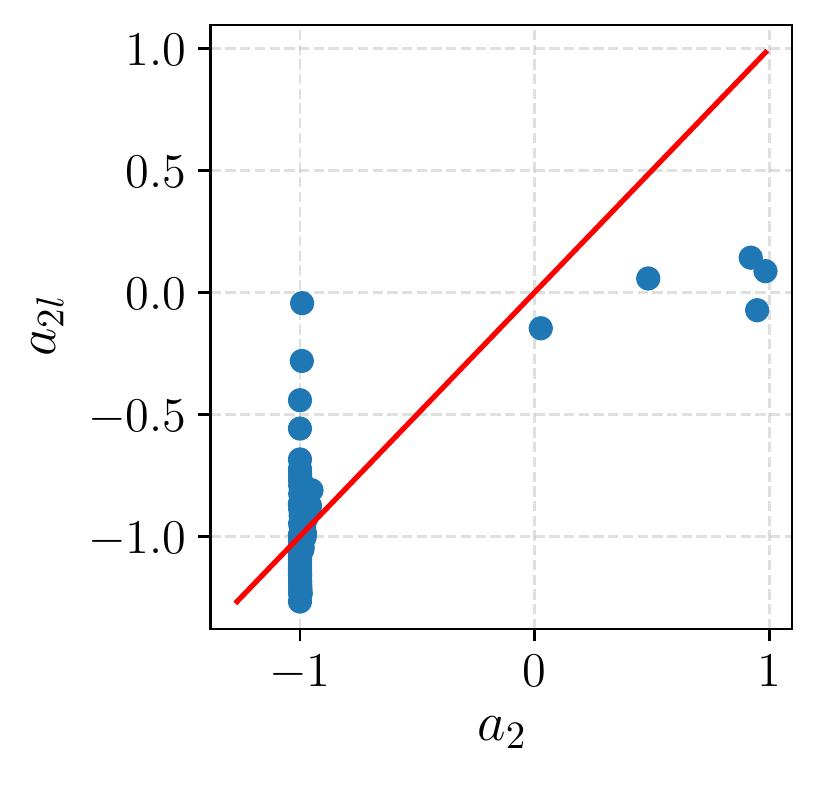} & \includegraphics[width = 0.22\textwidth]{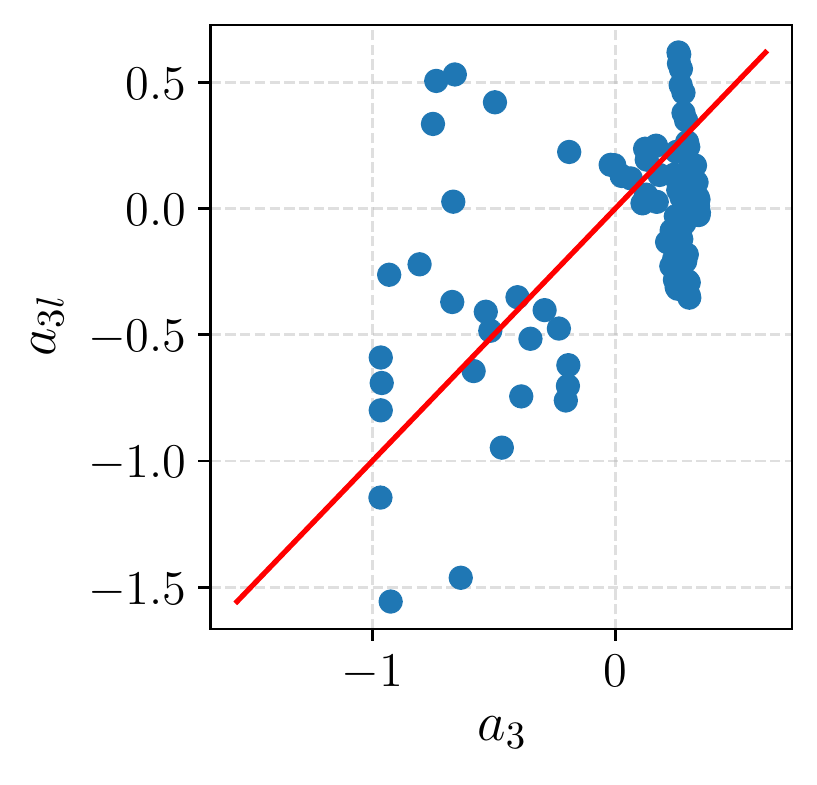} & \includegraphics[width = 0.22\textwidth]{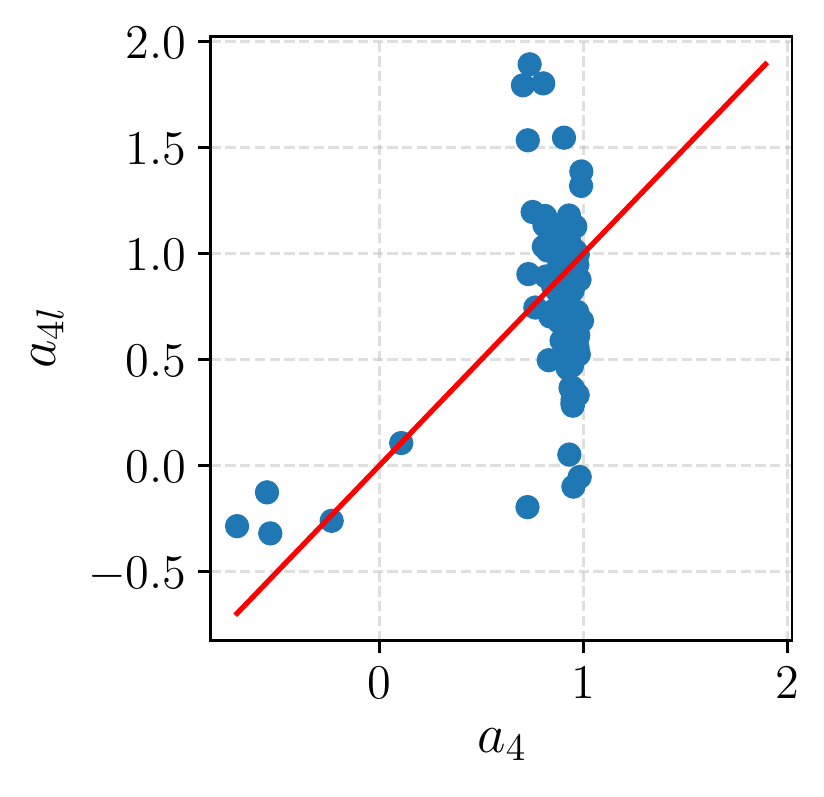}\\
		\multicolumn{4}{c}{\hspace{5mm}\textbf{GP}}\\\midrule
		\includegraphics[width = 0.22
		\textwidth]{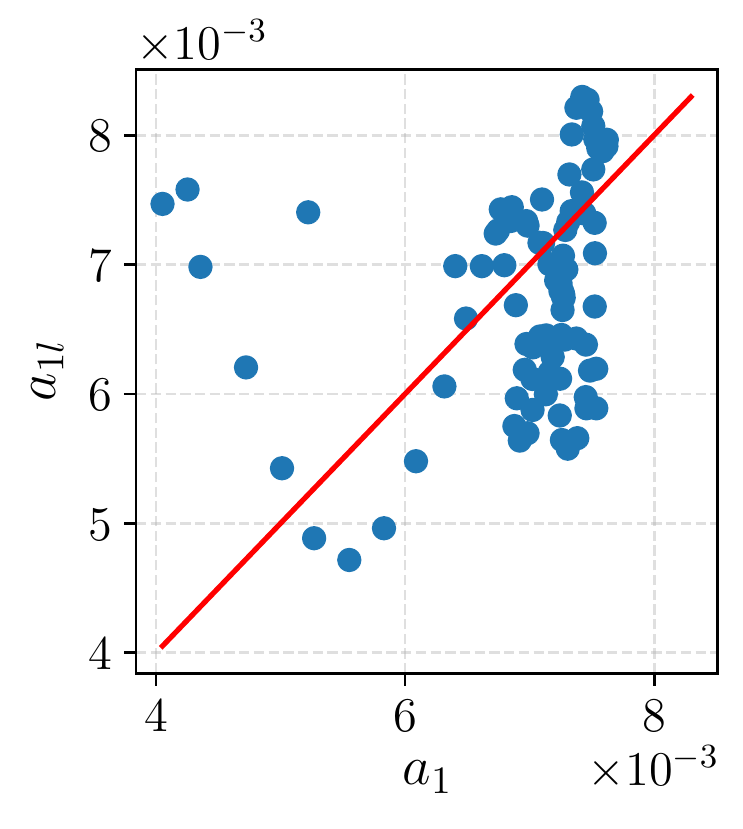} & \includegraphics[width = 0.22   \textwidth]{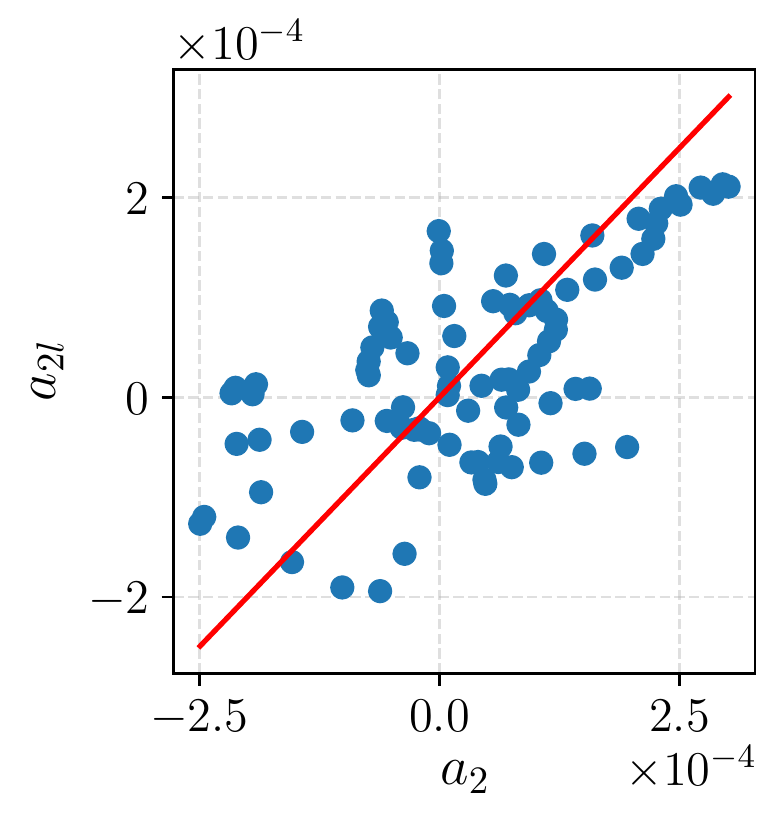} & \includegraphics[width = 0.22\textwidth]{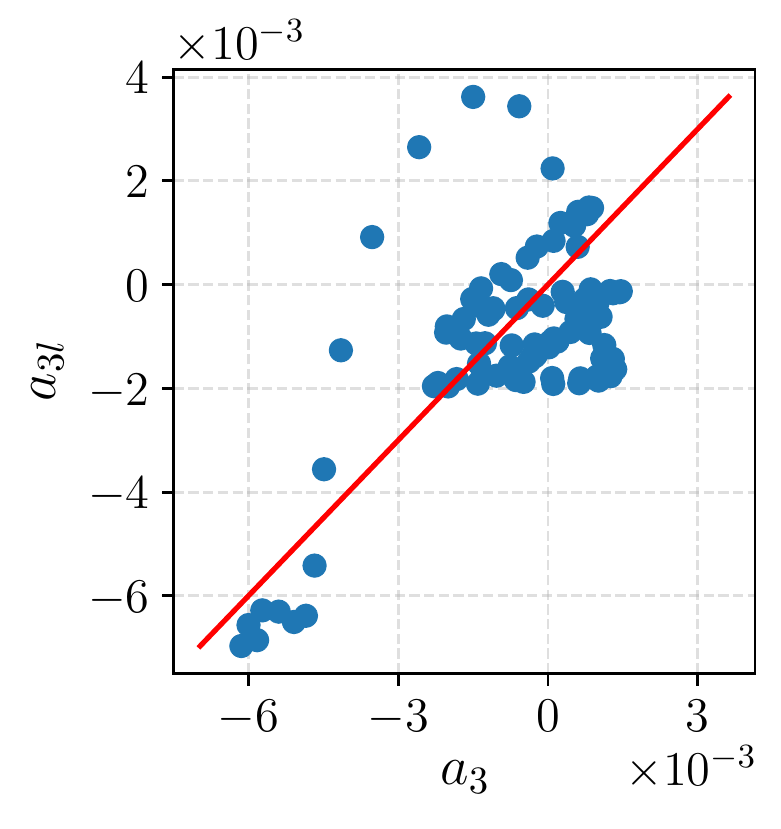} & \includegraphics[width = 0.22\textwidth]{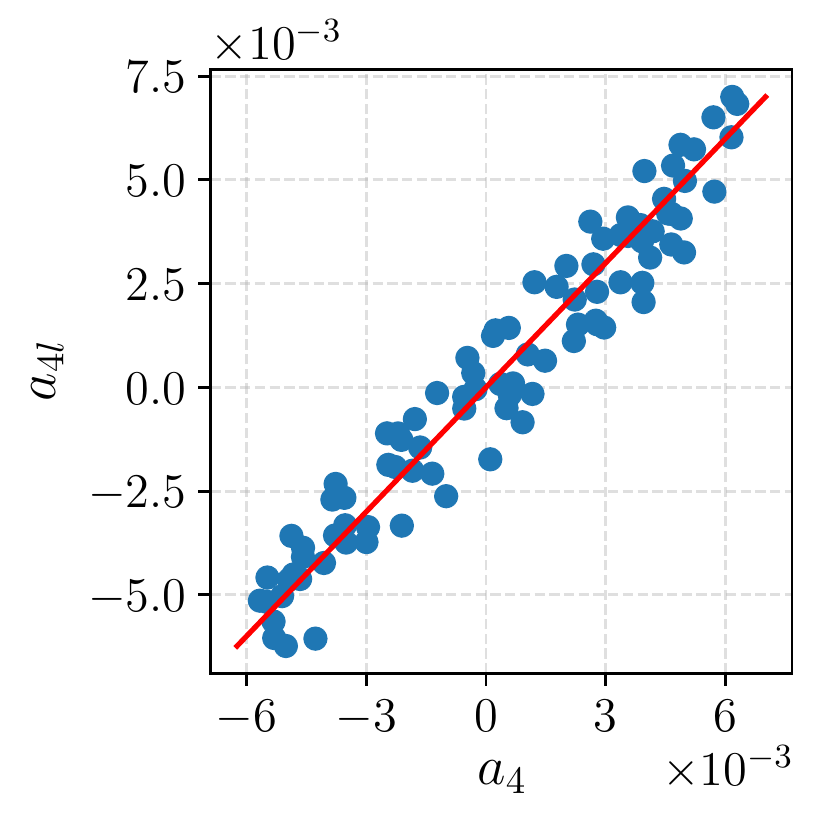}
	\end{tabular}
	\caption{Comparison of the optimal actions of the DDPG and GP (x axis) with their linearized version (y axis) for the four jets, the red line is the bisector of the first and third quadrant.}
	\label{comp_opt_action_vs_linearized_cylinder}
\end{table*}

\begin{table*}
	\centering
	\begin{tabular*}{\textwidth}{cc}
		\toprule  \textbf{Mean Value} & \textbf{Standard Deviation} \\ \midrule
		\multicolumn{2}{c}{\textbf{Baseline}} \\
		($\overline{C}_D=3.2$,\; $\overline{C}_L=-0.02$) & ($\sigma_{C_D}=0.2$,\; $\sigma_{C_L}=2$)
		\\\toprule
		\includegraphics[width = 0.5 \textwidth]{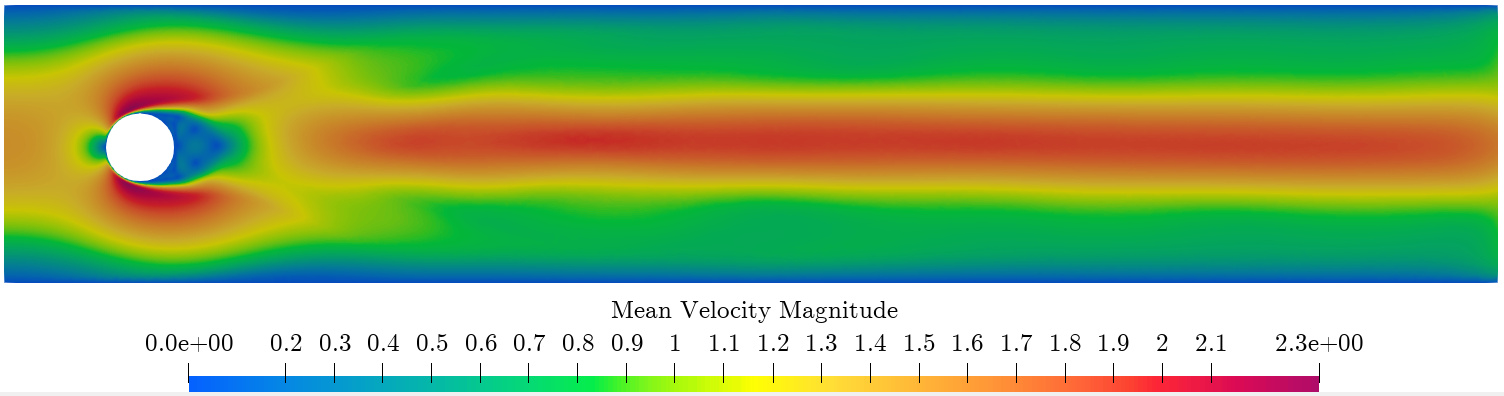} &  \includegraphics[width = 0.5 \textwidth]{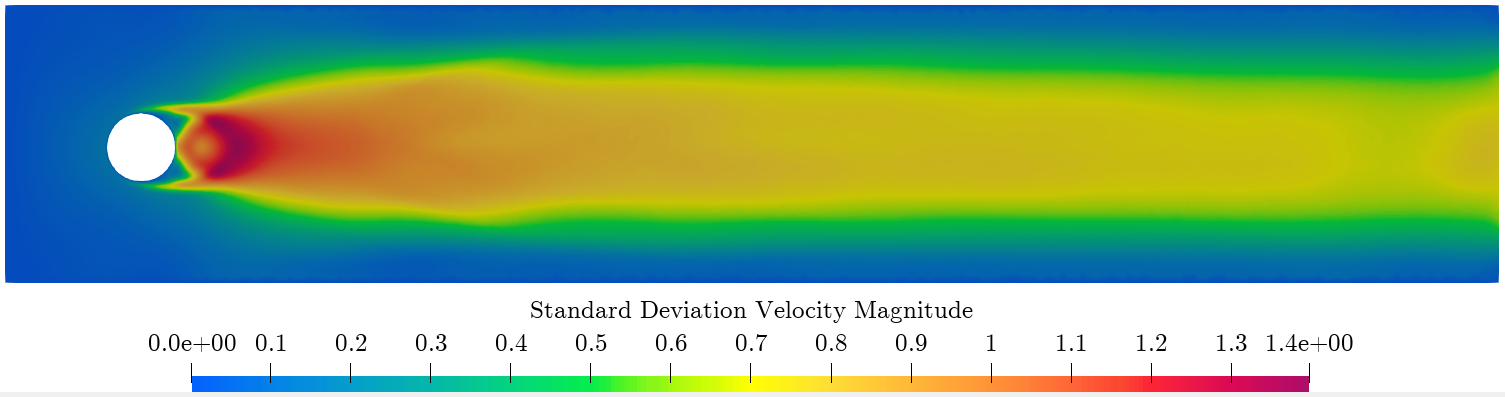} \\
		\multicolumn{2}{c}{\textbf{LIPO}} \\ 
		($\overline{C}_D=2.1$,\; $\overline{C}_L=0.9$) & ($\sigma_{C_D}=0.2$,\; $\sigma_{C_L}=1.1$)
		\\\toprule
		\includegraphics[width = 0.5 \textwidth]{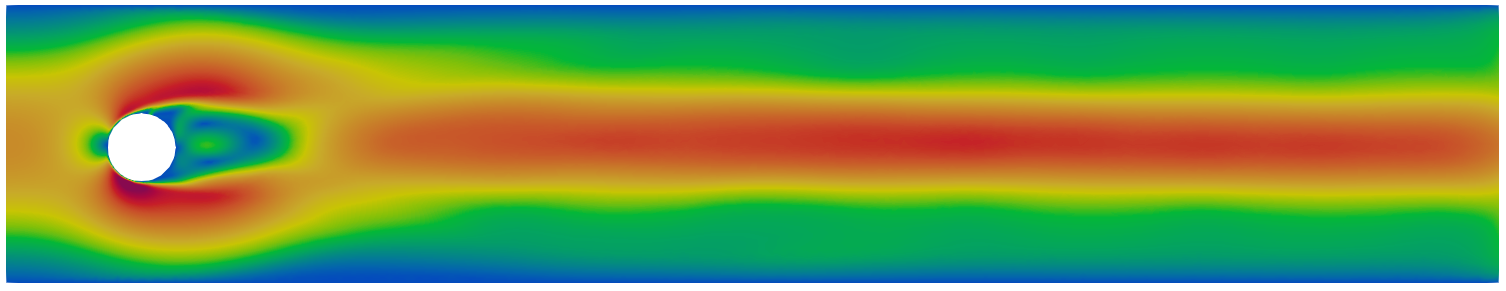} &  \includegraphics[width = 0.5 \textwidth]{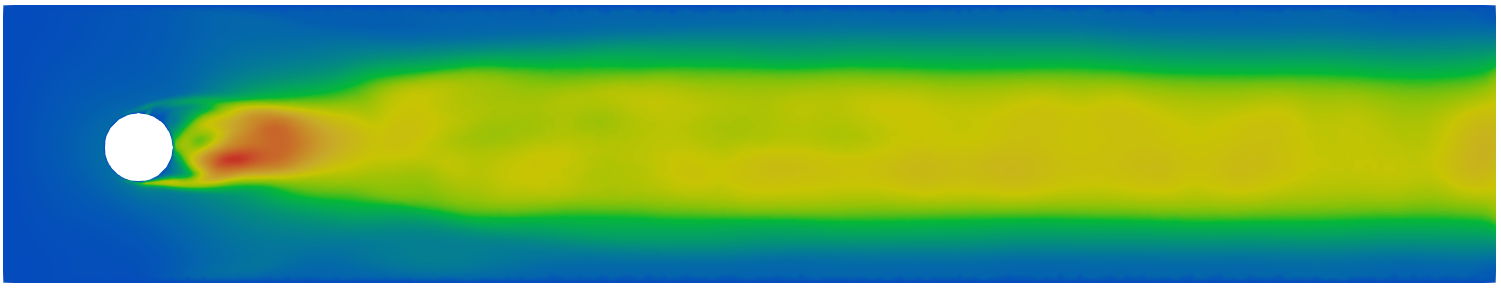} \\ 
		\multicolumn{2}{c}{\textbf{BO}} \\
		($\overline{C}_D=2.1$,\; $\overline{C}_L=1.13$) & ($\sigma_{C_D}=0.2$,\; $\sigma_{C_L}=0.9$)
		\\ \toprule
		\includegraphics[width = 0.5 \textwidth]{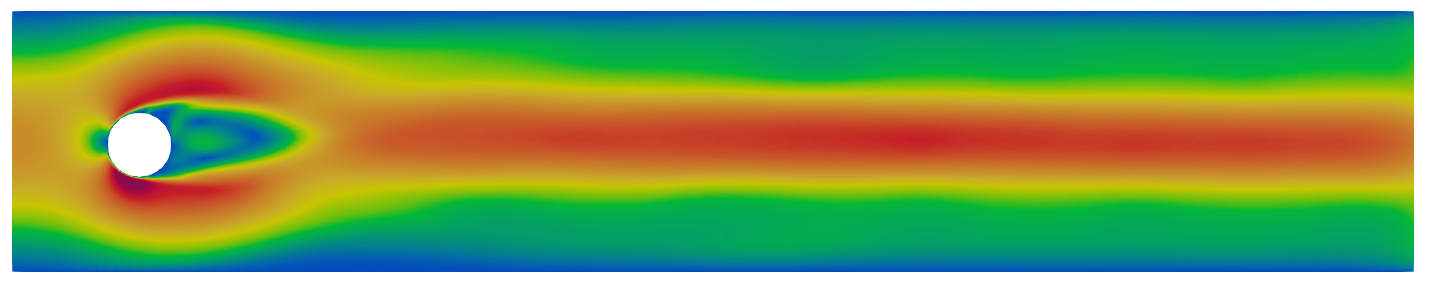} &  \includegraphics[width = 0.5 \textwidth]{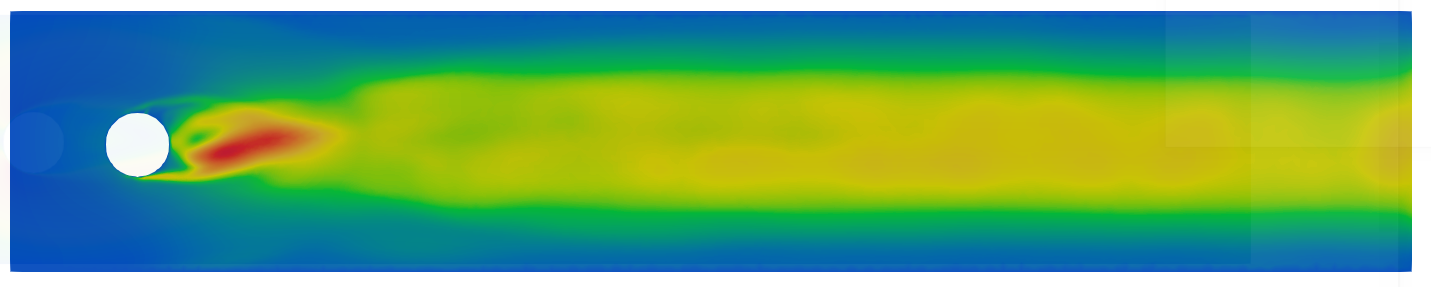} \\ 
		\multicolumn{2}{c}{\textbf{GP}} \\
		($\overline{C}_D=1.9$,\; $\overline{C}_L=0.6$) & ($\sigma_{C_D}=0.2$,\; $\sigma_{C_L}=0.6$)
		\\ \toprule
		\includegraphics[width = 0.5 \textwidth]{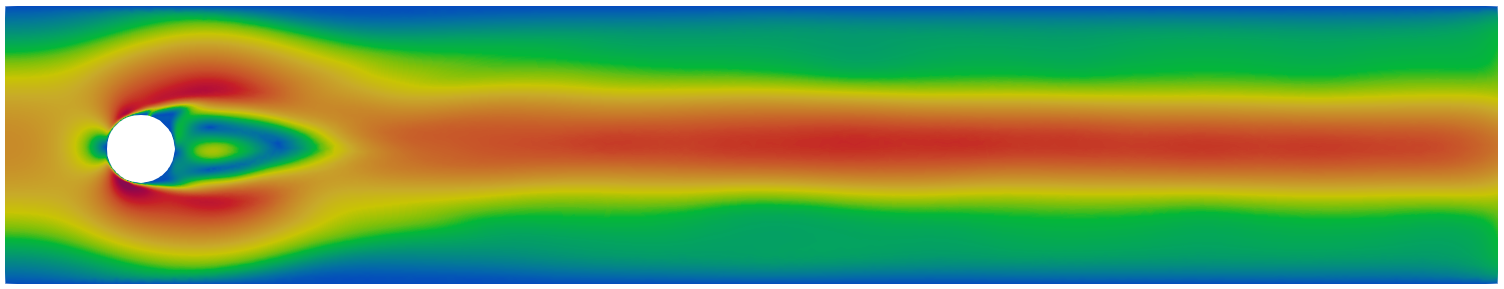} &  \includegraphics[width = 0.5 \textwidth]{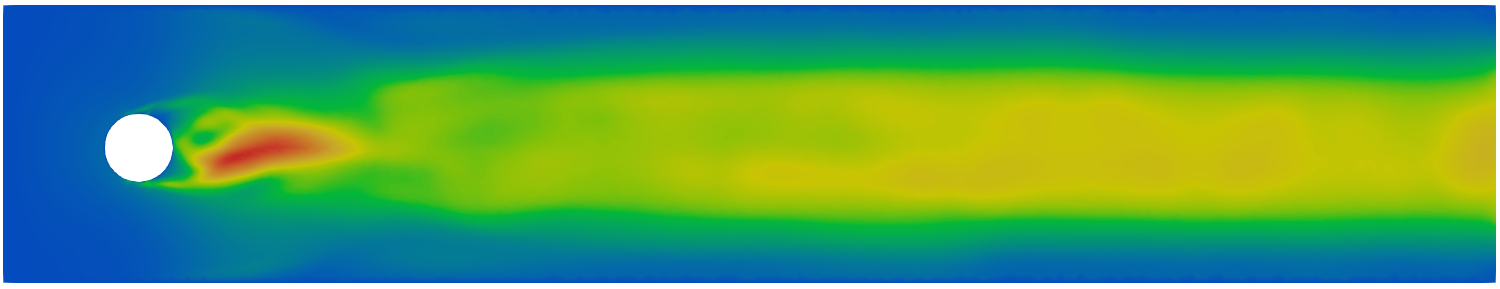} \\ 
		\multicolumn{2}{c}{\textbf{DDPG}} \\
		($\overline{C}_D=2.34$,\; $\overline{C}_L=-1.44$) & ($\sigma_{C_D}=0.29$,\; $\sigma_{C_L}=1.54$)
		\\ \toprule
		\includegraphics[width = 0.5 \textwidth]{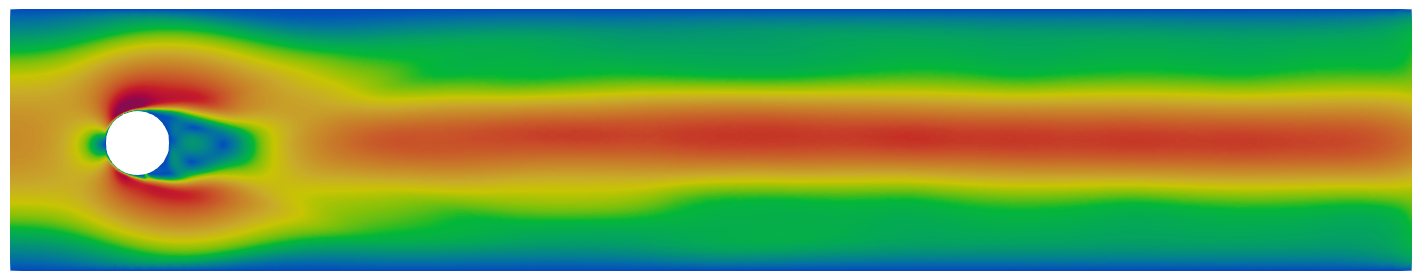} & \includegraphics[width = 0.5 \textwidth]{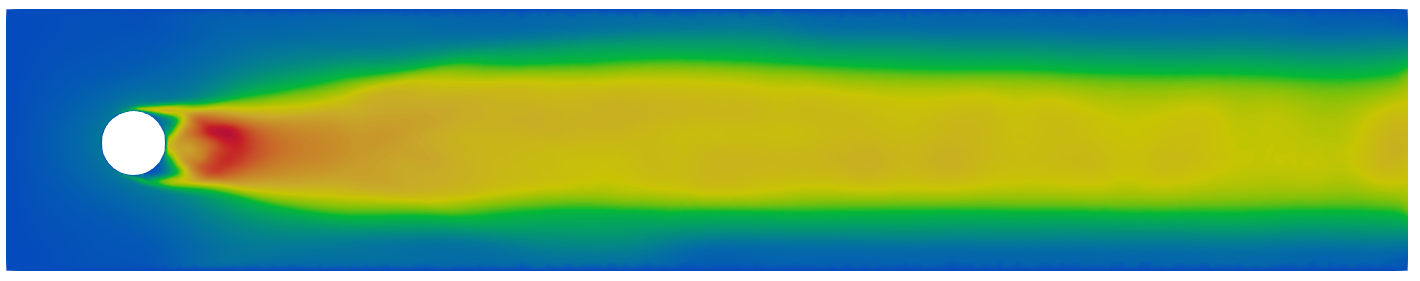} \\ 
		
	\end{tabular*}
	\caption{{\color{black}{Mean flow (left) and standard deviation (right)  using the best control action found by the different methods. The mean lift ($\overline{C}_L$) and drag ($\overline{C}_D$) are averaged over the last two uncontrolled vortex shedding periods.}} }
	\label{paraview_cylinder}
\end{table*}

On the other hand, the large exploration by the GP requires more than 300 episodes to outperform the other methods. LIPO and BO had similar trends, with an almost constant rate of improvement. This suggests that the surrogate models used in the regression are particularly effective in approximating the expected cumulative reward.

The DDPG follows a similar trend, but slightly worse performances and larger variance. The large model capacity of the ANN, combined with the initial exploratory phase, tend to set the DDPG on a bad initial condition. {\color{black}The exploratory phase is only partially responsible for the large variance, as one can see from the learning curve variance for $\mbox{ep}>20$ (see \eqref{Sec:IIIp3}), when the exploitation begins: although a step is visible, the variance remains high.} 

Despite the low variance in the reward, the BO and LIPO finds largely different weights for the linear control functions, as shown in Fig.\ref{cylinder_weights_comp_LIPO_BO}. This implies that fairly different strategies leads to comparable rewards, and hence the problem admits multiple optima.
In general, the identified linear law seeks to compensate the momentum deficit due to the vortex shedding by injecting momentum with the jets on the opposite side. {\color{black}For example, in the case of BO, the injection $q_4$ is strongly linked to the states $s_1$, $s_2$, $s_5$, laying on the lower half plane. In the case of LIPO, both ejections $q_1$ and $q_4$ are consistently linked to the observation in $s_5$, on the back of the cylinder, with the negligible uncertainty and highest possible weight. } 

Figure \ref{action_comparison_cylinder} show the time evolution of the four actions (flow rates) and (line red, the evolution of the instantaneous drag coefficient. Probably due to the short duration of the episode, none of the controllers identifies a symmetric control law. LIPO and BO, despite the different weights' distribution, find an almost identical linear combination. They both produce a small flow rate for the second jet and larger flow rates for the first, both in the initial transitory and in the final stages. As the shedding is reduced and the drag coefficient drops, all flow rates tends to a constant injection for both BO and LIPO, while the GP keep continuous pulsations in both $q_4$ and $q_3$ (with opposite signs). 

\begin{figure}
	\includegraphics[width=\textwidth]{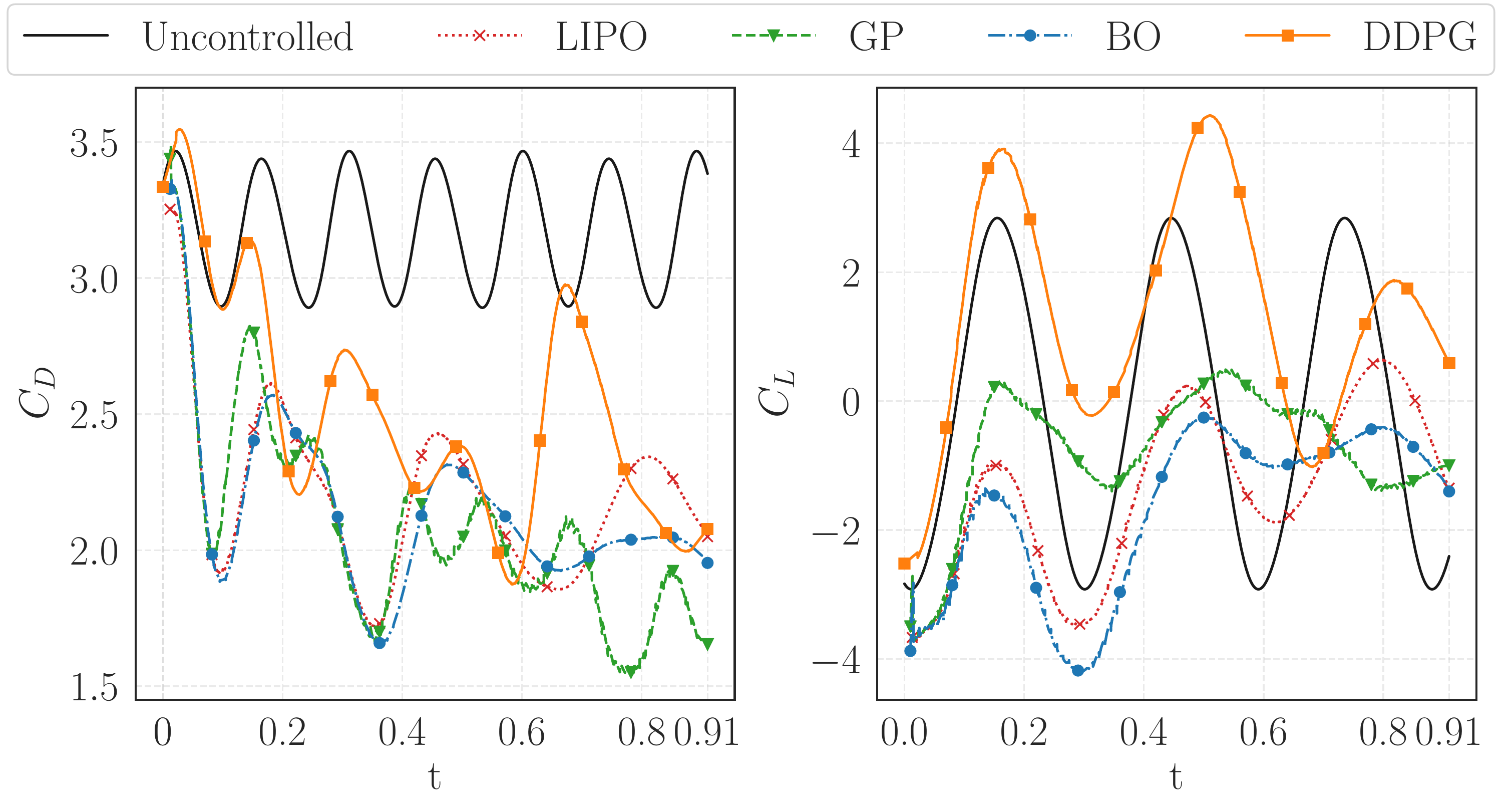}
	\caption{{\color{black}Comparison between the controlled and the uncontrolled $C_D$ and $C_L$ evolutions using the best policies found by the different methods.}}\label{Control_cyl}
\end{figure}

All the control methods leads to satisfactory performances, with a mitigation of the von Kármán street and a reduction of the drag coefficient, also visible by the increased size of the recirculation bubble in the wake. {\color{black}The evolution of the drag and lift coefficients are shown in Figure \ref{Control_cyl} for the uncontrolled and the controlled test cases.} The mean flow and standard deviation for the baseline and for the best strategy identified by the four techniques is shown in {\color{black}\underline{Table}~} \ref{paraview_cylinder}, which also reports the average drag and lift coefficients along with their standard deviation across various episodes. {\color{black} An animation of the flow field controlled by all agents is provided in the supplementary material.}


To analyze the degree of nonlinearity in the control laws derived by the GP and the DDPG,  we perform a linear regression with respect to the evolution of the states as performed for the PPO agent by \cite{Tang2020e} at the opening of this section. The results are shown in Table \ref{comp_opt_action_vs_linearized_cylinder}, which compares the action taken by the DDPG (first row) and the GP (second row), in the abscissa, with the linearized actions, in the ordinate, for the four injections. None of the four injections produced by the DDPG agent can be linearized and the open-loop behavior (constant action regardless of the states) is visible. Interestingly, the action taken by the GP on the fourth jet is almost linear.

\begin{figure*} \center
	\centering
	\includegraphics[width=0.5\textwidth]{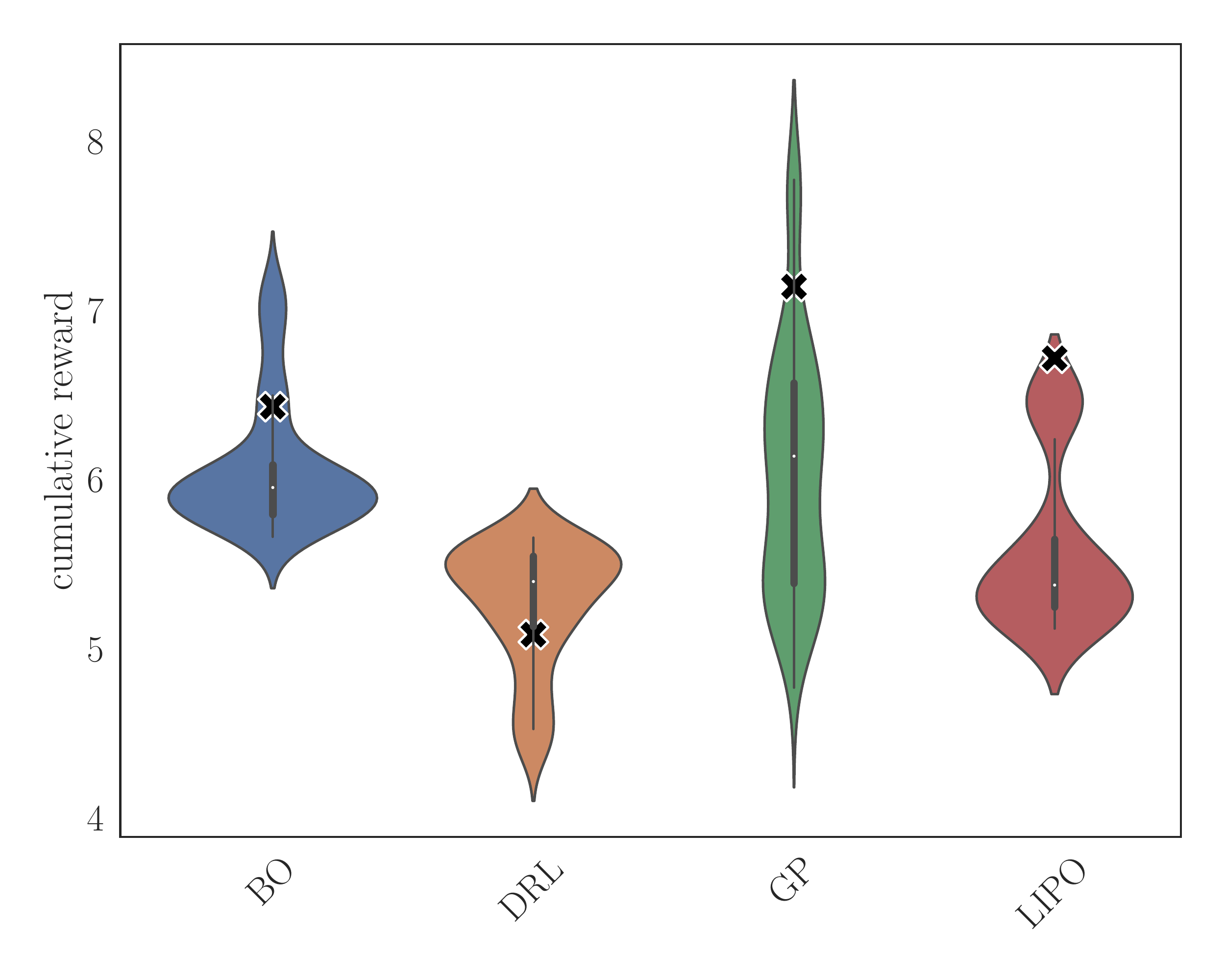}
	\caption{{\color{black}Robustness analysis of the optimal control methods with randomized initial conditions for the von
			Kármán street control problem. The violin plots represent the distribution of cumulative rewards obtained, whereas the black crosses show the best result of each controller at the end of the training phase.}}
	\label{cylinder_robusteness}
	
\end{figure*}

{\color{black}Finally, we close this section with the results of the robustness analysis tested on 100 randomly chosen initial conditions over one vortex shedding period. As for the previous test cases, these are collected in reward distribution for each agent in Figure \ref{cylinder_robusteness}. The mean results align with the learning performances (black crosses), but significantly differs in terms of variability. 
	
	Although the GP achieves the best control performances \emph{for some} initial conditions, the large distribution is a sign of overfitting, and multiple initial conditions should be included at the training stage to derive more robust controllers as done by \cite{castellanos2022machine}. While this lack of robustness might be due to the specific implementation of the multiple-output control, these results show that agents with higher model capacity in the policy are more prone to overfitting and require a broader range of scenarios during the training. As for the comparison between DDPG, BO and LIPO, who have run for the same number of episodes, it appears that the linear controller outperforms the DDPG agent both in performance and robustness. This opens the question of the effectiveness of complex policy approximators on relatively simple test cases and on whether this test case, despite its popularity, is well suited to show-case sophisticated machine learning control methods.}

\section{Conclusions and outlooks}\label{Sec:VI}

We presented a general mathematical framework linking machine learning-based control techniques and optimal control. The first category comprises methods based on `black-box optimization' such as Bayesian Optimization (BO) and Lipschitz Global Optimization (LIPO), methods based on tree expression programming such as Genetic Programming (GP), and methods from reinforcement learning such as Deep Deterministic Policy Gradient (DDPG).

We introduced the mathematical background for each method, in addition we illustrated their algorithmic implementation, {\color{black} in Appendix \ref{algo_psedo_code}}. Following the definition by \cite{Mitchell1997}, the investigated approaches are machine learning algorithms because they are designed to \emph{automatically} improve at a task (controlling a system) according to a performance measure (a reward function) with experience (i.e. data, collected via trial and errors from the environment). In its most classic formulation, the `data-driven' approach to a control problem is black-box optimization. The function to optimize measures the controller performance over a set of iterations that we call episodes. Therefore, training a controller algorithm requires (1) a function approximation to express the `policy' or `actuation law' linking the current state of the system to the action to take and (2) an optimizer that improves the function approximation episode after episode.

In Bayesian Optimization and LIPO, the function approximator for the policy is defined a priori. In this work, we consider linear or quadratic controllers, but any function approximator could have been used instead (e.g. RBF or ANN). These optimizers build a surrogate model of the performance measure and adapt this model episode by episode. In Genetic Programming, the function approximator is an expression tree, and the optimization is carried out using classic evolutionary algorithms. In {\color{black} Deep} Reinforcement Learning (DRL), particularly in the DDPG algorithm implemented in this work, the function approximation is an ANN, and the optimizer is a stochastic (batch) gradient-based optimization. In this optimization, the gradient of the cumulative reward is computed using a surrogate model of the Q-function, i.e. the function mapping the value of each state-action pair, using a second ANN.

In the machine learning terminology, we say that the function approximators available to the GP and the DDPG have a larger `model capacity' than those we used for the BO and the LIPO (linear or quadratics). This allows these algorithms to identify nonlinear control laws that are difficult to cast in the form of prescribed parametric functions. On the other hand, the larger capacity requires many learning parameters (branches and leaves in the tree expressions of the GP and weights in the ANN of the DDPG), leading to optimization challenges and possible local minima. 
\color{black}{Although it is well known that large model capacity is a key enabler in complex problems, this study shows that it might be harmful in problems where a simple control law suffices. This statement does not claim to be a general rule but rather a warning in the approach to complex flow control problems. Indeed, the larger model capacity proved particularly useful in the first two test cases but not in the third, for which a linear law proved more effective, more robust, and considerably easier to identify. In this respect, our work stresses the importance of better defining the notion of complexity of a flow control problem and the need to continue establishing reference benchmark cases of increasing complexity.} \color{black}

We compared the `learning' performances of these four algorithms on three control problems of growing complexity and dimensionality: (1) the stabilization of a nonlinear 0D oscillator, (2) the cancellation of nonlinear waves in the burgers' equation in 1D, and (3) the drag reduction in the flow past a cylinder in laminar conditions. The successful control of these systems highlighted the strengths and weaknesses of each method, although all algorithms identify valuable control laws in the three systems.

The GP achieve the best performances on both the stabilization of the 0D system and the control of the cylinder wake, while the DDPG gives the best performances on the control of nonlinear waves in the Burgers' equation. However, the GP has the poorest sample efficiency in all the investigated problems, thus requiring a larger number of interactions with the system, and has the highest learning variance, meaning that repeating the training leads to vastly different results. This behaviour is inherent to the population-based and evolutionary optimization algorithm, which has the main merit of escaping local minima in problems characterized by complex functionals. These features paid off in the 0D problem, for which the GP derives an effective impulsive policy, but are ineffective in the control of nonlinear waves in the Burgers' equation, characterized by a much simpler reward functional. 

On the other side of the scale, in terms of sample efficiency, are the black box optimizers such as LIPO and BO. Their performance is strictly dependent on the effectiveness of the predetermined policy parametrization to optimize. In the case of the 0D control problem, the quadratic policy is, in its simplicity, less effective than the complex policy derived by GP and DDPG. For the problem of drag reduction in the cylinder flow, the linear policy was rather satisfactory. To the point that it was shown that the PPO policy by \cite{Tang2020e} has, in fact, derived a linear policy. The DDPG implementation was trained using 5 sensors (instead of 236) and reached a performance comparable to the PPO by \cite{Tang2020e} in 80 episodes (instead of 800). Nevertheless, although the policy derived by our DDPG is nonlinear, its performances is worse than the linear laws derived by BO and LIPO. Yet, the policy by the DDPG is based on an ANN parametrized by $68361$ parameters ($4$ fully connected layers with $5$ neurons in the first, $256$ in the second and third and $4$ in the output) while the linear laws used by BO and LIPO only depend on $20$ parameters.

We believe that this work has shed some light (or open some paths) on two main aspects of the machine-learning-based control problem: (1) the contrast between the generality of the function approximator for the policy and the number of episodes required to obtain good control actions; (2) the need for tailoring the model complexity to control task at hand and the possibility of having a modular approach in the construction of the optimal control law. The resolution of both aspects resides in the hybridization of the investigated methods. 

Concerning the choice of the function approximator (policy parametrization or the 'hypothesis set' in the machine learning terminology), both ANN and expression {\color{black} trees} offer large modelling capacities, with the seconds often outperforming the first in the authors' experience. Intermediate solutions such as RBFs or Gaussian processes can provide a valid compromise between model capacity and dimensionality of their parameter space. They should be explored more in the field of flow control.

Finally, concerning the dilemma `model complexity versus task complexity', a possible solution could be increasing the complexity modularly. For example, one could limit the function space in the GP by first taking linear functions and then enlarging it modularly, adding more primitives. Or, in a hybrid formalism, one could first train a linear or polynomial controller (e.g. via LIPO or BO) and then use it to pre-train models of larger complexity (e.g. ANNs or expression trees) in a supervised fashion, or to assist their training with the environment (for instance by inflating the replay buffer of the DDPG with transitions learned by the BO/LIPO models). 

This is the essence of `behavioural cloning', in which a first agent (called 'demonstrator') trains a second one (called 'imitator') offline so that the second does not start from scratch. This is unexplored territory in flow control and, of course, opens the question of how much the supervised training phase should last and whether the pupil could ever surpass the master.

\appendix 

\section{Algorithms' pseudocodes}
\label{algo_psedo_code}
\subsection{BO pseudocode}
\label{BO_pseudocode}

Algorithm \ref{Alg_BO_1} reports the main steps of the Bayesian Optimization through Gaussian Process. Lines (1-9) defines the GPr predictor function, which takes in input the sampled points $\mathbf{W}^*$, the associated cumulative rewards $\mathbf{R}^*$, the testing points $\mathbf{W}$, and the Kernel function $\kappa$ in eq. \eqref{Matern}. This outputs the mean value of the prediction  $\mathbf{\mu_{*}}$ and its variance ${\Sigma_{*}}$. The algorithm starts with the initialization of the simulated weights $\mathbf{W}^*$ and rewards $\mathbf{R}^*$ buffers (line 10 and 11). Prior to start the optimization, 10 random weights $\mathbf{W}^0$ are tested (line 12 and 13). Within the optimization loop, at each iteration, 1000 random points are passed to the GPr predictor, which is also fed with the weight and rewards buffers (line 16 and 17) to predict the associated expected reward and variance for each weight combination. This information is then passed to an acquisition function (line 17) which outputs a set of values $\mathbf{A}$ associated to the weights $\mathbf{W}^+$. The acquisition function is then optimized to identify the next set of weights (line 19). Finally, the best weights are tested in the environment (line 20) and the buffers updated (line 21 and 22).

\subsection{LIPO pseudocode}
\label{LIPO_pseudocode}

The algorithm \ref{Alg_LIPO} reports the keys steps of the MaxLIPO+TR method. First, a function \textproc{globalsearch} function is defined (line 1). This performs a random global search of the parametric space if the random number selected from $S=\{x\in\mathbb{R}|\;0\leq x\leq1\}$ is smaller than $p$ (line 3), otherwise it proceeds with MaxLIPO. In our case $p=0.02$, hence the random search is almost negligible. The upper and lower bound ($\mathbf{U},\mathbf{L}$) of the search space are defined in line 10. A buffer object, initialized as empty in line 11, logs the weights $\mathbf{w}_i$ and their relative reward $R(\mathbf{w}_i)$ along the optimization. Within the learning loop (line 17), the second and third weights are selected randomly (line 19). Then, if the iteration number $k$ is even, the algorithm selects the next weights via \textproc{globalsearch} (line 23), else it relies on the local optimization method (line 31). If the local optimizer reaches an optimum within an accuracy of $\epsilon$ (line 33), the algorithm continues exclusively with \textproc{globalsearch}. At the end of each iteration, both the local and the global models are updated with the new weights $\mathbf{w}_{k+1}$ (line 38 and 39).

\subsection{GP pseudocode}
\label{GP_pseudocode}

Algorithm \ref{Alg_GP} shows the relevant steps of the learning process. First, an initial population of random individuals (i.e. candidate control policies) is generated and evaluated (lines 1 and 2) individually. An episode is run for each different tree structure.
The population, with their respective rewards (according to eq.\ref{REW}), is used to generate a set of $\lambda$ offspring individuals.
The potential parents are selected via \emph{tournament}, where new individuals are generated cross-over (line 9), mutation (line 12) and replication (line 15): each of the new member of the population has a probability $p_c$, $p_m$ and $p_r$ to arise from any of these three operations, hence $p_c+p_m+p_r=1$.

The implemented \textit{cross-over strategy} is the \emph{one-point} cross-over: two randomly chosen parents are first broken around one randomly selected cross-over point, generating two trees and two subtrees. Then, the offspring is created by replacing the subtree rooted in the first parent with the subtree rooted at the cross-over point of the second parent. Of the two offsprings, only one is considered in the offspring and the other is discarded. The \textit{mutation strategy} is a \emph{one-point} mutation, in which a random node (sampled with from a uniform distribution) is replaced with any other possible node from the primitive set. The \textit{replication strategy} consists in the direct cloning of one randomly selected parent to the next generation.

The tournament was implemented using the $(\mu+\lambda)$ approach,in which both parents and offsprings are involved; this is contrast with the $(\mu,\lambda)$, in which only the offsprings are involved in the process. The new population is created by selecting the best individuals, based on the obtained reward, among the old population $\mathbf{B}^{(i-1)}$ and the offspring $\tilde{\mathbf{B}}$ (line 19).

\subsection{DDPG pseudocode}
\label{DDPG_pseudocode}

We recall the main steps of the DDPG algorithm in algorithm \ref{Alg_DDPG}. After random initialization of the weights in both network and the initialization of the replay buffer (lines 1-3), the loop over episodes and time steps proceeds as follows. The agent begins from an initial state (line 5), which is simply the final state of the system from the previous episode or the last state from the uncontrolled dynamics. In other words, none of the investigated environments has a terminal state and no re-inizialitation is performed.

Within each episode, at each time step, the DDPG takes actions (lines 7-12) following \eqref{act} (line 8) or repeating the previous action (line 10). After storing the transition in the replay buffer (lines 13), these are ranked based on the associated TD error $\delta$ (line 14). This is used to sample a batch of $N$ transitions following a triangular distribution favouring the transitions with the highest $\delta$. The transitions are used to compute the cost functions $J^Q (\mathbf{w}^{Q})$ and $J^\pi (\mathbf{w}^{\pi})$ and their gradients $\partial_{\mathbf{w}^{Q}} J (\mathbf{w}^{\pi})$, $\partial_{\mathbf{w}^{\pi}} J (\mathbf{w}^{\pi})$ and thus update the weights following a gradient ascent (lines 17 and 19). This operation is performed on the `current networks' (defined by the weights $\mathbf{w}^{\pi}$ and $\mathbf{w}^{Q}$). However, the computation of the critic losses $J^{Q}$ is performed with the prediction $\mathbf{y}_t$ from the target networks (defined by the weights $\mathbf{w}^{\pi '}$ and $\mathbf{w}^{Q '}$). The targets are under-relaxed updates of the network weights computed at the end of each episode (lines 21-22). 

The reader should notice that, differently from the other optimization-based approaches, the update of the policy is performed \emph{at each time step} and not at the end of the episode.

In our implementation, we used the Adam optimizer for training the ANN's with a learning rate of $10^{-3}$ and $2 \cdot 10^{-3}$ for the actor and the critic, respectively. The discount factor was set to $\gamma = 0.99$ and the soft-target update parameters is $\tau = 5 \cdot 10^{-3}$. For what concerns the neural networks' architecture, the hidden layers used the rectified non-linear activation function, while the actor output was bounded relying on a hyperbolic tangent (tanh). The actor's network was $n_s$x$256$x$256$x$n_a$, where $n_s$ is the number of states and $n_a$ is the number of actions expected by the environment. Finally, the critic's network concatenates two networks. The first, from the action taken by the agent composed as $n_a$x$64$. The states are elaborated in two layers of size $n_s$x$32$x$64$. These are concatenated and expanded by means of two layers with $256$x$256$x$1$, neurons, where the output is the value estimated. 

\begin{algorithm}[!h]
	\caption{Bayesian Optimization using GPr, adapted from \cite{CarlEdward} and \cite{scikit-learn}}
	\begin{algorithmic}[1]
		\Function{predictor}{$\mathbf{W}^*,\mathbf{R}^*,\mathbf{W},\kappa$}
		\State Compute $\mathbf{K}\leftarrow \kappa(\mathbf{W},\mathbf{W})$
		\State Compute $\mathbf{K}_{**} \leftarrow \kappa(\mathbf{W}^*,\mathbf{W}^*)$
		\State Compute $\mathbf{K}_R\leftarrow \mathbf{K}_{**} + \sigma_{\mathbf{w}_*}\mathbf{I}$
		\State Compute Cholesky decomposition $\mathbf{L}\leftarrow \mathbf{K}_R$
		\State Compute $\alpha\leftarrow \mathbf{L}^{T}\mathbf{L}^{-1}R^{*}$
		\State Compute $\mathbf{v}\leftarrow \mathbf{L}\mathbf{K}^{-1}$
		\State \Return mean $\mu_{*}\leftarrow \mathbf{K}\alpha$ and variance $\Sigma_*\leftarrow \mathbf{K} - \mathbf{v}^T\mathbf{v}$
		\EndFunction
		\State Initialize weight buffer $\mathbf{W}^*$ as null
		\State Initialize function buffer $\mathbf{R}^*$ as null
		\State Initialize a set of 10 random weights $\mathbf{W}^0$ \State Collect reward from simulation $\mathbf{R}^0\leftarrow \mathbf{R}(\mathbf{W}^0)$
		\State Add rewards and weights to buffers $\mathbf{R}^*\leftarrow \mathbf{R}^0$ and $\mathbf{W}^*\leftarrow \mathbf{W}^0$
		\For {$k$ in (1,$N$)}
		\State Select 1000 random points $\mathbf{W}^{+}$
		\State Evaluate points $(\mu_{*},\Sigma_*)\leftarrow$ \textproc{predictor}$(\mathbf{W}^*,\mathbf{R}^*,\mathbf{W}^{+},\kappa)$
		\State Compute ($A,W^{+}$)$\leftarrow$\textproc{acqfunction}($(\mu_{*},\Sigma_*)$)
		\State $\mathbf{w}^k\leftarrow\underset{\mathbf{w}^{\dagger}}{\mathrm{argmin}}\;$\textproc{acqfunction}($\mathbf{w}^{\dagger}$)
		\State Collect reward from simulation $\mathbf{R}^k\leftarrow \mathbf{R}(\mathbf{w}^k)$
		\State Add result to buffers $\mathbf{R}^*\leftarrow \mathbf{R}^k$ and $\mathbf{W}^*\leftarrow \mathbf{w}^k$
		\EndFor
		\label{Alg_BO_1}
	\end{algorithmic}
\end{algorithm}

\begin{algorithm}[!htbp]
	\begin{algorithmic}[1]
		\Function{GloabalSearch}{}
		\If {{\color{black}$x\sim \mathcal{U}(S) > p$}}
		\State Select weights $\mathbf{w}$ based on MaxLIPO (Eq.\eqref{EXP_LIPO})
		\Else
		\State Select weights $\mathbf{w}$ randomly
		\EndIf
		\State Evaluate reward function $R(\mathbf{w})$
		\State \Return ($\mathbf{w}$, $R(\mathbf{w})$)
		\EndFunction 
		
		\State Define upper $\mathbf{U}$ and lower $\mathbf{L}$ weights' bounds
		\State Initialize buffer structure $\textbf{W}$ as empty
		\State Initialize weights as $\mathbf{w}_0 = (\mathbf{U} + \mathbf{L})/2$
		\State Evaluate reward function $R(\mathbf{w}_0)$
		\State Initialize the best weight and reward $(\mathbf{w}^*,R^*)\leftarrow (\mathbf{w}_0,R(\mathbf{w}_0))$
		\State Add weights and reward to the buffer $\mathbf{W}(\mathbf{w}_0,R(\mathbf{w}_0))$
		\State Initialize \textit{flag}$\leftarrow$False
		\For {$k$ in (1,$N_e$-1)}
		\If{$k$ < 3}
		\State Select weights $\mathbf{w}_{k}$ randomly
		\State Evaluate reward function $R(\mathbf{w}_{k})$
		\Else
		
		\If{\textit{flag} = True}
		\State $\mathbf{w}_{k}$, $R(\mathbf{w}_{k})\leftarrow$ \textproc{GloabalSearch}()
		\If{$R(\mathbf{w}_{k}) > R^*$}
		\State Set \textit{flag}$\leftarrow$False
		\EndIf    
		\Else
		\If{k mod 2 = 0}
		\State $\mathbf{w}_{k}$, $R(\mathbf{w}_{k})\leftarrow$ \textproc{GloabalSearch}()
		\Else
		\State Select weights $\mathbf{w}_{k}$ based on TR (Eq.\eqref{TR_approx})
		\State Evaluate reward function $R(\mathbf{w}_{k})$
		\If{$|R(\mathbf{w}_{k}) - R^*| < \epsilon$ (Eq.\eqref{disc_LIPO})}
		\State Set \textit{flag}$\leftarrow$True
		\State \textbf{continue}
		\EndIf
		\EndIf
		\State Update upper bound $U(\mathbf{w})$ with $\mathbf{w}_{k}$(Eq.\eqref{LIPO_SURR})
		\State Update TR ($m(\mathbf{w};\mathbf{w}^*)$  Eq.\eqref{TR_approx})
		\EndIf
		\EndIf
		\If{$R(\mathbf{w}_{k}) > R^*$}
		\State Update ($\mathbf{w}^*,R^*)\leftarrow(\mathbf{w}_{k},R(\mathbf{w}_{k}))$
		\EndIf\\
		\EndFor \\
		\textbf{EndFor}
		\caption{MaxLIPO + TR (Adapted from \cite{dlib09})}\label{Alg_LIPO}
	\end{algorithmic}
\end{algorithm}

\begin{algorithm}[!htbp]
	\begin{algorithmic}[1]
		\State Initialize population $\mathbf{B}^{(0)}$ with $\mu$ random individuals $\mathbf{a}_i$.
		\State Evaluate fitness $\mathbf{a}_i\leftarrow(\mathbf{w}_i,R(\mathbf{w}_i))$
		\For {$i$ in (1,$N_e$)}
		\State Initialize offspring population $\tilde{\mathbf{B}}$ with $\lambda$ individuals as empty.
		\For {$t$ in (1,$\lambda$)}
		\State Select random number $\zeta\in(0,1)$
		\If{$\zeta$ < $p_c$}
		\State Random sample two individuals ($\mathbf{a}_m$,$\mathbf{a}_n$) from $\mathbf{B}^{(i-1)}$
		\State  Compute offspring individual $\tilde{\mathbf{a}}_i$ $\leftarrow$ \textbf{Mate}($\mathbf{a}_m$,$\mathbf{a}_n$)
		\ElsIf{$\zeta$ < ($p_c + p_m$)}
		\State Random sample an individual ($\mathbf{a}_m$) from $\mathbf{B}^{(i-1)}$
		\State Compute offspring individual $\tilde{\mathbf{a}}_i$ $\leftarrow$ \textbf{Mutate}($\mathbf{a}_m$)
		\Else
		\State Random sample an individual ($\mathbf{a}_m$) from $\mathbf{B}^{(i-1)}$
		\State Compute offspring individual $\tilde{\mathbf{a}}_i$ $\leftarrow$ $\mathbf{a}_m$
		\EndIf
		\EndFor
		
		\State Evaluate fitness of mated and mutated $\mathbf{\tilde{a}}_i\leftarrow(\mathbf{w}_i,R(\mathbf{w}_i))$
		\State Update population $\mathbf{B}^{(i)}$ $\leftarrow$ \textbf{Select}($\mathbf{B}^{(i)}$,$\tilde{\mathbf{B}}$,$\mu$)
		\EndFor
		\caption{GP ($\mu,\lambda$)-ES (Adapted from \cite{article_GP_algo})}\label{Alg_GP}
	\end{algorithmic}
\end{algorithm}

\begin{algorithm}[!htbp]
	\begin{algorithmic}[1]
		\State Initialize $Q(\mathbf{s},\mathbf{a};\mathbf{w}^q)$ and $\pi(\mathbf{s};\mathbf{w}^{\pi})$ with random $\mathbf{w}^q$ and $\mathbf{w}^{\pi}$.
		\State Initialize targets $\mathbf{w}^{Q'}\leftarrow \mathbf{w}^{Q}$ and $\mathbf{w}^{\pi'}\leftarrow \mathbf{w}^{\pi}$.\
		\State Initialize replay Buffer $\mathcal{R}$ as empty. 
		\For {\mbox{ep} in (1,$n_E$)}
		\State Observe initial state $\mathbf{s}_0$
		\For {$t$ in (1,T)}
		\If{$t=1 $ or $mod(t, K)=0$}
		\State $\mathbf{a}_t=\mathbf{a}(\mathbf{s}_t;\mathbf{w}^{\pi})+\eta(\mbox{ep})\mathcal{N}(t;\theta,\sigma)$
		\Else
		\State $\mathbf{a}_{t}=\mathbf{a}_{t-1}$
		\EndIf
		\State Execute $\mathbf{a}_t$, get $r_t$ and observe $\mathbf{s}_{t+1}$
		\State Store the transitions $(\mathbf{s}_t,\mathbf{a}_t,r_t,\mathbf{s}_{t+1})$ in $R$
		\State Rank the transition by TD error $\delta$
		\State Select $N$ transitions in $R$, favouring the highest $\delta$
		\State Compute $y_t=r_t+\gamma Q'(\mathbf{s}_t,\pi(\mathbf{s}_t,\mathbf{w}^{\pi '}))$ 
		\State Compute $J^Q=\mathbb{E}(y_t-Q(\mathbf{s}_t,\pi(\mathbf{s}_t,\mathbf{w}^{\pi})))$ and $\partial_{\mathbf{w}^{Q}} J^Q$ 
		\State Update $\mathbf{w}^{Q} \leftarrow \mathbf{w}^{Q} +\alpha_q\partial_{\mathbf{w}^{Q}} J^Q $
		\State Compute $J^{\pi}(\mathbf{w}^{\pi'} )$ and $\partial_{\mathbf{w}^{\pi'} }
		J^{\pi}$
		\State Update $\mathbf{w}^{\pi}\leftarrow \mathbf{w}^{\pi} +\alpha_q\partial_{\mathbf{w}^{\pi'}} J^{\pi} $
		\State Update targets in Q:  $\mathbf{w}^{Q'}\leftarrow \tau \mathbf{w}^{Q'}+(1-\tau)\mathbf{w}^{Q'}$
		\State Update targets in $\pi$: $\mathbf{w}^{\pi '}\leftarrow \tau \mathbf{w}^{\pi'}+(1-\tau)\mathbf{w}^{\pi'}$
		\EndFor
		\EndFor
		\caption{DDPG (Adapted from \cite{Lillicrap2015})}\label{Alg_DDPG}\end{algorithmic}
\end{algorithm}

\FloatBarrier

\newpage

\section{Weights identified by the BO and LIPO}

The tables below collects the weights for the linear and nonlinear policies identified by LIPO and BO for the three investigated control problems. The reported value represents the mean of ten optimization with different random conditions and the uncertainty is taken as the standard deviation.

\begin{landscape}
	
	\begin{table*}
		\centering 
		\begin{tabular*}{\linewidth}{@{\extracolsep{\stretch{1}}}*{11}{c}@{}}
			\toprule     & $w_1$         & $w_2$        & $w_3$         & $w_4$        & $w_5$         & $w_6$         & $w_7$         & $w_8$         & $w_9$         & $w_{10}$        \\\midrule
			LIPO & \textbf{1.13}$\;\pm$2.23 & \textbf{0.26}$\;\pm$2.28 & \textbf{1.94}$\;\pm$1.67 & \textbf{2.73}$\;\pm$0.39 & \textbf{0.6}$\;\pm$2.47 & \textbf{0.5}$\;\pm$2.25 & \textbf{-0.03}$\;\pm$2.66 & \textbf{-0.24}$\;\pm$2.12 & \textbf{0.23}$\;\pm$2.39 & \textbf{-1.52}$\;\pm$1.57\\
			
			BO   & \textbf{0.2}$\;\pm$1.83 & \textbf{-0.1}$\;\pm$1.58 & \textbf{1}$\;\pm$1.26  & \textbf{2}$\;\pm$0.77 & \textbf{-0.4}$\;\pm$1.36  & \textbf{0}$\;\pm$1.61 & \textbf{0.7}$\;\pm$1.55  & \textbf{0}$\;\pm$2.14  & \textbf{0.2}$\;\pm$1.66 & \textbf{0.1}$\;\pm$ 1.04 \\\bottomrule
			\\
			\addlinespace[0.05cm]\toprule
			& $w_{11}$        & $w_{12}$       & $w_{13}$        & $w_{14}$       & $w_{15}$        & $w_{16}$        & $w_{17}$        & $w_{18}$        & $w_{19}$        & $w_{20}$        \\\midrule
			LIPO & \textbf{-0.18}$\;\pm$2.3 & \textbf{-0.7}$\;\pm$2.43 & \textbf{-0.32}$\;\pm$1.6 & \textbf{0.27}$\;\pm$2.25 & \textbf{-0.58}$\;\pm$2.14 & \textbf{0.19}$\;\pm$2.26 & \textbf{-0.12}$\;\pm$2.56 & \textbf{1.05}$\;\pm$2.26 & \textbf{-0.17}$\;\pm$1.98 & \textbf{-0.57}$\;\pm$2.72 \\
			
			BO   & \textbf{0.2}$\;\pm$1.72 & \textbf{0}$\;\pm$1.26 & \textbf{-0.8}$\;\pm$0.75 & \textbf{0.4}$\;\pm$1.69 & \textbf{-0.1}$\;\pm$1.45 & \textbf{-0.6}$\;\pm$0.8 & \textbf{0.5}$\;\pm$1.5 & \textbf{1.8}$\;\pm$1.25  & \textbf{-0.3}$\;\pm$1.73  & \textbf{0.3}$\;\pm$2.19\\\bottomrule
		\end{tabular*}
		\caption{Mean value and half standard deviation of the 0D feedback control law coefficients}
		\label{0D_weights_solutions_LIPO_BO}
		\vspace{9mm}
		\centering 
		\begin{tabular}{c@{\hspace{0.5cm}}c@{\hspace{0.4cm}}c@{\hspace{0.4cm}}c}
			\toprule    & $w_1$         & $w_2$        & $w_3$    \\\midrule
			LIPO & \textbf{-0.02}($\;\pm$0.01) & \textbf{0.03}($\;\pm$0.03) & \textbf{-0.03}($\;\pm$0.02)\\
			BO   & \textbf{-0.02}($\;\pm$0.00) & \textbf{0.02}($\;\pm$0.01) & \textbf{-0.03}($\;\pm$0.00)\\\bottomrule
		\end{tabular}
		\caption{Mean value and half standard deviation of the Burgers' feedback control law coefficients}
		\label{burgers_weights_solutions_LIPO_BO}
		\vspace{9mm}
		\begin{tabular*}{\linewidth}{@{\extracolsep{\stretch{1}}}*{11}{c}@{}}
			\toprule     & $w_1$         & $w_2$        & $w_3$         & $w_4$        & $w_5$         & $w_6$         & $w_7$         & $w_8$         & $w_9$         & $w_{10}$        \\\midrule
			LIPO & \textbf{-0.29}$\;\pm$0.69 & \textbf{-0.48}$\;\pm$0.67 & \textbf{-0.12}$\;\pm$0.74 & \textbf{0.40}$\;\pm$0.62 & \textbf{0.23}$\;\pm$0.84 & \textbf{0.30}$\;\pm$0.80 & \textbf{-0.38}$\;\pm$0.71 & \textbf{-0.47}$\;\pm$0.76 & \textbf{-0.64}$\;\pm$0.47 & \textbf{-0.21}$\;\pm$0.70\\
			
			BO  & \textbf{-0.64}$\;\pm$0.44 & \textbf{-0.2}$\;\pm$0.78 & \textbf{0.17}$\;\pm$0.81 & \textbf{0.7}$\;\pm$0.43 & \textbf{0.46}$\;\pm$0.8 & \textbf{0.59}$\;\pm$0.57 & \textbf{-0.26}$\;\pm$0.62 & \textbf{-0.28}$\;\pm$0.72  & \textbf{-0.4}$\;\pm$0.67  & \textbf{-0.32}$\;\pm$0.75\\\bottomrule
			\\
			\addlinespace[0.05cm]\toprule
			& $w_{11}$        & $w_{12}$       & $w_{13}$        & $w_{14}$       & $w_{15}$        & $w_{16}$        & $w_{17}$        & $w_{18}$        & $w_{19}$        & $w_{20}$        \\\midrule
			LIPO & \textbf{0.38}$\;\pm$0.67 & \textbf{0.89}$\;\pm$0.2 & \textbf{-0.15}$\;\pm$0.72 & \textbf{0.47}$\;\pm$0.55 & \textbf{-0.11}$\;\pm$0.80 & \textbf{-0.22}$\;\pm$0.65 & \textbf{-0.99}$\;\pm$0.04 & \textbf{-0.53}$\;\pm$0.39 & \textbf{0.76}$\;\pm$0.32 & \textbf{1.0}$\;\pm$0.0 \\
			
			BO   & \textbf{0.44}$\;\pm$0.75 & \textbf{0.61}$\;\pm$0.69 & \textbf{0.11}$\;\pm$0.81 & \textbf{0.31}$\;\pm$0.62 & \textbf{-0.26}$\;\pm$0.75 & \textbf{-0.5}$\;\pm$0.73 & \textbf{-0.4}$\;\pm$0.92 & \textbf{0.23}$\;\pm$0.73  & \textbf{0.73}$\;\pm$0.43  & \textbf{0.4}$\;\pm$0.92\\\bottomrule
		\end{tabular*}
		\caption{Mean value and half standard deviation of the von Karman vortex street feedback control law coefficients}
		\label{cylinder_weights_solutions_LIPO_BO}
	\end{table*}
\end{landscape}

\bibliographystyle{jfm}
\bibliography{Fabio_et_al_2022}

\end{document}